%% file: wd_ire-v16.tex
\documentclass[useAMS,usenatbib]{mn2e}
\usepackage{graphicx,lscape,amssymb}

\newcommand{\Teff}{\mbox{$\mathrm{T}_{\mathrm{eff}}$}}
\newcommand{\Logg}{\mbox{$\log(g)$}}
\newcommand{\Mwd}{\mbox{$\mathrm{M}_{\mathrm{wd}}$}}
\newcommand{\Rwd}{\mbox{$\mathrm{R}_{\mathrm{wd}}$}}
\newcommand{\Msun}{\mbox{$\mathrm{M}_{\odot}$}}
\def\gsim{ \lower .75ex \hbox{$\sim$} \llap{\raise .27ex \hbox{$>$}} }
\def\lsim{ \lower .75ex \hbox{$\sim$} \llap{\raise .27ex \hbox{$<$}} }
\newcommand{\SDS}{7444} 
\newcommand{\SDP}{9341} 
\newcommand{\UKS}{1990} 
\newcommand{\UKP}{1771} 
\newcommand{\UKDA}{1275} 
\newcommand{\Comp}{95.4} 
\newcommand{\Effi}{62.3} 
\newcommand{\QSOrem}{89.6} 

\title{DA white dwarfs in SDSS DR7 \textit{and} a search for infrared
excess emission}

\author[J.Girven et al.]{
J. Girven$^1$, B. T. G\"ansicke$^1$, D. Steeghs$^1$, D. Koester$^2$\\
$^{1}$ Department of Physics, University of Warwick, Coventry CV4 7AL,
UK \\
$^{2}$ Institut f\"{u}r Theoretische Physik und Astrophysik, University of Kiel,
24098 Kiel, Germany\\ %
}

\begin{document}

\date{Started 2009}

\maketitle

\label{firstpage}

\begin{abstract}
We present a method which uses colour-colour cuts on SDSS photometry to select
white dwarfs with hydrogen rich (DA) atmospheres without the recourse to
spectroscopy. This method results in a sample of DA white dwarfs that is
95\% complete at an efficiency of returning a true DA white dwarf of 62\%. The
approach was applied to SDSS Data Release 7 for objects with and without SDSS
spectroscopy. This lead to 4636 spectroscopicially confirmed DA white
dwarfs with $g\leq19$; a $\sim70\%$ increase compared to
\citeauthor{eisensteinetal06-1}'s \citeyear{eisensteinetal06-1} sample.
Including the photometric-only objects, we estimate a factor of 3 increase
in DA white dwarfs. We find that the SDSS spectroscopic follow-up is
$44\%$ complete for DA white dwarfs with $\Teff\gtrsim8000$K. We further
cross-correlated the SDSS sample with Data Release 8 of the UKIDSS Large
Area Survey. The spectral energy distributions of both subsets, with and
without SDSS spectroscopy, were fitted with white dwarf models to
determine the fraction of DA white dwarfs with low-mass stellar companions
or dusty debris discs via the detection of excess near-infrared emission.
From the spectroscopic sample we find that $2.0\%$ of white dwarfs have an
excess consistent with a brown dwarf type companion, with a firm lower
limit of $0.8\%$. From the white dwarfs with photometry only, we find that
$1.8\%$ are candidates for having brown dwarf companions. Similarly, both
samples show that $\sim1\%$ of white dwarfs are candidates for having a
dusty debris disc.
\end{abstract}

\begin{keywords}
Stars: individual: white dwarfs -- infrared excess
\end{keywords}

\section{Introduction}
\label{s-int}

White dwarfs are the most common stellar remnants in the Galaxy, having
descended from main sequence stars with $0.8\Msun\lsim M\lsim8\Msun$. Of
these, hydrogen rich (DA) white dwarfs make up the vast majority. Large
samples of white dwarfs are particularly useful in many studies, for
example constraining the luminosity function, which in turn can be used to
determine the ages of many Galactic populations
\citep[e.g.][]{wingetetal87-1, oswaltetal96-1, degennaroetal08-1}. The low
luminosity of white dwarfs also makes them ideal targets for searches of
low-mass companions, such as pioneered by \citet{probst+oconnel82-1}. The
companion mass distribution is thought to drop near the low-mass end
($\sim0.1\Msun$), and the fraction of FGK stars with substellar companions
estimated from radial velocity surveys is $\la1$\%
\citep[e.g.][]{marcy+butler00-1, grether+lineweaver06-1}, though
\citet{metchev+hillenbrand09-1} suggest that substellar companions are
more frequent at larger orbital separations. Because white dwarfs are the
progeny of main-sequence stars with masses of up to 8\,\Msun, studies of
white dwarf binaries can probe the companion mass function over a wide
range of (initial) host star masses. Currently, only four white dwarfs are
confirmed to have (non-interacting) substellar companions
(\citealt{becklin+zuckerman88-1, farihi+christopher04-1, maxtedetal06-1,
steeleetal09-1}, but see \citealt{lumanetal11-1} for a very low-mass
candidate), and the fraction of white dwarfs with brown dwarf companions
appears to be consistent with the low number found around FGK stars
\citep{farihietal05-1, hoardetal07-1}.

While white dwarfs are also excellent candidates for searches of Jovian planets
\citep[e.g.][]{ignace01-1, burleighetal02-1}, no planet has yet been
unambiguously discovered around a white dwarf (\citealt{burleighetal08-1,
hoganetal09-1, mullallyetal09-1}). However, $\sim15$ white dwarfs are known to
exhibit infrared flux excesses that can not be explained by the presence of
low-mass (planetary or substellar) companions \citep{farihietal09-1}. The first
such infrared excess was found around G\,29-38 and tentatively associated with a
brown dwarf companion \citep{zuckerman+becklin87-1}. Using the fact that
G\,29-38 is a pulsating white dwarf, \citet{grahametal90-1} demonstrated that
the observations are best explained with the presence of circumstellar dust
heated by the white dwarf, and proposed the disruption of a large asteroid as
the origin of the dust. \citet{koesteretal97-1} detected metals in the
photosphere of G\,29-38, demonstrating that the white dwarf is accreting
circumstellar material. The scenario of the tidal disruption of rocky asteroids
has been refined by \citet{jura03-1}, and is now the generally accepted
interpretation of the growing number of metal-polluted white dwarfs with
infrared excess. \citet{farihietal09-1} estimate that $1-3$\% of all white
dwarfs with cooling ages below 0.5\,Gyr should exhibit infrared excess related
to the tidal disruption of asteroids. The location of the debris material within
the white dwarf tidal disruption radius for a rocky asteroid has ultimately been
confirmed by \citet{gaensickeetal06-3} who detected Doppler-broadened Ca\,II
emission lines from a gaseous component of the circumstellar disc in
SDSSJ122859.93+104032.9, a metal-polluted white dwarf with infrared excess
\citep{brinkworthetal09-1}.

In this paper, we use the Sloan Digital Sky Survey (SDSS) to first
identify a method to select hydrogen-rich (DA) white dwarfs using
photometric criteria, and then determine those with infrared flux excess
in the UKIRT Infrared Deep Sky Survey (UKIDSS). Our method is
sensitive to unresolved M and L-type companions and to warm debris discs.
The long-term goal of this project is to increase the number of such
objects available for detailed follow-up studies, and to provide a more
detailed understanding of the frequency of both substellar companions to
and planetary debris discs around white dwarfs.

The structure of the paper is as follows: In Sect.\,\ref{s-surv} we
discuss the three surveys that are utilised in this paper; the Sloan
Digital Sky Survey, the UKIRT Infrared Deep Sky Survey and the Wide-field
Infrared Survey Explorer (WISE). The method for selecting DA white dwarfs
from SDSS DR7 is outlined in Sect.\,\ref{ss-damet} along with a discussion
of the spectroscopic completeness of SDSS  with respect to DA white
dwarfs. The procedure for cross-matching objects from SDSS with UKIDSS is
demonstrated in Sect.\,\ref{s-uki}. Identifying which white dwarfs show an
IR excess and the method for fitting these excesses is given in
Sect.\,\ref{s-ire}. Further to this, Sect.\,\ref{ss-comp} discusses
comparisons between the spectroscopic and photometric fitting methods,
including how contaminants were dealt with. A summary of the overall
numbers of objects at each stage of the process is given in
Sect.\,\ref{ss-sum}. Some notes on interesting objects discovered in this
work are given in Sect.\,\ref{s-wdwire}. A comparison between the work
herein and the white dwarf--main sequence star catalogue of
\citet{rebassa-mansergasetal10-1} is made in Sect.\,\ref{s-wdms}. Objects
that are detected in WISE are discussed in Sect.\,\ref{s-wise}. The
implications of the DA white dwarf selection and the IR excess selection
methods are discussed in Sect.\,\ref{s-dis}. Finally, the conclusions of
this work are drawn together in Sect.\,\ref{s-con}.

\section{Large area surveys}
\label{s-surv}

The Sloan Digital Sky Survey (SDSS, \citealt{yorketal00-1,
  stoughtonetal02-1}) and the UKIRT Infrared Deep Sky Survey (UKIDSS,
\citealt{lawrenceetal07-1}) are currently the deepest large-area
optical and infrared surveys that are publicly available, and we
summarise below details of both surveys relevant for our work.

\subsection{SDSS}
\label{ss-sdss}

We have made use of the SDSS Data Release seven (DR7,
\citealt{abazajianetal09-1}), which represents the final DR of the
SDSS\,II project, including the low-latitude extension SEGUE
\citep{yannyetal09-1}. SDSS DR7 provides $ugriz$ photometry for 357
million objects spanning a magnitude range $\sim15-22$ and covering
$11500\,\deg^2$, approximately one-quarter of the celestial sphere, as
well as follow-up low-resolution ($R\simeq1800$, $3800-9200$\,\AA)
spectroscopy for 1.44 million galaxies, quasars, and stars. We limit
our search to point-like sources, and to apparent magnitudes $g\le19$,
which leaves $\simeq24$ million unique photometric and 87,000 unique
spectroscopic objects. An additional important resource within DR7 are
proper motions computed from the USNO-B and SDSS positions.

\subsection{UKIDSS}
\label{ss-ukidss}

\begin{figure}
\includegraphics[angle=90,width=\columnwidth]{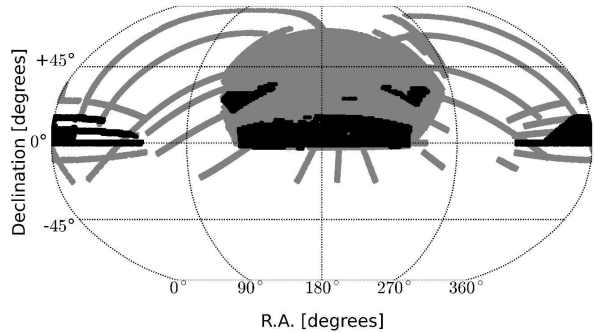}
\caption{\label{f-coverage} Coverage of the SDSS DR7 (grey) and UKIDSS DR8
LAS (black) in equatorial coordinates.}
\end{figure}

UKIDSS is a set of five near infrared surveys being undertaken with
the Wide Field Camera (WFCAM) instrument on the United Kingdom
Infrared Telescope (UKIRT) in Hawaii.  One of the five sub-surveys,
the Large Area Survey (LAS), aims to be the IR counterpart to the
SDSS. Here we match SDSS hydrogen-dominated (DA) white dwarfs with
UKIDSS LAS as opposed to 2MASS used in similar studies
\citep[e.g.][]{hoardetal07-1} because a large majority of the SDSS
white dwarfs are too faint in the IR for 2MASS to pick up. UKIDSS LAS
will eventually provide imaging over $4028\,\deg^2$ in four broad band
colours, $Y$, $J$, $H$, and $K$, with limiting (Vega) magnitudes of
20.2, 19.6, 18.8 and 18.2, respectively, which adds a significant
increase in depth over 2MASS. Here, we made use of UKIDSS DR8 (see
\citealt{dyeetal06-1}). The overlap between SDSS DR7 and UKIDSS/LAS
DR8 is illustrated in Fig.\,\ref{f-coverage}, and amounts to
$\sim2700$ square degrees.

\subsection{WISE}
\label{ss-wise}

The Wide-field Infrared Survey Explorer (WISE) is a NASA Medium-class
Explorer mission designed to survey the entire sky in four infrared
wavelengths, $3.4$, $4.6$, $12$, and $22\mu$m \citep{wrightetal10-1}.
WISE consists of a 40 cm telescope that images all four bands
simultaneously every 11 s. It covers nearly every part of the sky a
minimum of eight times, ensuring high source reliability, with more
coverage at the ecliptic poles. Astrometric errors are less than 0.5
arcsec with respect to 2MASS \citep{wrightetal10-1}. The preliminary
estimated S/N = 5 point source sensitivity on the ecliptic is 0.08, 0.1,
0.8, and 5 mJy in the four bands \citep[assuming eight exposures per
band;][]{wrightetal10-1}. Sensitivity improves away from the ecliptic due
to denser coverage and lower zodiacal background. We took advantage of
the preliminary data release (PDR). The SDSS white dwarfs that are the
subject of this work are largely too faint to be seen in the $12$, and
$22\mu$m bands and therefore we generally only discuss the $3.4$ and
$4.6\mu$m fluxes.

\section{Selecting DA white dwarfs in SDSS DR7}
\label{ss-damet}

\begin{table*}
\caption{\label{e-poly} Colour selection for finding DA white dwarfs in
$ugriz$ space. Objects were selected to be primary objects and point
sources. Flags are shown in Hexagesimal notation. These are the
standard SDSS ``good photometry'' flags, as documented on SDSS
\textsc{CASJOBS} \citep{li+thakar08-1}, and were chosen so that
the object was detected in BINNED1, and did not have any of the following:
EDGE, NOPROFILE, PEAKCENTER, NOTCHECKED, PSF\_FLUX\_INTERP, SATURATED, or
BAD\_COUNTS\_ERROR.}
\begin{tabular}{lcl}
\hline
Colour & constraint & \\
\hline
$(u-g)$ & $\ge$ & $-20.653 \times (g-r)^5 + 10.816 \times (g-r)^4 + 15.718
\times
(g-r)^3 - 1.294 \times (g-r)^2 - 0.084 \times (g-r) + 0.300$\\
$(u-g)$ & $\le$ & $-24.384 \times (g-r)^5 - 19.000 \times (g-r)^4 + 3.497
\times
(g-r)^3 + 1.193 \times (g-r)^2 + 0.083 \times (g-r) + 0.610$\\
$(g-r)$ & $\le$ & $-0.693 \times (r-i)^2 + 0.947 \times (r-i) + 0.192$\\
$(g-r)$ & $\ge$ & $ -1.320 \times (r-i)^3 + 2.173 \times (r-i)^2 + 2.452
\times (r-i) - 0.070$\\
$(r-i)$ & $\ge$ & $-0.560$\\
$(r-i)$ & $\le$ & $0.176 \times (i-z) + 0.127$\\
$(r-i)$ & $\le$ & $-0.754 \times (i-z) + 0.110$\\
$g$     & $\le$ & 19 \\
$0$     & $!=$  & flags \& 0x10000000 \\
$0$     & $=$   & flags \& 0x8100000c00ac \\
\hline
\end{tabular}
\end{table*}

The latest catalogue of spectroscopicially identified SDSS white dwarfs
was based on DR4 \citep{eisensteinetal06-1}, which comprised roughly
half of the sky coverage of DR7. Here, we exploit the much
larger footprint of SDSS~DR7, and also extend the white dwarf sample
to photometric objects without follow-up spectroscopy. We restricted
our ambitions to DA white dwarfs for a number of reasons. Firstly, the
vast majority of all known white dwarfs belong to the DA class
\citep{mccook+sion99-1}. Secondly, determining the atmospheric
parameters of DA white dwarfs, \Teff\ and \Logg, from fitting
atmosphere models to either spectroscopy \citep{bergeronetal92-1} or
photometry \citep{koesteretal79-1} is a well-established and robust
procedure. This is essential for the purpose of identifying white
dwarfs with infrared excess, as we need to accurately extrapolate the
white dwarf flux to the $J, H$, and $K$ bands. Thirdly, optical
spectra of DA white dwarfs are characterised by strong Balmer
absorption lines on a blue continuum, and the strong dependence of the
Balmer line equivalent widths results in DA white dwarfs occupying a
distinct region in colour space.

We have developed a two-pronged approach to identify as many DA white
dwarfs with spectroscopy within DR7, and subsequently to select white
dwarf candidates which have $ugriz$ photometry but were not
spectroscopically followed-up by SDSS.

\begin{figure*}
\begin{minipage}{2\columnwidth}
\includegraphics[angle=0,width=\columnwidth]{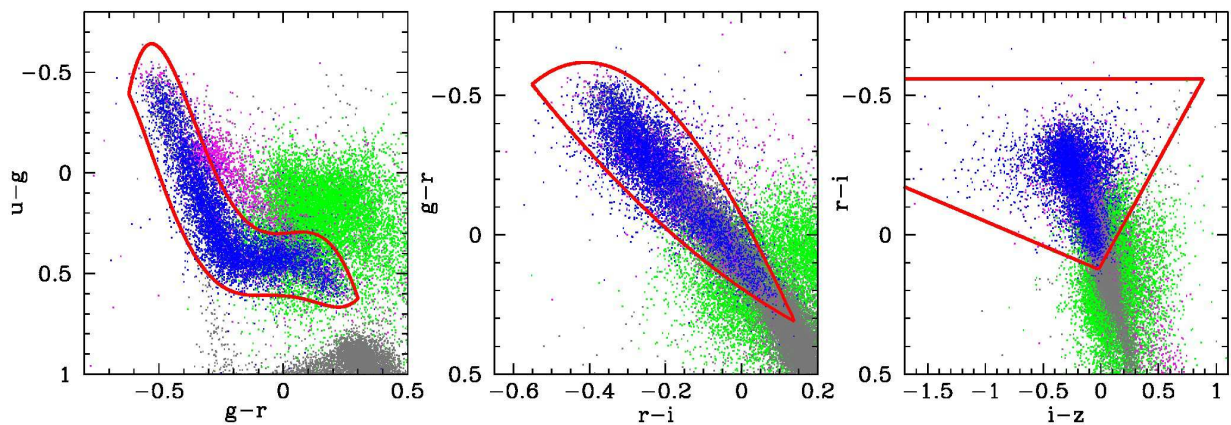}
\caption{\label{f-colsel} Colour-colour plots illustrating the location of
the SDSS spectroscopic objects. DA white dwarfs, non-DA white dwarfs,
quasars and main-sequence stars are shown as blue, magenta, green and grey
dots respectively. The polynomial colour cuts from Table\,\ref{e-poly} are
overlaid as red lines.}
\end{minipage}
\end{figure*}

As a start, we retrieved the DR7 spectra and $ugriz$ photometry for
all white dwarfs with $g\leq19$ and classified by
\citet{eisensteinetal06-1} as DA or DA\_auto, corresponding to
visually confirmed and automatically classified hydrogen-dominated
white dwarfs, respectively. This totals 2889 unique objects, 938 being
classified as DA and 1951 DA\_auto. All spectra were visually
inspected to corroborate their DA classification, and we found 99.4\%
agreement with the classification for white dwarfs by
\citet{eisensteinetal06-1}. The 0.6\% disagreement primarily comes
about from non-DA white dwarfs that were classified as DA\_auto by
\citeauthor{eisensteinetal06-1}'s classification routine. The sample
of spectroscopically confirmed DA white dwarfs was then used to trace
the locus of DA white dwarfs in the $(u-g,g-r)$, $(g-r,r-i)$, and
$(r-i,i-z)$ colour-colour planes. The population of DA white dwarf
follows a boomerang-shape in $(u-g,g-r)$ colours, where it is clearly
separated from the main sequence, but intersects the quasar population
(Fig.\,\ref{f-colsel}).

This relatively complex shape was approximated by the intersection of
two 5th-order polynomials (Table\,\ref{e-poly}). In $(g-r,i-r)$ the DA
white dwarfs lie along a relatively narrow band, overlapping with the
blue end of the main sequence and, to some extent with quasars, which
we approximated by the combination of a 2nd and 3rd order
polynomial. Finally, in $(r-i,i-z)$ , the DA white dwarfs are again
located at the blue end of the main sequence, but display a relatively
large spread in $i-z$. Visual inspection of the SDSS spectra of the
outliers confirms them as DA white dwarfs, though the majority of them
are near the faint end of the \citeauthor{eisensteinetal06-1}
sample. We decided to include those outliers in our colour-cut, and
hence approximated the DA locus by the intersection of three linear
relations in $(r-i,i-z)$.

We applied a magnitude cut of $g\leq19$ as a (conservative) measure, so
that each of the SDSS white dwarfs within the UKIDSS footprint would have
a significant $K$-band detection. Finally, we applied a set of
recommended data quality flags to the SDSS photometry to minimise
contamination by instrumental artifacts and blended stars. Applying the
constraint set summarised in Table\,\ref{e-poly} to DR7 resulted in the
selection of \SDS\ unique spectroscopic objects, which were then visually
classified as DA white dwarfs, other (non-DA) white dwarfs, quasars,
narrow line hot stars (see below), and other objects
(Table\,\ref{t-effg}).

\subsection{Narrow Line Hot Stars}
\label{s-nlhs}

The optical spectra of ultra-low mass white dwarfs and very cool DAs,
particularly those at low signal-to-noise ratio, can look rather similar
to early type main sequence stars, subdwarfs, extreme horizontal branch
stars or very metal-poor halo stars, which we all lump together as
contaminants with the designation ``narrow line hot stars'' (NLHS). These
are particularly dominant at the bright end of the white dwarf banana
($g\leq16$, see Table\,\ref{t-effg}). While there is noticeable interest
in ultra-low mass white dwarfs \citep[e.g.][]{liebertetal04-1,
kilicetal07-2, marshetal10-1}, they represent a tiny fraction of all DAs.
Given that our aim is to study the bulk population of DA white dwarfs, we
make no attempt to accurately classify ultra-low mass white dwarfs. In
addition, our colour selection is only complete to effective temperatures
above $\sim8000$K. Attempting to include much cooler white dwarfs would
result in significant contamination from NLHS.

\subsection{Overall Completeness and Efficiency}

Completeness and efficiency were the key parameters in designing our
selection algorithm (Table\,\ref{e-poly}), where we define
\textit{completeness} as the fraction of
\citeauthor{eisensteinetal06-1}'s DA white dwarfs recovered by our
constraints, and \textit{efficiency} as the ratio of spectroscopically
confirmed DA white dwarfs to all objects in our colour-magnitude
selection. In the case of the spectroscopic DA sample, one may argue
that completeness is more important than efficiency because
contaminants can be removed through visual spectral classification of all
candidate objects. On the other hand, the photometric-only DA sample
requires a high level of efficiency to minimise the number of
contaminants. We optimised the colour boundaries to maximise both
completeness and efficiency, and the constraint set in
Table\,\ref{e-poly} results in a completeness of \Comp\% and an
efficiency of \Effi\% (Table\,\ref{t-effg}). From SDSS DR7, a total
$4636$ unique spectroscopic DA white dwarfs are contained within the
colour-magnitude cuts. This represents a $70\%$ increase in spectroscopic
DA white dwarfs with $g\leq19$ over the \citet{eisensteinetal06-1} sample.
The photometric-only DA white dwarf candidates sample similarly contains
9341 objects. Assuming \Effi\% efficiency of the selection, $\sim6000$
extra white dwarfs are contained within the sample and therefore the total
increase over \citet{eisensteinetal06-1} is approximately a factor or
three. The efficiency is however only a lower limit for the photometric
sample because the SDSS is by design almost spectroscopicially complete
for QSOs, one of our main contaminants. Therefore, QSOs are only a minor
contaminant in the photometric-only sample.

The spectral classification and completeness are given as a function of
$g$-magnitude in Table\,\ref{t-effg}. The completeness drops slightly
towards larger apparent magnitude because of the larger scatter in the
colour-colour diagrams. The fraction of NLHS contaminants is largest at
the bright end of our sample, which is a natural consequence of the much
larger intrinsic brightness of subdwarfs and early-type main-sequence
stars, and the fraction of quasar contaminants increases towards larger
apparent magnitudes. It is possible to eliminate a fair fraction of the
contaminants in the photometric-only sample by using additional
information such as proper motions and infrared colours (see
Sect.\,\ref{ss-comp}).

\begin{table*}
\setlength{\tabcolsep}{0.9ex}
\caption{\label{t-effg} Efficiency and completeness of the polynomial
  colour-colour cuts (Table\,\ref{e-poly}) as a function of SDSS $g$
  magnitude. The total number of spectroscopic objects that matched
  our selection and were visually classified was \SDS. The
  classification 'Other white dwarfs' contains white dwarfs of type
  DAB, DAO, DB, DC, DQ, DZ, magnetic white dwarfs and white dwarf+MS
  binaries. The classification 'Other' includes CVs, galaxies,
  peculiar objects and unidentified spectra. The final two right hand
  columns show the number of photometric-only objects in each $g$
  magnitude bin (N$_{\mathrm{Tot}}$) and the number of predicted DA
  white dwarfs, calculated by multiplying this by the efficiency
  (N$_{\mathrm{DA}}$). The bottom line of the table gives the total
  number of DA white dwarfs with ($g\leq19$) in
  \citet{eisensteinetal06-1}, and the number of them included in our
  colour cut.}
\begin{tabular}{llllllllllllllll}
\hline
 & All WD & \multicolumn{2}{l}{DA WD} & \multicolumn{2}{l}{NLHS} &
\multicolumn{2}{l}{QSO} & \multicolumn{2}{l}{Other WD} &
\multicolumn{2}{l}{Other} & Efficiency & Completeness &
\multicolumn{2}{l}{Photometric-only} \\
$g$ & Candidates & N & \% & N & \% & N & \% & N & \% & N & \% &  &  &
N$_{\mathrm{Tot}}$ & N$_{\mathrm{DA}}$ \\ \hline
$\leq{16}$ & 258  & 79   & 30.6 & 161 & 62.4 & 0    & 0.0  & 13  & 5.0 & 5
 &
1.9 & 30.6\% & 95.8\% & 893 & 273 \\
16-17      & 581  & 326  & 56.1 & 185 & 31.8 & 22   & 3.8  & 44  & 7.6 & 4
 &
0.7 & 56.1\% & 96.7\% & 980 & 550 \\
17-18      & 1719 & 1092 & 63.5 & 225 & 13.1 & 230  & 13.4 & 165 & 9.6 & 7
 &
0.4 & 63.5\% & 96.5\% & 2278 & 1447 \\
18-19      & 4886 & 3139 & 64.2 & 269 & 5.5  & 1028 & 21.0 & 439 & 9.0 &
11 &
0.2 & 64.2\% & 95.0\% & 5190 & 3332 \\
Total      & \SDS & 4636 & 62.3 & 840 & 11.3 & 1280 & 17.2 & 661 & 8.9 &
27 &
0.4 & 62.3\% & 95.5\% & 9341 & 5819 \\
\hline
 & Total DA WD & \multicolumn{6}{l}{Photometrically selected DAs}
& & & & & 
&
Completeness &  & \\ \hline
Eisenstein & 2889 & \multicolumn{2}{l}{2757} &  &  &  &  &  &  &  &  &  &
\Comp\% &  & \\
\hline
\end{tabular}
\end{table*}

\subsection{Completeness of SDSS spectroscopy for DA white dwarfs}
\label{s-comp}

\begin{figure*}
\begin{minipage}{2\columnwidth}
\includegraphics[width=\columnwidth]{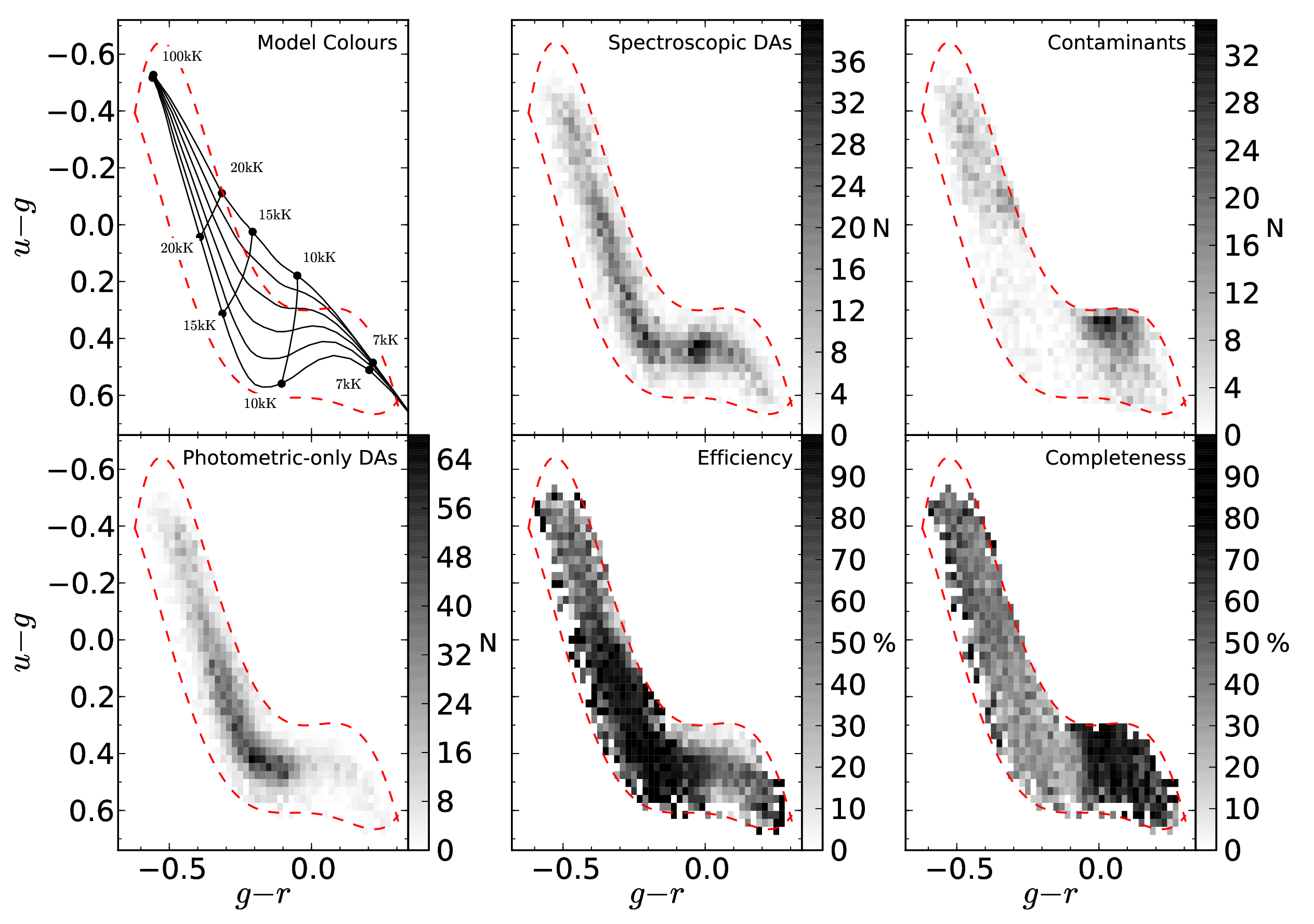}
\caption{\label{f-cc_comp} The spectroscopic completeness of DA white
dwarfs in SDSS~DR7 within the $(u-g, g-r)$ colour-colour plane. In the top
left panel, the colour selection from Table\,\ref{e-poly} is shown as a
red dashed line, overlaid with the DA white dwarf cooling tracks of
\citet{holberg+bergeron06-1}. From the bottom up, these curves represent
$\Logg=7.0-9.5$ in steps of 0.5. The number of spectroscopically confirmed
DA white dwarfs and contaminants within our colour selection are shown in
the upper middle and right panels, respectively. Two distinct regions of
high contamination  are visible, with NLHS and quasars being concentrated
at the bluest and reddest colours of the DA ``banana'', respectively. The
number of DA white dwarfs without SDSS spectroscopy is shown in the
bottom-left panel, calculated as the number of photometric-only objects
weighted by the colour-dependent efficiency of our selection algorithm
(lower middle panel). Finally, the spectroscopic completeness of SDSS for
DA white dwarfs, i.e. the ratio of spectroscopic DAs to the total number
of DAs, is shown in the bottom right panel. Cool white dwarfs have a very
high completeness thanks to their colour-proximity to ultraviolet-excess
quasars, which were intensively targeted by SDSS.}
\end{minipage}
\end{figure*}

The sample produced here provides an excellent opportunity to investigate
the spectroscopic completeness of SDSS for DA white dwarfs. We used the
cuts in Table\,\ref{e-poly} to select both the spectroscopic and
photometric objects within DR7. From these two sets of data, we then
calculate the spectroscopic completeness within the $(u-g, g-r)$
colour-colour plane (Fig.\,\ref{f-cc_comp}). The upper middle and right
panels display the density of spectroscopically confirmed DA white dwarfs
(middle) and contaminants (such as NLHS and quasars, right hand side
panel). For comparison, the upper left hand panel shows the DA white dwarf
cooling tracks from
\citet{holberg+bergeron06-1}\footnote{See
  http://www.astro.umontreal.ca/$\sim$bergeron/CoolingModels for an
  updated grid.}. The
efficiency of our colour cuts is obtained for each bin within the $(u-g,
g-r)$ plane as the ratio of the number of the DA white dwarfs to the total
number of objects in the bin (lower centre panel). This clearly displays a
reduced efficiency of selecting both the hottest and coldest white dwarf
because of the increased numbers of contaminants. Our selection
method however has an extremely high efficiency when selecting white
dwarfs with temperatures between $\sim10,000-20,000$K. The number of DA
white dwarfs without SDSS spectroscopy is predicted by scaling the number
of photometric-only objects with the efficiency (resulting in the lower
left panel). Finally, the DA white dwarf spectroscopic completeness was
calculated as the ratio of spectroscopically confirmed DA white dwarfs to
the total number of DA white dwarfs, with and without spectroscopy (lower
right panel). The overall spectroscopic completeness is $44.3\%$ down to
$g=19$. As mentioned in Sect.\,\ref{s-nlhs}, this analysis is limited to
white dwarfs with $\Teff\gtrsim8000$K. The preference of SDSS spectroscopy
to target quasars is clearly seen in the lower right corner of the
spectroscopic completeness colour-colour diagram, where the vast majority
of quasars lie. This corresponds to a very high spectroscopic completeness
for cool DA white dwarfs. In contrast, the spectroscopic completeness for
white dwarfs with $\Teff\ga12000$K is significantly lower.

\section{Cross-matching with UKIDSS}
\label{s-uki}

\begin{figure}
 \includegraphics[width=\columnwidth]{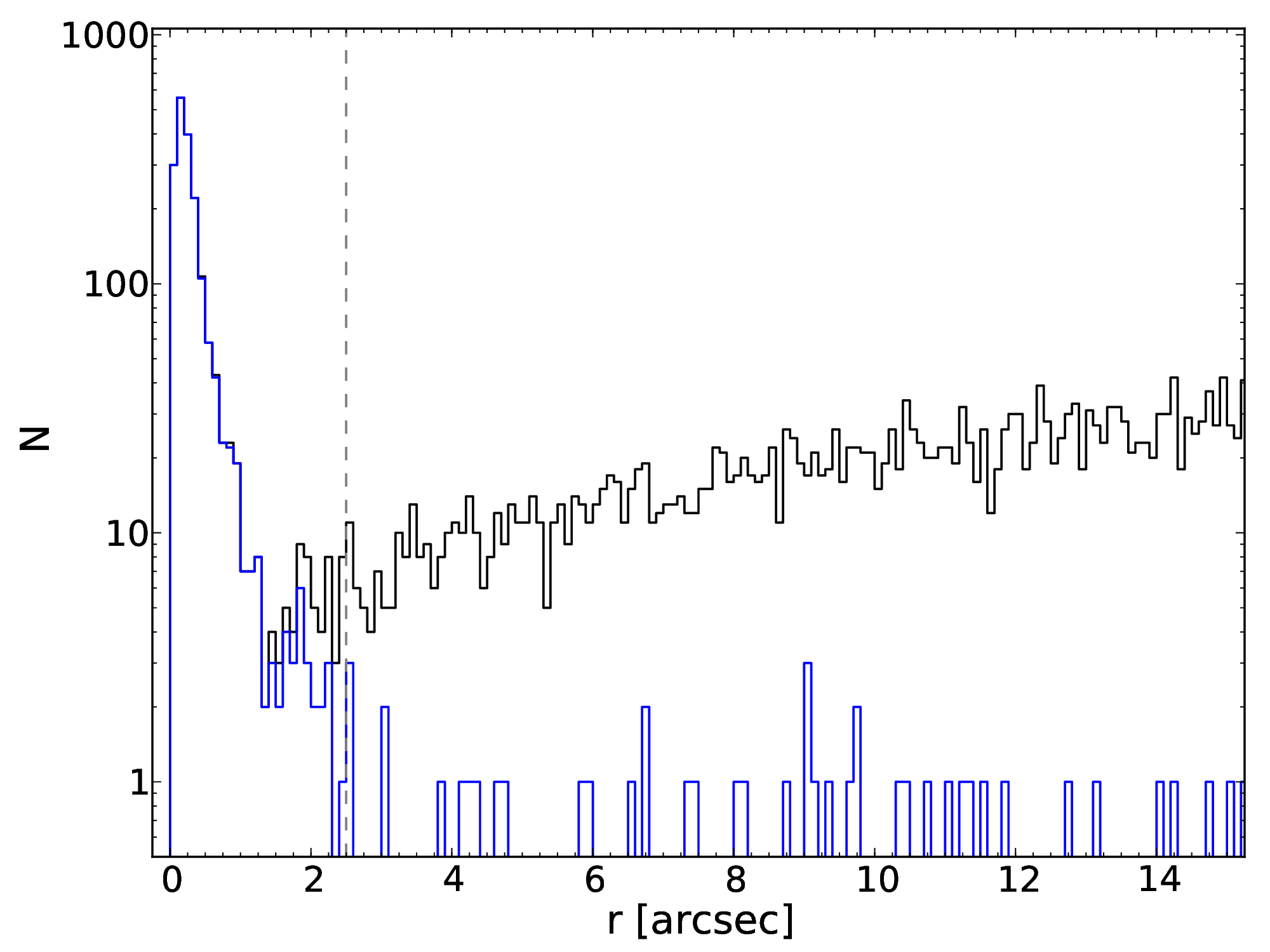}
 \caption{\label{f-dd} Spatial offsets of the SDSS and UKIDSS
   positions for a sample of 5000 spectroscopically confirmed DA white
   dwarfs randomly selected from our constraint set
   (Table~\ref{t-effg}). The blue histogram shows the distance to the
   closest neighbour in UKIDSS, the black histogram plots the
   distances to all possible matches. The bins have a width of
   $0.1\arcsec$.}
\end{figure}

All (spectroscopic and photometric-only) objects from the DA selection
in SDSS DR7 were matched with the UKIDSS database using the CrossID
function. To decide upon a matching radius, a sample of 5000 randomly
selected spectroscopically confirmed DA white dwarfs were matched to
the UKIDSS database with a $60\arcsec$ search radius $r$. The
distribution of the distance between the SDSS objects and the UKIDSS
matches is shown in Fig.\,\ref{f-dd}. The number of all possible
matches within $60\arcsec$ (black histogram) grows approximately as
$r^2$, as would be expect for chance coincidence, whereas true matches
are primarily within $r<3\arcsec$. Selecting only the closest match
(blue histogram), the majority of these random mismatches are
removed. The blue and black distributions agree well at small
distances ($r\la2.5\arcsec$), indicating that crowding is not a major
problem. $2.5\arcsec$ is much larger than the quoted astrometric
accuracies of both SDSS and UKIDSS (of order a few tenths of an arc
second), but the large proper motions of the white dwarfs and the
potentially large ($\sim$few years) time interval between both surveys
can lead to positional shifts up to a few arc seconds. We
adopt $r=2.5\arcsec$ for the final cross-matching of our spectroscopic
and photometric-only SDSS DA samples with UKIDSS, which limits the
number of spurious matches, and will exclude only a handful of (halo)
objects with extremely high proper motions.  Any remaining positional
mismatches are flagged in the examination of the SDSS and UKIDSS
images carried out later. We restrict our analysis to unresolved
systems, as the physical association of spatially resolved companions
to white dwarfs will be difficult to demonstrate with the available data.
Consequently, objects that are flagged as partially resolved in the UKIDSS
images are removed form the sample.

A total of \UKS\ of the SDSS objects with spectra were found to
have at least one measured magnitude in the UKIDSS database. \UKDA\ of
these are spectroscopically confirmed DA white dwarfs. Similarly, \UKP\ of
the photometric-only objects had at least one match in the UKIDSS
database.

\begin{table}
\caption{\label{t-detec} The number of all SDSS DR7 objects satisfying our
  constraint set (Table\ref{e-poly}), and of various subsets with
  different UKIDSS bands.}
\begin{tabular}{lll}
\hline\hline
Detections & Spectroscopic & Photometric \\
in band & Objects & Objects \\ \hline
Total SDSS & \SDS & \SDP \\
Any UKIDSS & \UKS & \UKP \\
$Y$ & 1815 & 1614 \\
$J$ & 1787 & 1549 \\
$H$ & 1503 & 1281 \\
$K$ & 1108 & 840 \\
$H$ \& $K$ & 1075 & 809 \\
$J$, $H$ \& $K$ & 979 & 720 \\ \hline
\end{tabular}
\end{table}

\section{Identifying infrared excess objects}
\label{s-ire}

\subsection{DA White Dwarf Fitting}
\label{ss-dafit}

A grid of synthetic DA white dwarf spectra was calculated with the
model atmosphere code described by \citet{koester10-1} and using
the latest line profiles of \citet{tremblay+bergeron09-1}. These cover
$\Teff=6,000-100,000$K in 131 steps nearly equidistant in
$\log(\Teff)$, and $\Logg=5.0-9.5$ in steps of $0.25$dex.

\subsubsection{Fitting the SDSS spectroscopy}
\label{sss-dafitspect}

We fitted the SDSS spectra of all DA white dwarfs found within our colour
cuts (Table\,\ref{e-poly}) following the method described in
\citet{rebassa-mansergasetal07-1}. A $\chi^2$ minimisation is used to find
a best fit from our grid of DA white dwarf model spectra, providing \Logg\
and \Teff. Using the cooling models of \citet{holberg+bergeron06-1}, \Mwd,
\Rwd\ and the distance $d$ can be calculated for each object.

\begin{figure}
\includegraphics[width=\columnwidth]{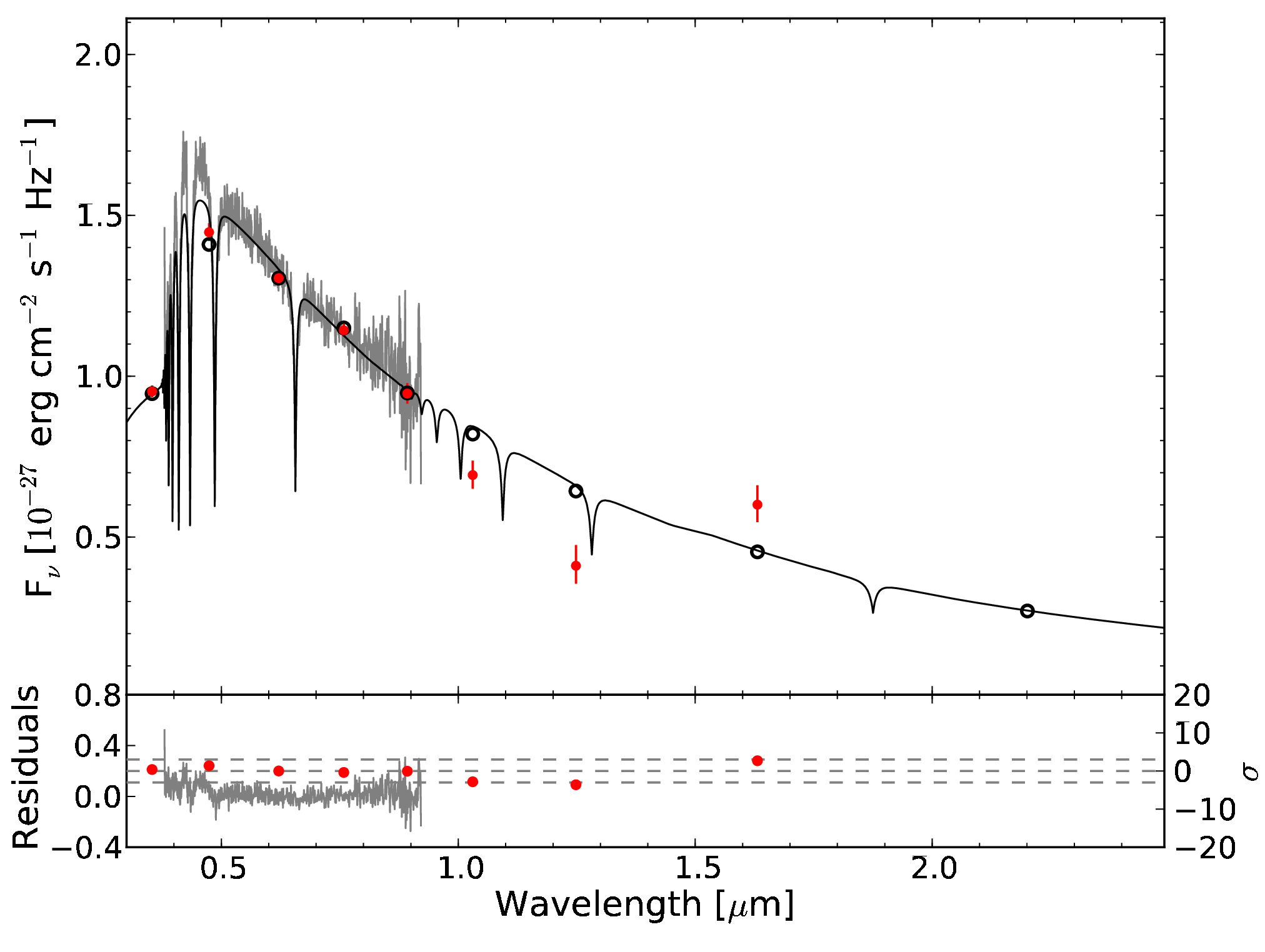}
\caption{\label{f-1218} The SED of SDSS\,J1218+0042, an example of a
  spectroscopically confirmed DA white dwarf with a possible IR flux
  excess (classified as DAire: in Table~\ref{t-spa}),
  SDSS\,J1218+0042. The best-fit model to the $ugri$ photometry is
  shown ($\Teff=10000$K, $\Logg=7.75$), which results in an $H$-band
  excess just below $3\sigma$. Adopting the slightly higher
  temperature ($\Teff=11173$K) from the analysis of the SDSS spectrum
  boosts the excess to just above $3\sigma$, flagging the object as an
  IR excess candidate. However, given that $Y$ and $J$-band magnitudes
  fall significantly below the model, further IR data are necessary to
  confirm or refute the IR excess of this white dwarf.}
\end{figure}

\subsubsection{Fitting the SDSS photometry}
\label{sss-dafitphot}

We also fitted \textit{all} photometric objects found with our colour
cuts, including all objects that do have SDSS spectroscopy.
We also fitted objects known \textit{not} to be DA white dwarfs,
to allow us to investigate the properties of the contaminants
among the photometric-only DA candidates.

Photometric objects were fitted by comparing the SDSS $u$, $g$, $r$ and
$i$ magnitudes to the white dwarf model grid, again based upon a
smallest $\chi^2$. Model magnitudes were calculated for each of the \Logg\
and \Teff\ values by folding the theoretical spectra through the SDSS and
UKIDSS $ugrizYJHK$ filter curves.  Magnitudes redder than $i$ were not
included since we are searching for composite systems. This ensures we can
recognise objects with an excess already showing in $z$, such as
white dwarf plus M-dwarf binaries. This does not significantly
affect the spectroscopic fitting method because we fit line
profiles. This method is most sensitive at shorter wavelengths, where the
companion does not significant contribute. For the majority of photometric
DA candidates, the 4-band photometry did not provide sufficient
constraints to accurately determine the surface gravity. For objects with
calculated effective temperatures in the range $9000-20000$K, where the
spectral line widths are narrow and therefore do not significantly affect
the SED, we adopted a canonical value of $\Logg=8.0$ for those objects.

\begin{figure*}
\includegraphics[width=\columnwidth]{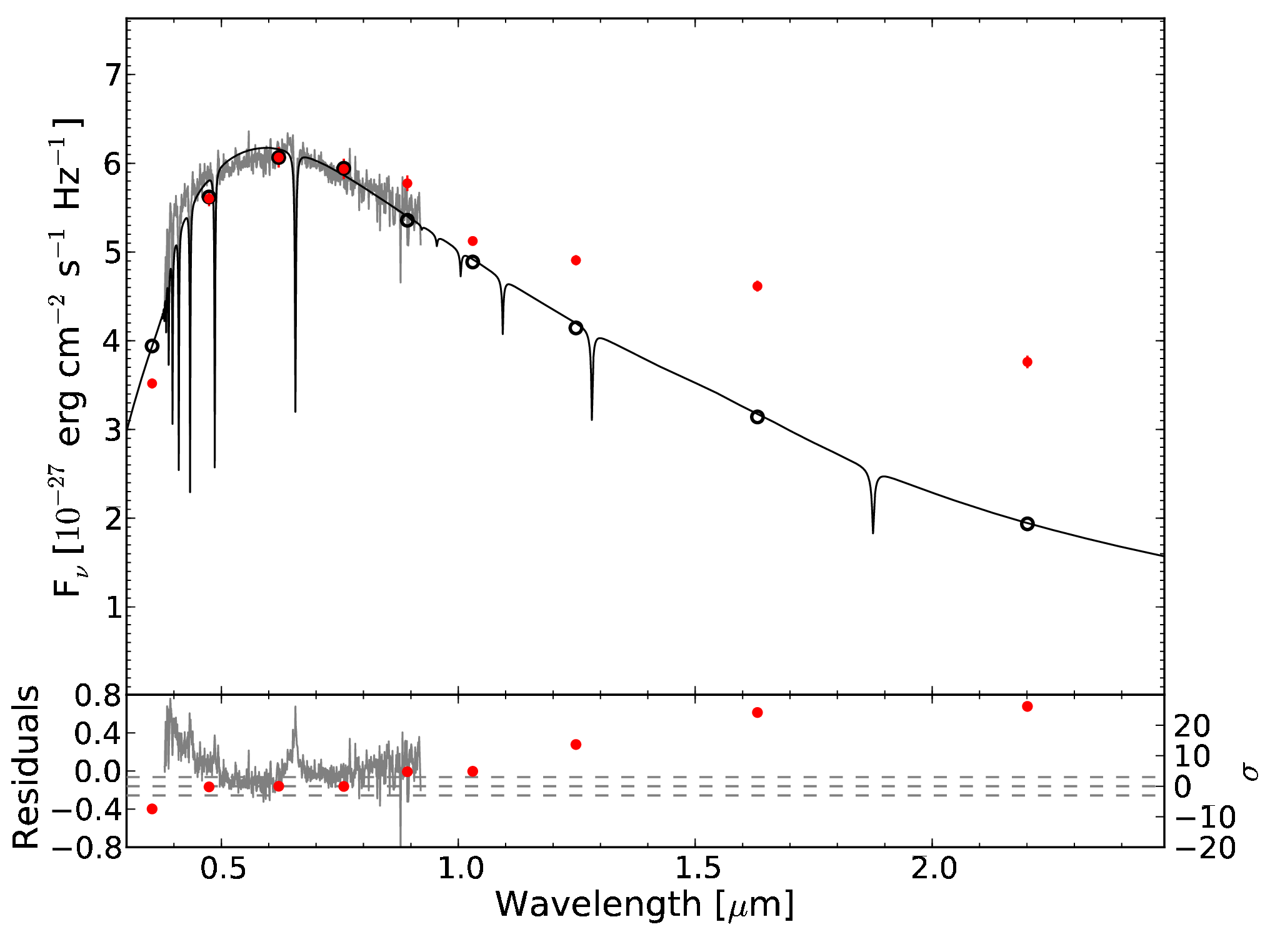}
\includegraphics[width=\columnwidth]{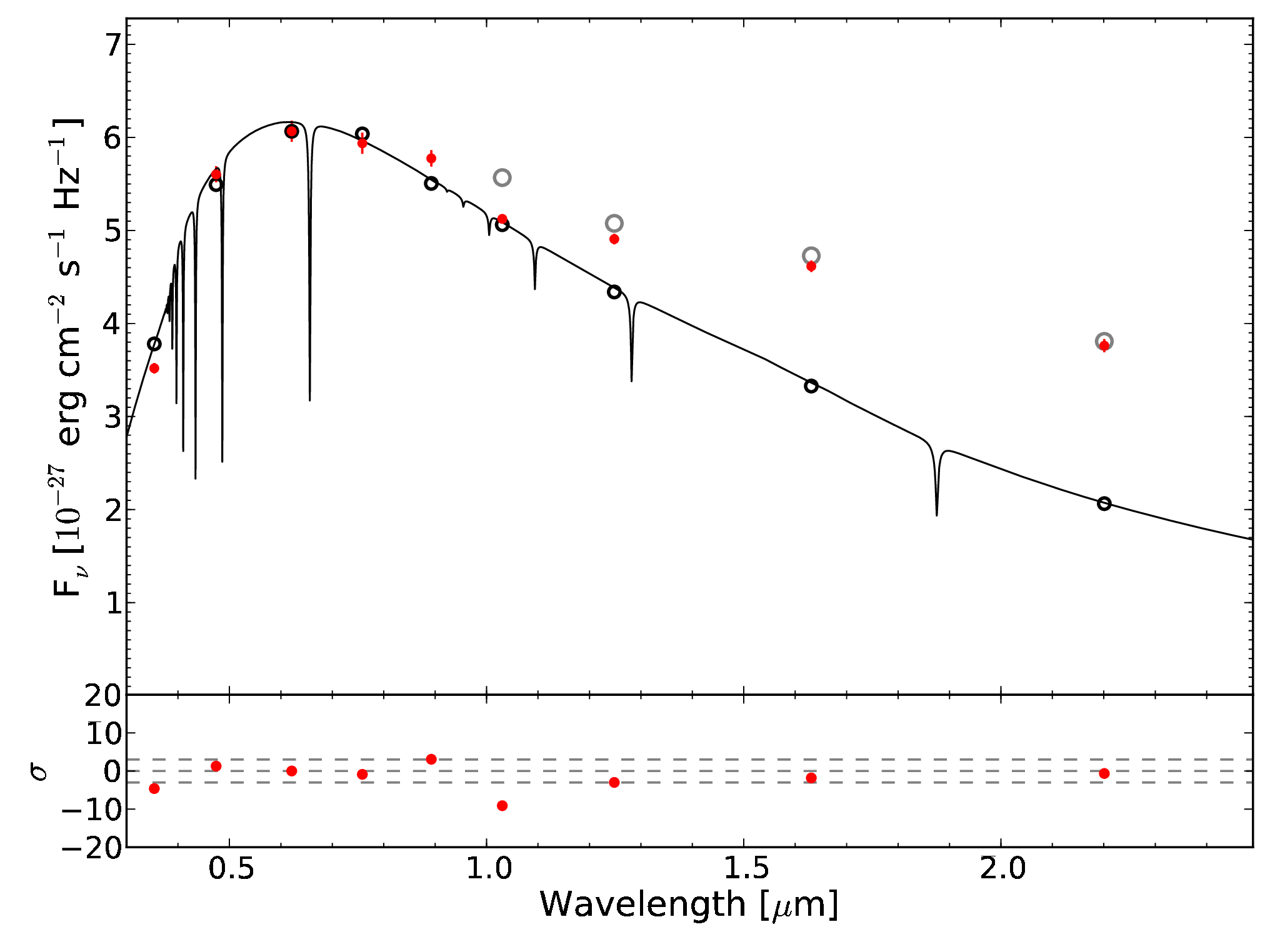}
\caption{\label{f-0135} SDSS\,J0135+1445; a cool white dwarf with a
  probable low mass companion. The fit is a $\Teff=7467\pm18$K,
  $\Logg=7.34\pm0.04$ spectra at a distance of $69\pm2$pc. The
  calculated mass is $\Mwd=0.29\pm0.01\Msun$. The photometry is
  best fit with a white dwarf of $\Teff=8000\pm^{20}_{10}$K, $\Logg=8.0$.
  Based upon the fit to the photometry, the addition of an L8-type
  companion to the white dwarf magnitudes is the best fit to the IR excess
  (Table\,\ref{t-s+p}) and is plotted on the right hand figure as open
  grey circles. The model $Y$-band magnitude does not match the UKIDSS $Y$
  measurement, but we found that the $Y$-band data did relatively often
  disagree with SDSS $z$  and UKIDSS $J$.}
\end{figure*}

\begin{figure*}
\begin{minipage}{2\columnwidth}
\includegraphics[width=\columnwidth]{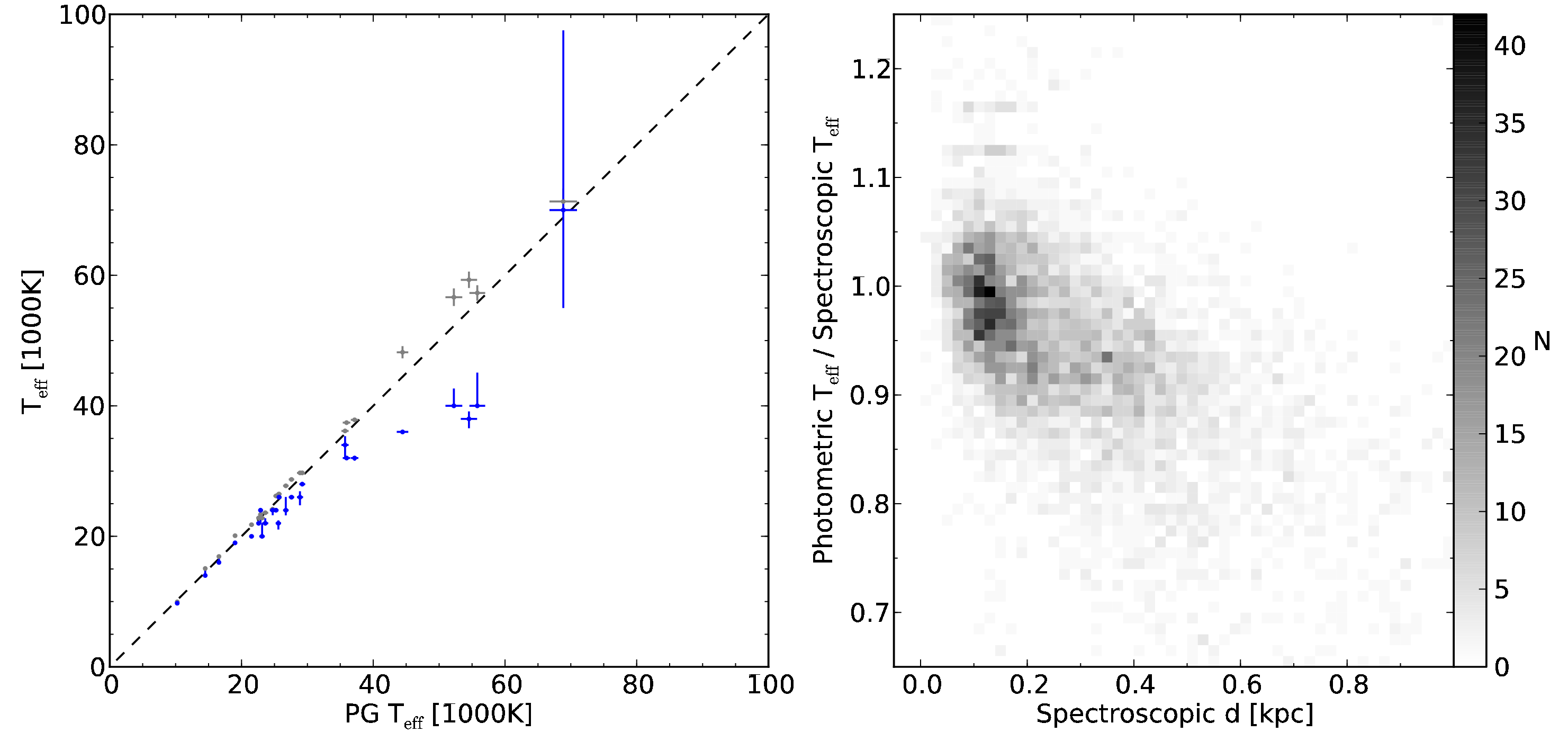}
\caption{\label{f-tc} A demonstration of the systematically lower
measured photometric effective temperatures. The left hand panel shows
the \citet{liebertetal05-1} PG survey white dwarfs with corresponding
stars in our sample (25 objects). This figure compares the effective
temperature calculated by \citet{liebertetal05-1} with that from our
spectroscopic (grey) and photometric (blue) fitting techniques. When using
the photometric method, we significantly underestimate the temperature
of hot white dwarfs. The hottest white dwarf at $70,000$K has a
large error on the photometric fit because the continuum slope of the
white dwarf at these temperatures is effectively Rayleigh-Jeans and
therefore contains no temperature information. In the right hand panel we
show the ratio of the effective temperatures calculated using the
photometric and spectroscopic fitting methods for each of the objects in
the spectroscopic DA white dwarf sample, as a function of distance. Again,
we underestimate the effective temperature of white dwarfs at large
distances.}
\end{minipage}
\end{figure*}

The temperatures measured from the SDSS photometry were found to be
systematically lower than those from the fitting of line profiles
(e.g. Table\,\ref{t-s+p}; objects with SDSS spectroscopy, but fitted with
the photometric method). Our sample of SDSS white dwarfs overlaps with the
Palomar Green (PG) sample, and Fig.\,\ref{f-tc} shows a comparison between
our spectroscopic and photometric temperatures with those of
\citet{liebertetal05-1}, which were determined from independent data,
spectral models, and fitting routines. We find good agreement between the
results of \citet{liebertetal05-1} and our spectroscopic method. However,
our photometric temperatures are systematically too low, a trend that is
strongly correlated with either white dwarf temperature or distance. At
200pc (500pc), the photometric temperatures are on average $5\%$ ($10\%$)
below our and Liebert's spectroscopic values.  This could suggest
that interstellar reddening is, at least in part, the culprit for reduced
temperatures.
While reddening would not significantly effect the shape of the line
profiles, it could noticeably change the slope of the continuum (see also
\citealt{holbergetal08-2}). No clear correlation is, however, seen when
comparing the mismatch in temperatures to the \citet{schlegeletal98-1}
values of $E(B-V)$ at the positions of the white dwarfs. The
\citeauthor{schlegeletal98-1} maps probe interstellar reddening through
the entire Galaxy, whereas the white dwarfs in our sample lie at a few
hundred parsecs at most. Typical (total Schlegel) reddening along the
lines-of-sight towards our white dwarfs is $E(B-V)\sim0.05$. De-reddening
the SDSS photometry with that total $E(B-V)$, and re-fitting the
photometric white dwarf sample indeed leads, as expected, to a large
over-correction of the white dwarf temperatures. Analysing the sample of
spectroscopic DAs, we estimate that the typical reddening in front of the
white dwarfs is $E(B-V)\sim0.01-0.02$. However, we can not systematically
correct for the effect of reddening for the sample of photometric-only DA
candidates.
We note that for hot white dwarfs, $\Teff \ga 45000$K, non-LTE
effects become important, which may also lead to some systematic
differences in the fit parameter for the hottest stars in our sample.
Therefore, the temperatures calculated from the photometry alone have an
additional systematic uncertainty, on top of the statistical uncertainty
from the fit, and the true temperatures are likely to be a few thousand
Kelvin higher. In the context of our search for infrared flux excess
(Sect.~\ref{ss-ire_det} below), changing the white dwarf temperature by a
few thousand Kelvin does not have a significant impact on the level of
excess detected (see Sect.\,\ref{s-wdwire} for examples).

\subsection{IR Excess Detection}
\label{ss-ire_det}

In both the spectroscopic and photometric fitting methods, the best fit
model extends into the IR and all objects with UKIDSS data were examined
for an excess by comparing the IR white dwarf model flux with the observed
$YJHK_s$ magnitudes. Objects with a $3\sigma$ excess in any band over the
white dwarf model were defined as a robust excess candidate (``DAire'' and
``DA:ire'' for the spectroscopically confirmed DAs and the
photometric-only DA candidates, see Tables~\ref{t-spa}, \ref{t-poa} and
\ref{t-s+p}). Further to this, objects that appeared to have a best fit
model which over-estimated the flux in $Y$, $J$ and $H$, but showed only a
$\sim2\sigma$ excess in $K$ were also flagged as tentative excess
candidates (``DAire:'' and ``DA:ire:'', as above, Tables~\ref{t-spa},
\ref{t-poa} and \ref{t-s+p}), see Fig.\,\ref{f-1218} for an example.
Further IR data is definitely needed to confirm these marginal IR excess
candidates. Similarly, spectroscopically confirmed DAs (photometric DA
candidates) with close to $3\sigma$ excess that by eye require further
data to confirm the excess were also marked as ``DAire:'' (``DA:ire:'').
For the photometric-only objects the uncertainty on the model parameters
is generally larger compared to the spectroscopically confirmed DAs. This
was accounted for by not flagging objects with a marginal IR excess
\textit{and} a large uncertainty on effective temperature as excess
candidates.

Spurious excesses were often caused by spatially close background or
foreground objects to the white dwarf and bad SDSS or UKIDSS images. We
visually inspected all flagged sources and discounted resolved and
partially resolved systems because the physical association of the two
objects could not be demonstrated based on the available data.

A total of 42 white dwarfs were found to have an excess from the
spectroscopic fitting method, and 105 infrared excess candidates were
found from the photometric fitting method (Table\,\ref{t-lmc}). The
excesses have a variety of spectral shapes, generally consistent with
various spectral type companions. Table\,\ref{t-spa} and \ref{t-poa} list
all the spectroscopic and photometric IR excess candidates, respectively.

\subsection{IR Excess Modelling}
\label{ss-ire_mod}

\begin{figure}
\includegraphics[width=\columnwidth]{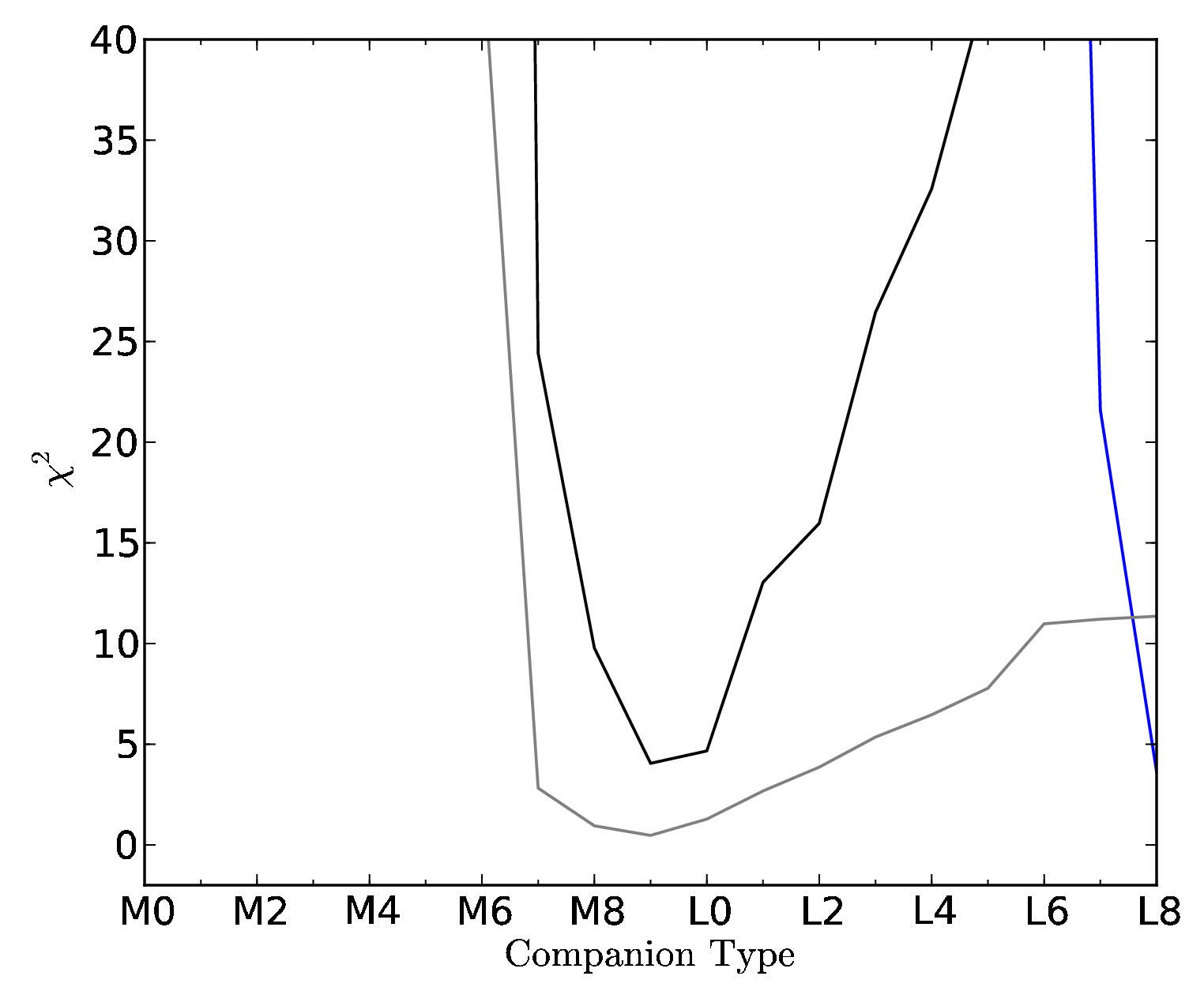}
\caption{\label{f-c2} Reduced $\chi^2$ as a function of companion type for
three white dwarfs with well constrained companions using the photometric
method. $\chi^2$ is calculated from comparing the UKIDSS $JHK$ magnitudes
with those of the low-mass companions of \citet{hoardetal07-1}.
SDSS\,J0135+1145, SDSS\,J0842+0004 and SDSS\,J0925-0140 are plotted as
blue, grey and black respectively. We find that a good fit has a reduced
$\chi^2 \la 10$, a bad fit (DAire: or DA:ire: in Tables\,\ref{t-spa} and
\ref{t-poa}) has $\chi^2 \sim 10 - 100$ and a QSO has
$\chi^2 \sim 100 - 1000$.}
\end{figure}

\begin{figure}
\includegraphics[width=\columnwidth]{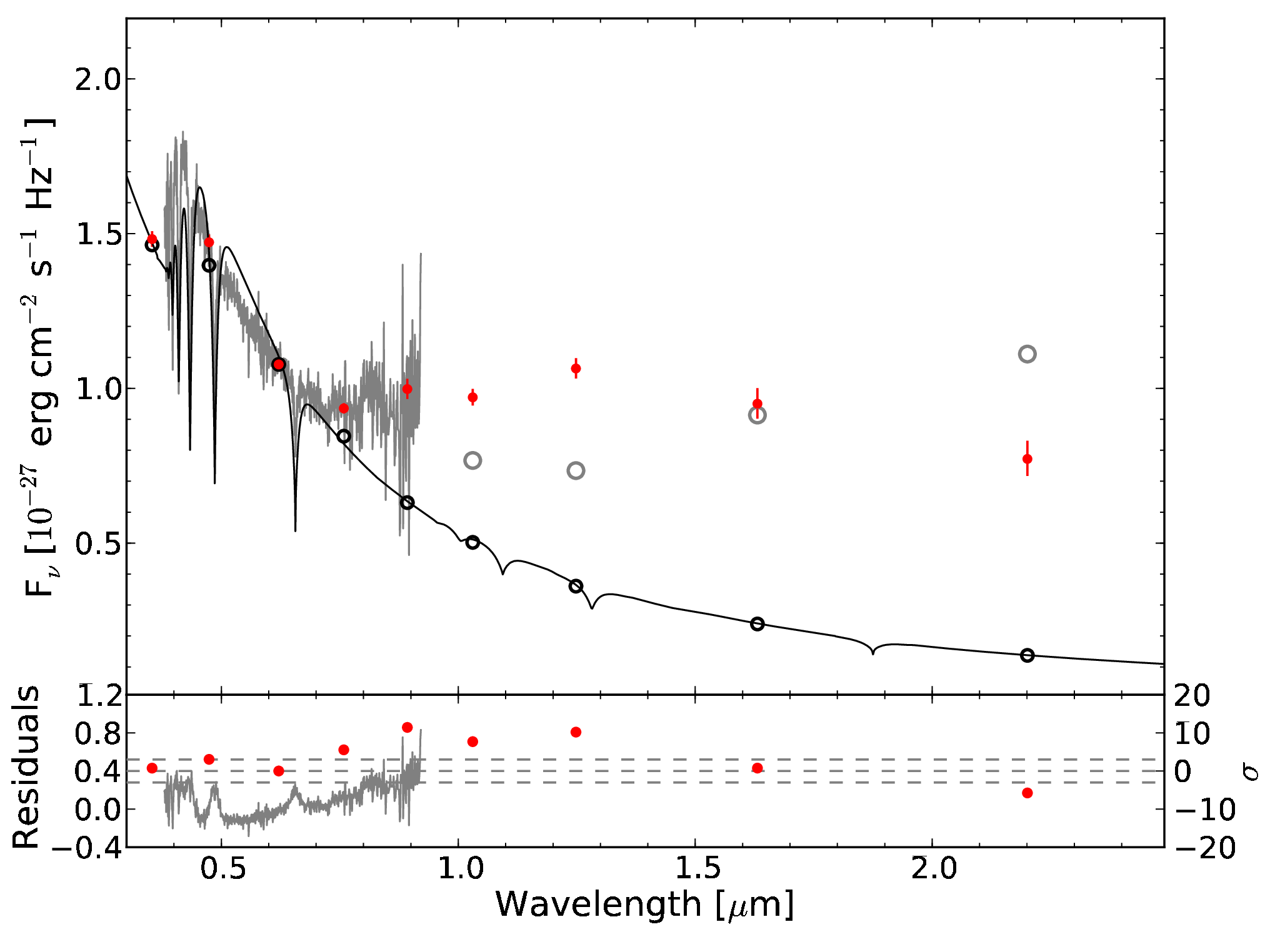}
\caption{\label{f-1619} The SED of SDSS\,J1619+2533, an example of a
  spectroscopically confirmed DA white dwarf where the photometric
  method substantially underestimates the white dwarf temperature.
  The best fit to the spectrum is a $\Teff=25595$K and $\Logg=7.21$ DA
  white dwarf model ($\Mwd=0.33\pm0.04$), where as the photometric
  fitting method finds the best solution at $\Teff=18000$K and
  $\Logg=9.5$, the latter is plotted here in black.  Within the
  photometric method, the lower temperature (and higher gravity) leads
  to the distance being substantially underestimated, and therefore
  the flux of the companion being overestimated. Fitting the SED with
  the two-component model described in Sect.\,\ref{ss-ire_mod}
  compensates for the low distance by choosing a companion with a
  larger absolute magnitude, i.e. a later spectral type. For
  SDSS\,J1619+2533, the spectroscopic method results in a M4 type
  companion, the photometric method in an L6 companion. Because the
  optical-IR SED can not be well fitted with any companion type at the
  photometric distance, this object is flagged as ``bad fit'' in
  Table~\ref{t-s+p}. }
\end{figure}

The $YJHK_s$ magnitudes of objects that were found to exhibit an
infrared excess were fitted with a composite model consisting of the
best-fit white dwarf plus a set of low-mass companions with spectral
types M0 through to L8. We used the 2MASS $JHK_s$ magnitudes of
\citet{hoardetal07-1}, which we converted into the UKIDSS filter
system adopting the equations given in
\citet{dyeetal06-1}. \citeauthor{hoardetal07-1}'s absolute magnitudes
 of the low-mass companions were scaled to apparent magnitudes using
the distance modulus calculated from the white dwarf fit.  For
photometrically fitted objects, the white dwarf distance modulus was
calculated as an average of the difference between the best-fit
absolute magnitudes and the SDSS apparent magnitudes in each of the
$u$, $g$, $r$ and $i$ bands. Finally, the model magnitudes of the
composite system were computed from the combined white dwarf and
companion star fluxes.
The best composite fit to the $JHK$ IR photometry was
calculated using a least $\chi^2$ search and subsequently confirmed by
visual inspection. Fig\,\ref{f-c2} shows reduced $\chi^2$ as a function of
companion type for three white dwarfs with well constrained companions
using the photometric method as an example. It was seen that a white dwarf
with a good fit to the companion has a reduced $\chi^2 \la 10$, an example
of a the corresponding SED is shown in Fig.\,\ref{f-0135}. Composite fits
with $\chi^2 \sim 10 - 100$ were flagged as ``bad fits'' and flagged as
DAire: or DA:ire in Table\,\ref{t-spa} and \ref{t-poa}, respectively, as
the nature of the IR excess remains somewhat unclear (see
Fig.\,\ref{f-1619} and Sect.\,\ref{sss-da}). Finally, QSOs stand out
because of their very high $\chi^2$, $\sim100-1000$ (e.g.
Fig.\,\ref{f-0046}). 
A summary of companion types for each method is given in
Table\,\ref{t-lmc}.

\begin{table*}
\caption{\label{t-lmc} Objects with an IR excess split by estimated
  low-mass companion type. The first column shows the IR excess
  candidates among the spectroscopically confirmed DAs
  (Table\,\ref{t-spa}) and the final column contains the information
  on the photometric-only sample (Table\,\ref{t-poa}), where there is
  no SDSS spectrum available for classification. The section in the
  middle shows all objects with SDSS spectroscopy for which fitting
  their $ugri$ photometry resulted in an IR flux excess
  (Table\,\ref{t-s+p}), split by their spectroscopic
  classification. This set of objects is useful to gauge the
  contamination of photometric-only IR excess DA candidates by other
  types of objects. It also allows the like-for-like comparison of the
  infrared excess detection/modelling for genuine DA white dwarfs. To
  aid this comparison of the two methods, the column ``DA (w SIRE)''
  lists the spectroscopically confirmed DA white dwarfs that exhibit
  an IR flux excess when fitted with both the spectroscopic and
  photometric method, and the column ``DA (wo SIRE)'' lists 6
  spectroscopically confirmed DA white dwarfs that are found to have
  an IR excess only when analysed with the photometric method. A ``-''
  mark denotes that the corresponding entry is not possible. ``No Phot
  IR Excess'' means that no photometric excess was found, even though
  a spectroscopic excess is seen.}
\begin{tabular}{lc@{~~~~~}ccccccc@{~~~~~}c}
\hline\hline
 & Spectroscopic & \multicolumn{7}{c}{Spectroscopic using} &
Photometric \\
 &  & \multicolumn{7}{c}{Photometric Method} & \\ \hline
Primary Type $\Rightarrow$ & DA & \multicolumn{2}{c}{DA} & NLHS & WDMS &
DAH &
QSO & Total &
- \\
Companion Type $\Downarrow$   &    & w SIRE & wo SIRE  &      &      &    
&    
&       &  
 \\
\hline 
M-type            & 16 & 9      & 1        & 10   & 2    & 0   & 0   & 22 
  &
15  \\
L-type            & 19 & 11     & 0        & 12   & 1    & 0   & 0   & 24 
  &
19  \\
$\geq L7$ or disc & 7  & 6      & 2        & 0    & 0    & 0   & 0   & 8  
  & 
4   \\
QSO               & -  & 0      & 0        & 0    & 0    & 0   & 323 & 323
  &
40  \\
Bad fit           & 0  & 4      & 3        & 9    & 1    & 3   & 0   & 20 
  &
27  \\
No Phot IR Excess & -  & 12     & -        & -    & -    & -   & -   & 12 
  &
-  \\\hline
$\Sigma$          & 42 & 42     & 6        & 31   & 4    & 3   & 323 & 409
  &
105 \\
\hline
\end{tabular}
\end{table*}

\subsection{Comparison of the spectroscopic and photometric methods}
\label{ss-comp}

The results from using either the spectroscopic or purely photometric
methods to find IR excess can be directly compared because \textit{all}
the objects with SDSS spectra also have SDSS photometry. We discuss in the
following sub-sections how well the spectroscopic and photometric method
compares for genuine DA white dwarfs (Sect.\,~\ref{sss-da}), what can be
learned from the spectroscopic contaminants that show an IR excess when
fitted with the photometric method (Sect.\,\ref{sss-qso}, \ref{sss-nlhs}),
and how additional information such as IR colours and proper motion could
be used to suppress contaminants in the photometric-only sample
(Sect.\,\ref{sss-qso}).

\subsubsection{DA White Dwarfs}
\label{sss-da}
Of the 42 spectroscopic DA white dwarf IR excess candidates, 30
($\sim71\%$) are recovered by the photometric fitting method as well
(column ``DA w SIRE'' in Table\,\ref{t-lmc}, 9 M-type, 11 L-type, 6 $\geq
L7$-type companions/debris discs and 4 ``bad fits'', which are marked as
``DAire:'' in Table\,\ref{t-poa}), and 12 do not show an excess when using
the photometric method. These 12 objects are close to the $\sim3\sigma$
limit in the spectroscopic method and were not recovered. They were not
recovered by the photometric method because of the larger uncertainty on
effective temperature leading to the excesses being within the combined
model and flux errors. 6 objects with DA white dwarf SDSS spectra are
found to have an IR excess in the photometric method, but do not exhibit
an IR excess when using the spectroscopic method (column ``DA wo SIRE'' in
Table\,\ref{t-lmc}). Three of these have excesses which are accentuated by
a slightly hotter photometric fit compared to the spectroscopic one. They
fall just outside the criteria (Section\,\ref{ss-ire_det}) for having an
excess in the spectroscopic method. For the other three objects the SDSS
spectra are too poor to obtain reliable \Teff\ and \Logg\ measurements
from fitting their Balmer lines. Therefore an excess was not recognised in
the spectroscopic method. However, the photometric data were good enough
and thus these three objects are classified as photometric IR excess
candidates. We may therefore expect $\sim9-17\%$ (3/36--6/36) spurious IR
excess candidates from the photometric method. All objects that display an
infrared excess either in the spectroscopic, or the photometric method,
and have an SDSS spectrum are listed in Table\,\ref{t-s+p}.

Table\,\ref{t-lmc} also lists the IR excess candidates split by modelled
companion type. Among the 30 objects which are defined as IR excess
candidates in both methods, the distribution of companion type has a
similar form. $\sim80\%$ of the companions have M or early L spectral
types (split evenly between the two classes), and the remaining $\sim20\%$
have companions $\geq L7$, i.e. brown dwarf or debris disc candidates.

As briefly outlined in Sect.\,\ref{ss-dafit}, the temperatures resulting
from the photometric fits are systematically too low, and this will
introduce a bias in the spectral type of the companion. The flatter SED of
a cooler white dwarf will reduce/distort the IR flux excess relative to
the white dwarf, and therefore a later type companion will provide
sufficient flux to account for the excess. An additional effect is that a
cooler white dwarf will suggest a lower distance when fitting the $ugri$
magnitudes.  Underestimating the distance will lead to an overestimate of
the absolute flux of the companion. To compensate for this, the fit to the
companion will resort to a companion with a larger absolute magnitude,
i.e. lead to a companion spectral type that is too late. A moderately
extreme example of these effects is shown in Fig.\,\ref{f-1619}
(SDSS\,J1619+2533, see Table\,\ref{t-s+p}). Assuming that the
spectroscopic fit parameters are correct, the white dwarf temperature is
underestimated by $\sim8000$K, which leads to an L6 companion in the
photometric method, as opposed to an M6 companion resulting from the
spectroscopic analysis. This object is marked in the notes column as
having a bad fit to the companion in the photometric method, where a ``bad
fit'' is defined as an object having an excess which is inconsistent with
any companion type (at the photometric white dwarf distance).

\subsubsection{Quasar Elimination}
\label{sss-qso}

\begin{figure}
\includegraphics[width=\columnwidth]{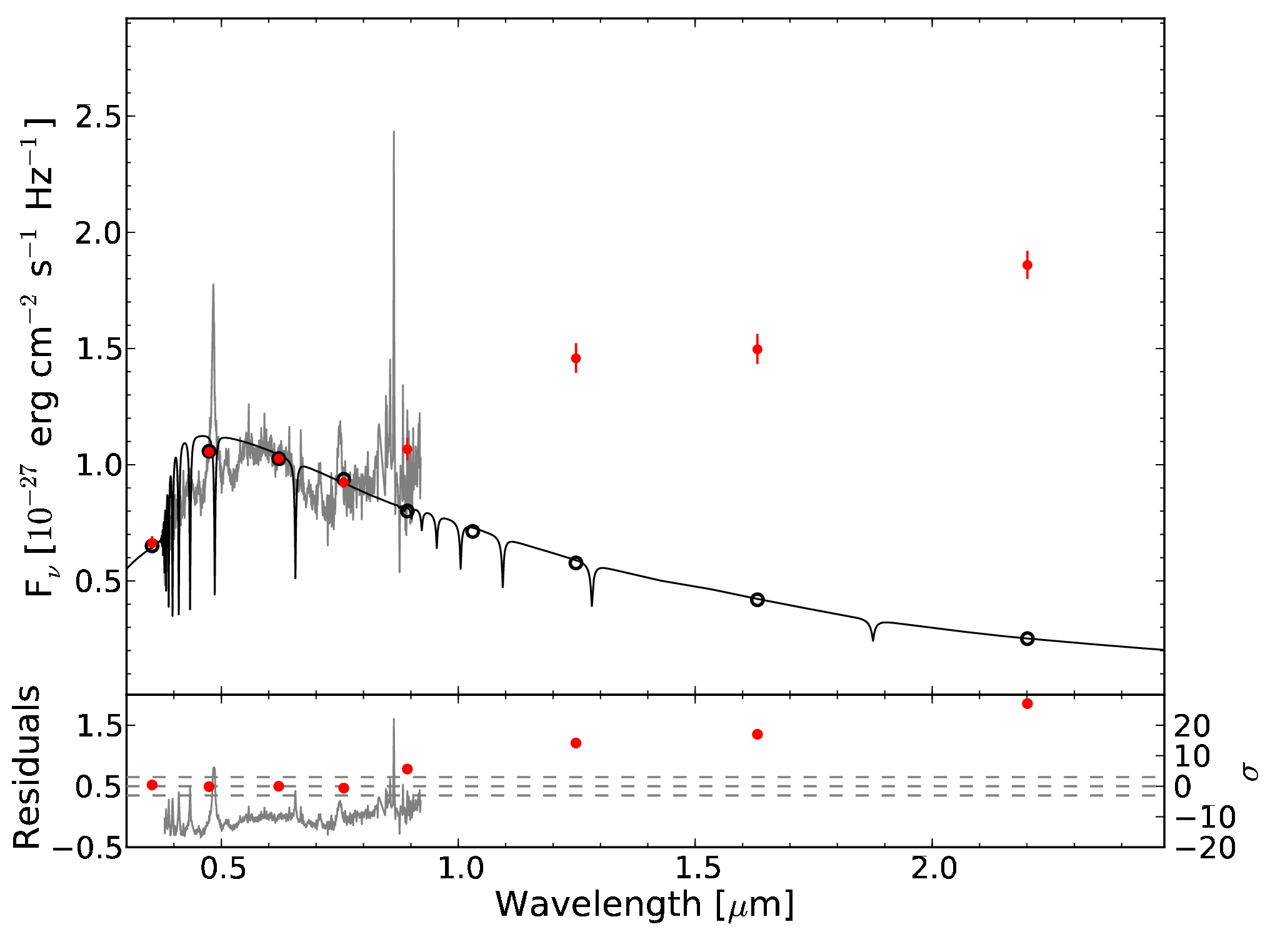}
\caption{\label{f-0046} The SED of SDSS\,J0046$-$0044, an example of a
  quasar, that was selected as a candidate DA white dwarf (based on
  its colours, Table\,\ref{e-poly}) with possible IR excess
  (``DA:ire:'') by the photometric method. The best-fit to the $ugri$
  photometry is found for $\Teff=9000$K and $\Logg=7.0$ at a distance
  of $37$pc and shown as black line. The excess over the model
  keeps rising steeply into the mid-IR and can not be modelled by any
  companion type at the distance of the (photometric) DA fit,
  identifying this object as a quasar contaminant.}
\end{figure}

As can be seen in Table\,\ref{t-lmc}, a significant fraction (323) of the
IR excess candidates from the photometric method with spectra are quasars.
This is almost the entire population of spectroscopically confirmed
quasars in the DA white dwarf sample with UKIDSS magnitudes. The remaining
5 quasars do not have sufficient IR data to show their quasar nature, but
equally are not flagged as IR excess candidates. This is caused by the
flat SED exhibited by QSOs \citep{coveyetal07-1}. When modelled with a DA
white dwarf, they generally have an effective temperature of $\sim8000$K
and have an IR excess that is much higher than physically possible for an
M-dwarf or brown dwarf companion at the distance of the (photometric) DA
fit (e.g. Fig.\,\ref{f-0046}). Because of the (apparent) low effective
temperatures, the quasars also generally have low (apparent) distances.
These properties, however, overlap with those of genuine cool nearby white
dwarfs and cannot be used to directly distinguish between white dwarfs and
quasar. Another common sign of QSOs are large jumps in brightness between
adjacent magnitudes (caused by emission lines) that are not seen in any of
the genuine DA white dwarfs with infrared excess. Given that we correctly
classified, on the base of their optical-IR SED, $99\%$ of (spectroscopic)
quasars contaminants that were found as DA:ire by the photometric method,
we are confident that we can identify the vast majority of quasars among
the photometric-only sample.

Among the photometric-only sample of IR excess candidates, we find
38 objects whose SEDs are very similar to those of our 323-strong
spectroscopicially confirmed quasar sample and which we therefore believe
to be quasars as well. Optical spectroscopy is needed to confirm their
nature. These have been removed from Table\,\ref{t-poa} and can be found
in the online QSO tables available via CDS.

\subsubsection{Contamination by NLHS and Non-DA White Dwarfs}
\label{sss-nlhs}

Contaminant NLHS and non-DA white dwarfs are more difficult to identify
and remove from the photometric-only DA candidate sample, as their overall
SEDs are all rather similar. Closer inspection of the SDSS spectroscopy of
the NLHS and cross-checking them in Simbad suggests that a large fraction
of them are subdwarf B (sdB) stars.  The bulk of subdwarfs are believed to
have formed in binary interactions \citep{hanetal03-1, heber09-1} and
therefore it is expected that a large majority will still have companions.
Such companions would cause an IR flux excess over the Rayleigh-Jeans tail
of the subdwarf, and it is hence likely that our photometric-only sample
of candidate DA white dwarfs with infrared excess contains a significant
contamination from sdB plus low-mass companion binaries.

Given that we fit all photometric objects with DA model spectra, we may
expect some imperfections in the fits to the photometric NLHS objects.
Nevertheless, the detection of a near-IR flux excess over the
Rayleigh-Jeans tail of the model is likely to be correct for many of the
NLHS objects among the photometric sample. However, a DA fit to the
photometry of a physically much larger NLHS object will dramatically
underestimate its radius, and hence its distance. Consequently, fitting
the companion with the composite model (Sect.\,\ref{ss-ire_mod}) will
result in a spectral type of the companion that is much too late.

\begin{figure*}
\begin{minipage}{2\columnwidth}
\includegraphics[width=\columnwidth]{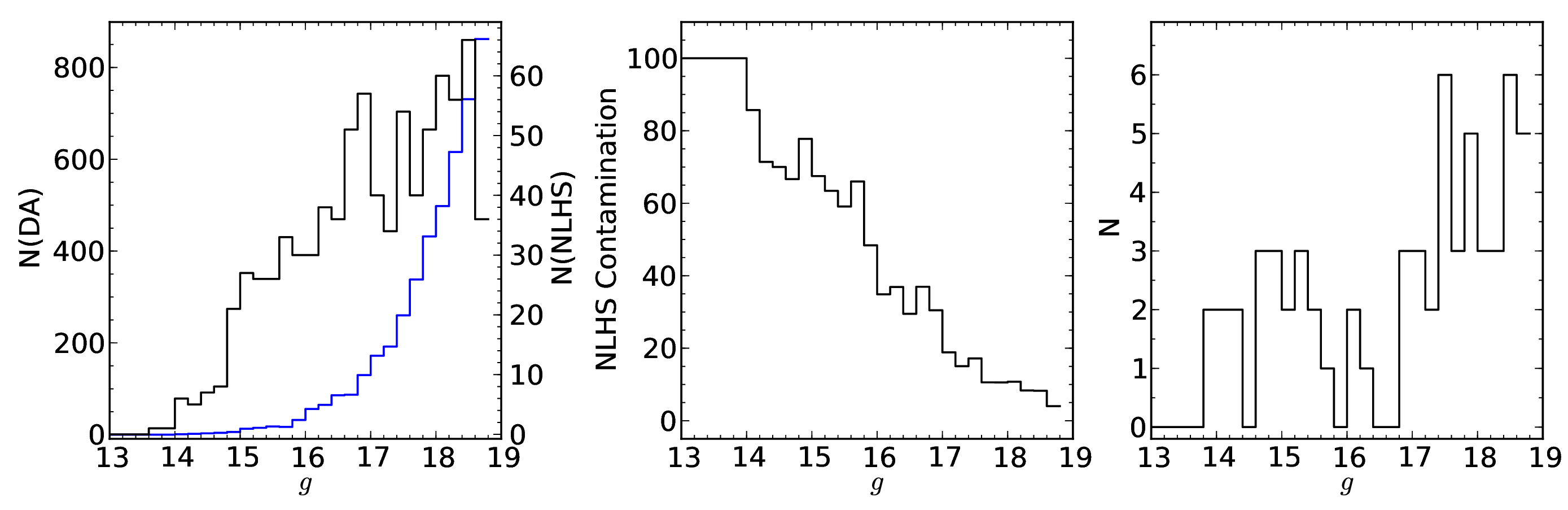}
\caption{\label{f-Pnlhs} Left panel: The distribution of DA white
  dwarfs (blue, left hand scale) and NLHS (black, right hand scale) as
  a function of $g$-band magnitude. Middle panel: the fraction of NLHS
  as a function of $g$-band magnitude. The probability of
  contamination by NLHS clearly drops with increasing $g$-band
  magnitude. Right panel: distribution of the photometric-only IR
  excess candidates, the brightest of these group are most likely NLHS
  rather than DA white dwarfs.}
\end{minipage}
\end{figure*}

From the spectroscopic sample, we find that the frequency of NLHS inside
the DA colour selection (Table\,\ref{e-poly}, Fig.\,\ref{f-colsel}) is
$14.6\%$ (Table\,\ref{t-effg}), however the level of contamination is
strongly magnitude-dependent. Figure\,\ref{f-Pnlhs} (left) shows the
distribution of NLHS and DA white dwarfs as a function of $g$-band
magnitude. The ratio of these two, and thus the expected level of
contamination of the photometric-only sample, is shown in the middle panel
(this assumes that the majority of quasar have been removed because of the
characteristic shape of their SED). The contamination of the
photometric-only DA candidate sample by NLHS drops significantly towards
fainter $g$ magnitudes.  Subdwarfs are $\sim100$ times brighter than white
dwarfs and therefore apparent magnitudes of $g\sim18-19$ sample distances
of many kpc. This is several times the exponential scale height of the
Galactic thick-disc population, and hence the number of sdBs at such large
distances is relatively small. Figure\,\ref{f-Pnlhs} (right panel) shows
the photometric-only IR excess candidates as a function of $g$ magnitude,
where we can assume that most objects with $g\lesssim16$ are likely to be
NLHS.

\begin{figure*}
\begin{minipage}{2\columnwidth}
\includegraphics[width=\columnwidth]{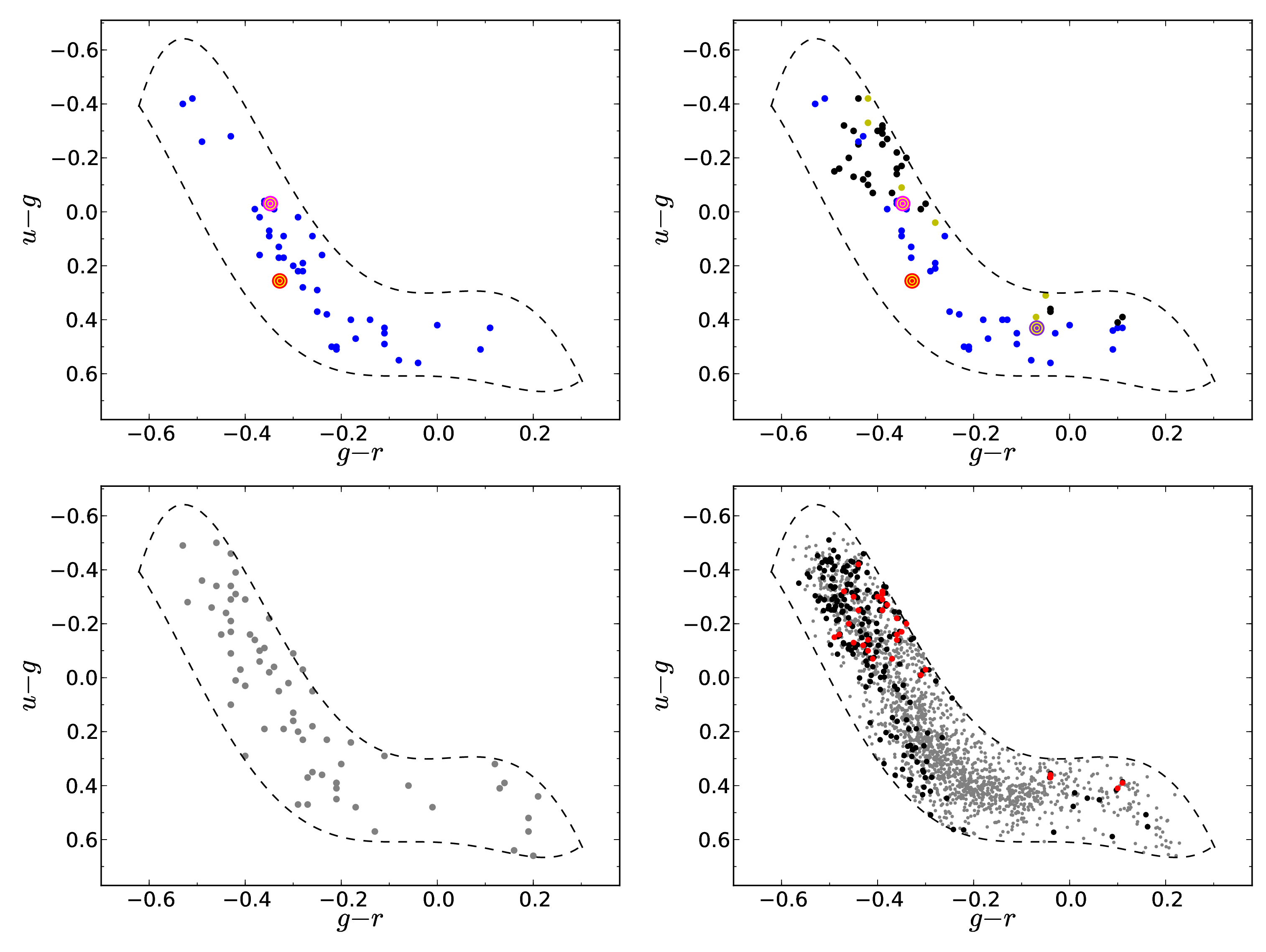}
\caption{\label{f-ire} Top left panel: the distribution of the 42
  spectroscopically confirmed DA white dwarfs with infrared excess
  (Table\,\ref{t-spa}) in the $(u-g,g-r)$ colour-colour space.  The
  black dashed line outlines the $(u-g,g-r)$ DA colour selection
  (Table\,\ref{e-poly}, ig.\,\ref{f-colsel}). Top right panel: the
  distribution of the 74 IR excess objects found by photometric
  method (excluding quasars and quasar candidates), with DA white
  dwarfs, non-DA white dwarfs and NLHS shown in blue, yellow, and
  black respectively.  SDSS\,J1228+1040 and SDSS\,J1043+0855, two DA
  white dwarfs known to have debris discs, are indicated by pink and
  red bulls-eye symbol, respectively (top left and right panel, see
  Sect.\,\ref{ss-wd1228}, \ref{ss-wd1043}). SDSS\,J1212+0136, a known
  magnetic white dwarf with a brown dwarf companion is shown as purple
  bulls-eye symbol (right panel only, see Sect.\,\ref{ss-wd1212}).
  Bottom left panel: the distribution of the 67 photometric-only
  infrared excess candidates. Bottom right panel: all photometric-only
  objects satisfying our DA constraint set from Table\,\ref{e-poly}
  (\UKP\ objects, gray dots), spectroscopically classified NLHS (209
  objects, black), and spectroscopically classified NLHS with an
  infrared excess (31 objects, red). Based on the analysis of the
  spectroscopic sample (Sect.\,\ref{sss-nlhs}), we estimate that
  $\sim12\%$ of the photometric-only objects are NLHS, with a strong
  concentration towards the (blue) top-end of the DA ``banana''.}
\end{minipage}
\end{figure*}

An additional clue on the NLHS vs DA classification of the
photometric-only objects comes from their location in the $(u-g,g-r)$
colour-colour diagram (Fig.\,\ref{f-ire}, bottom right panel), where the
majority of the spectroscopically confirmed NLHS objects are concentrated
at the blue end of the ``DA'' banana. We can therefore assume that NLHS
are the primary contaminants of the photometric sample in this region as
well (Fig.\,\ref{f-ire}, top right panel). 

Assuming that the distribution of NLHS contaminants is similar between the
spectroscopic and photometric-only sample, we expect $\sim259$ of the
\UKP\ photometric objects to be NLHS. When fitting the photometric sample,
31 ($12\%$) spectroscopically confirmed NLHS were found to have an IR
excess (red dots in the bottom right panel of Fig.\,\ref{f-ire}). Again,
assuming that the contamination among the spectroscopic and photometric
samples is similar, we would expect that $\sim37$ (12\% of 259) of our
photometric DA candidates with infrared excess (DA:ire and DA:ire:) are in
reality NLHS, primarily sdB stars with main sequence companions. These are
still interesting in their own right (e.g. Sect.\,\ref{s-pg0014+067}), but
not the primary focus of the IR excess search (a detailed investigation of
subdwarfs with main-sequence star companions will be published elsewhere,
Girven et al. (2011, in prep).

In contrast to the above discussion on NLHS, fitting non-DA white dwarfs
with DA white dwarf models is likely to provide a reasonably good estimate
of all system parameters, including the companion type~--~the downside
being that on the base of photometry only, it is nearly impossible to
differentiate between DA and non-DA white dwarfs. However, based on the
statistics of the \citet{eisensteinetal06-1}, we only expect a small level
of contamination by non-DA white dwarfs (Table\,\ref{t-effg}).

\subsubsection{Independent checks: proper motions and infrared colours}

Our method of identifying infrared excess candidates follows
\citet{tremblay+bergeron07-1}, i.e. fitting model spectra to SDSS
spectroscopy and photometry. In this section, we carry out an
independent investigation of our sample using colour-colour diagrams,
such as previously explored by e.g. \citet{wachteretal03-1} and
\citet{hoardetal07-1}, and test our classification of the
photometric-only systems by making use of proper motions.

\begin{figure*}
\begin{minipage}{2\columnwidth}
\centerline{\includegraphics[width=0.8\columnwidth]{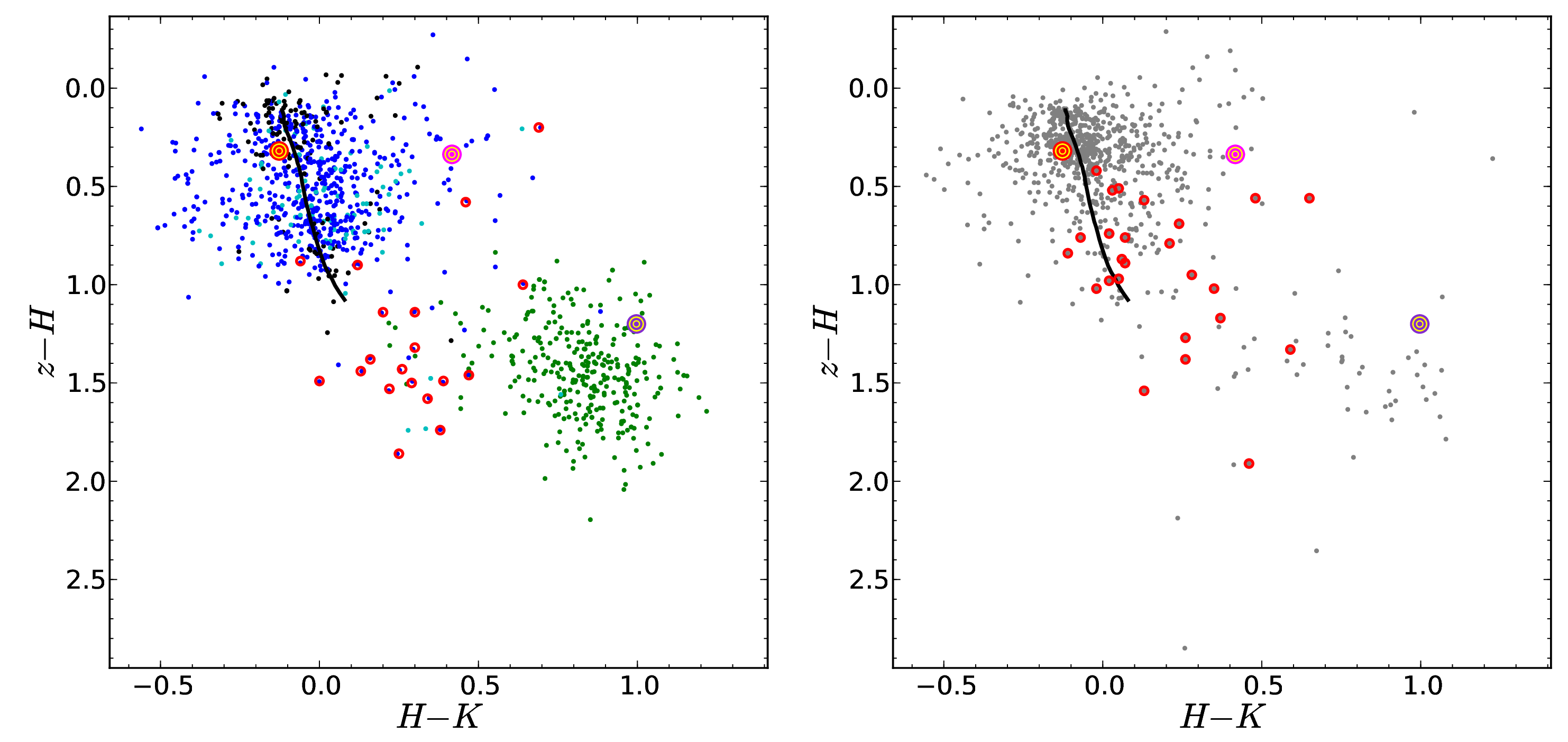}}
\caption{\label{f-zhhk} Location of the SDSS/UKIDSS sample in
  $(z-H,H-K)$ colour space. The left panel shows the objects with SDSS
  spectra, DA white dwarfs, NLHS, other white dwarfs and quasars in
  blue, black, cyan and green, respectively.  The 23 robust ``DAire''
  infrared excess objects from Table\,\ref{t-spa} are shown as red
  open circles. The positions of, SDSS\,J1228+1040, SDSS\,J1043+0855
  and SDSS\,J1212+0136, two white dwarfs in the sample known to have a
  debris disc and a DAH white dwarf plus brown dwarf binary (see
  Sect~\ref{ss-wd1228}, \ref{ss-wd1043}, \ref{ss-wd1212}), are
  indicated by pink, red and purple bulls-eye  symbols,
  respectively. A track of model DA white dwarf colours for
  $6000\mathrm{K}\leq\Teff\leq100000\mathrm{K}$ and $\Logg=8.0$ is
  plotted as a solid black line. High temperature white dwarfs are
  found at the top left end of the line. The scatter about this line
  cannot be explained by a spread in \Logg\ alone, but by
  uncertainties on the magnitudes (primarily those from UKIDSS) and
  background contamination. The photometric-only objects are shown in
  the right hand panel, with the 29 robust ``DA:ire'' infrared
  excess candidates (Table\,\ref{t-poa}) again highlighted in red.}
\end{minipage}
\end{figure*}

Figure\,\ref{f-zhhk} shows the distribution of the spectroscopic
SDSS/UKIDSS sample in the $(z-H,H-K)$ colour-colour space (left panel).
Using model white dwarf colours, it can be seen that the DA cooling
sequence runs from top left to bottom right through the white dwarf group.
The NLHS and DA white dwarfs are clearly separated from the quasars, which
can be understood as stars are on the Rayleigh-Jeans tail in the infrared,
whereas quasar follow a flatter power law \citep{coveyetal07-1}.
Therefore, as an additional test of how reliable our classification of the
photometric-only quasar candidates works, we inspected the infrared
colours of the photometric-only sample. Choosing an empirical cut of
$H-K>0.627$ selects \QSOrem\% of the quasar contaminants in the
spectroscopic sample. Adopting the same $H-K$ cut for the photometric-only
sample flags 38 objects as quasar candidates. This includes 31 of the 38
photometric-only objects identified as quasar candidates on the base of
their SED (Sect.\,\ref{sss-qso} (see online QSO table), of which only 37
have both $H$ and $K$ measurements). In addition, two of the $H-K$
selected QSO candidates correspond to the two ``weak'' photometric-only
quasar candidates listed in Table\,\ref{t-poa}.

A third concentration of objects can be seen in Fig.\,\ref{f-zhhk} in the
region between the DA white dwarfs and the quasars, a significant fraction
of which are DA infrared excess candidates. The presence of an infrared
excess over the stellar flux distribution results in a flatter spectral
slope, and hence moves these objects closer to the quasar locus. On closer
inspection, some white dwarfs in this region are found to be blended
sources or suffer from background contamination from a nearby galaxy and
therefore were not included as infrared excess candidates. 

In summary, adopting $H-K>0.627$ would efficiently remove the bulk of
contaminating quasars from the photometric-only sample, however, such a
cut would also remove a handful of genuine white dwarfs with the largest
infrared excess emission (such as e.g. SDSS\,J1228+1040, see
Sect.\,\ref{ss-wd1228}).

\begin{figure*}
\begin{minipage}{2\columnwidth}
\centerline{\includegraphics[width=\columnwidth]{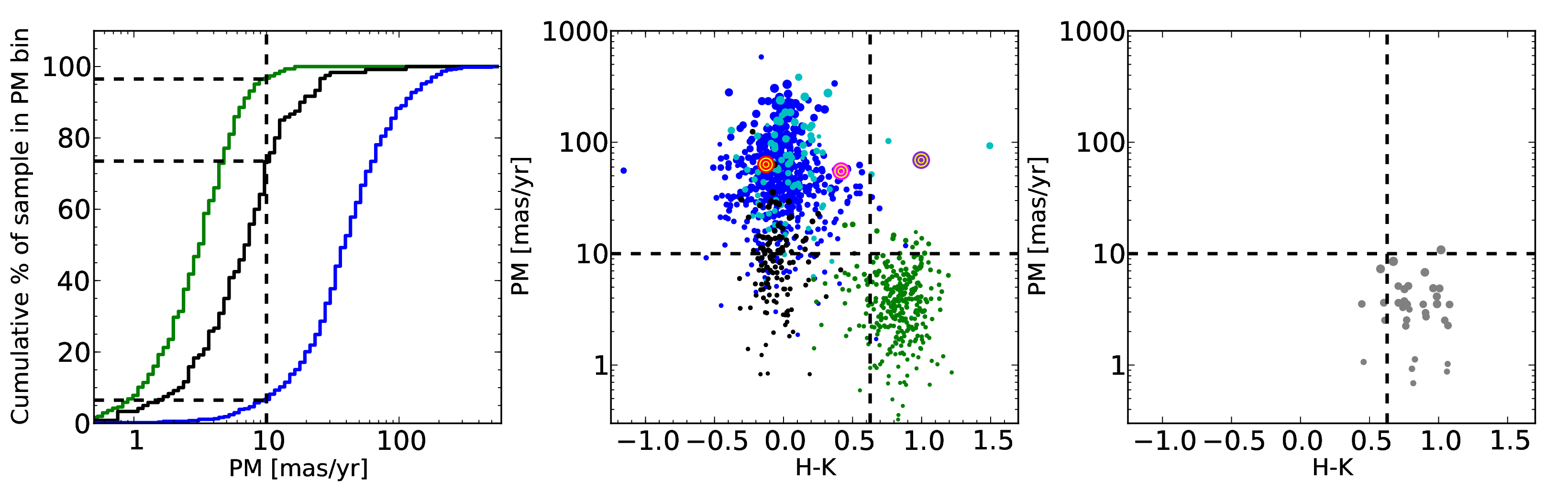}}
\caption{\label{f-pm} Left panel: The cumulative histogram of
spectroscopically confirmed DA white dwarfs (blue), NLHS (black) and
quasar (green) as a function of proper motion (PM). Black dashed lines
show a proper motion of $10\mathrm{mas/yr}$ and corresponding population
levels. $7\%$ of DA white dwarfs are contained within the bin of
$\mathrm{PM}\leq10\mathrm{mas/yr}$. Similarly, $74\%$ of NLHS and and
$97\%$ of quasar. Middle panel: distribution of proper motions as a
function of $H-K$ colour, using the same colour-coding, and showing in
addition non-DA white dwarfs in cyan. The significance of the proper
motion is encoded in the size of the points, where larger points
denote more significant proper motions, clipped at a maximum
of $30\sigma$. Black dashed lines show $\mathrm{PM}=10\mathrm{mas/yr}$ and
$H-K=0.627$. The position of the three benchmark objects SDSS\,J1228+1040,
SDSS\,J1043+0855 and SDSS\,J1212+0136 (see Sect~\ref{ss-wd1228},
\ref{ss-wd1043}, \ref{ss-wd1212}) are indicated by pink, red and purple
bulls-eye symbols respectively. Right panel: same as the middle panel, but
showing the (35) quasar candidates identified on the basis of
their SED shape (see online QSO table) which have both $H$ and $K$ band
data in UKIDSS and proper motions in SDSS\,DR7.}
\end{minipage}
\end{figure*}

Given that white dwarfs are nearby low-luminosity objects, they are
expected to exhibit larger proper motions than the more luminous NLHS, and
quasars are not expected to show any significant proper motion at all.
Thus, proper motions can be used to distinguish between white dwarfs and
quasars in the region with $H-K>0.627$. We have retrieved proper motions
from DR7 for all objects in the SDSS/UKIDSS sample. Figure\,\ref{f-pm}
(left panel) shows a cumulative proper motion distribution for the white
dwarfs, quasars and NLHSs. Based upon this, we chose a cut in proper
motion of $\mathrm{p.m.}\leq10$\,mas/yr to define low proper motion
objects such as quasars. In the spectroscopic sample, this cut selects
$7\%$, $74\%$ and $97\%$ of the DA white dwarfs, NLHS and quasar
respectively, efficiently eliminating the majority of the quasar without
removing too many white dwarfs.

Figure\,\ref{f-pm} (middle panel) plots the magnitude of the proper motion
as a function of $H-K$. The statistical significance of the proper motion
is encoded in the size of the points, where larger points denote
more significant proper motions. As expected, the spectroscopically
objects classified as NLHS stars and quasars show very small proper
motions, which are in most cases consistent with zero. The right hand side
panel of Fig.\,\ref{f-pm} shows the location of 35 photometric-only
objects classified as quasar because of the characteristic shape of their
SEDs (see online QSO table) that have both $H-K$ colours and proper
motions. The vast majority of these objects are contained within
$H-K>0.627$ and $\mathrm{p.m.}\leq10$\,mas/yr, corroborating our SED-based
classification.

A small number of objects with $H-K>0.627$, i.e. within the ``quasar''
region, display large and statistically significant proper motions
(Fig.\,\ref{f-pm}, middle panel). These are listed in
Table\,\ref{t-qsopm}. Among those objects are three spectroscopically
confirmed DA white dwarfs. At closer inspection, the UKIDSS magnitudes of
SDSS\,J1244+0402 may be contaminated by a nearby background object, which
would lead to a spurious $H-K$ colour. The other two spectroscopic DA
white dwarfs, SDSS\,J0753+2447 and SDSS\,J1557+0916 have very red $H-K$
colours and high proper motions. They are therefore excellent IR excess
candidates. Applying the same procedure to the photometric-only DA
candidate sample, SDSS\,J0959$-$0200 is the strongest infrared excess
candidate among the three photometric-only objects. SDSS\,J1440+1223 and
SDSS\,J1509+0539 also appear to suffer from background contamination.

Another four spectroscopic objects with quasar-like IR colours and high
proper motions are classified as three magnetic white dwarfs (DAH) and one
DZ white dwarf. We would expect that the IR spectra of all of these
objects should be close to a Rayleigh-Jeans distribution, suggesting that
the inferred IR excess is probably real. In fact, one of the DAH,
SDSS\,J1212+0136 is a well-studied DAH plus brown dwarf binary
\citep{schmidtetal05-3}, which exhibits a genuine infrared excess
\citep{debesetal06-1} (see Sect.\,\ref{ss-wd1212}). As discussed in the
introduction, the metals seen in DAZ white dwarf atmospheres are from
recent or ongoing accretion \citep[e.g.][]{dupuisetal93-1,
koester+wilken06-1}. The same is true for cool DZ white dwarfs
\citep[e.g.][]{aannestadetal93-1}, and \citet{farihietal10-2} discuss the
potential connection between the large number of DZ white dwarfs, and the
DAZ white dwarfs with dusty debris discs thought to originate from the
tidal disruption of rocky asteroids. The DZ found here, SDSS\,J1342+0522,
exhibits a very red $H-K$ colour.  The excess in $K$ over the model
spectrum is slightly under $3\sigma$ and therefore is, in our
classification scheme, only a marginal candidate for having an IR excess.
This DZ white dwarf and the two new DAH white dwarfs IR excess candidates
warrant further investigation.

Finally, eight quasars have statistically significant proper motions
($\sim3-5\sigma$; see Table\,\ref{t-qsopm}), which highlights the fact
that the SDSS vs USNO-B proper motions have to be considered with a pinch
of salt: among a total of 328 quasars with SDSS spectra and UKIDSS data,
we would expect only one to have a $3\sigma$ significant proper motion,
and none at $4\sigma$.

We conclude that dissecting the white dwarf sample selected with our
constraint set (Table\,\ref{e-poly}) using colour-colour diagrams and
proper motions leads to mutually consistent results when compared to our
primary methodology (Sect.\,\ref{ss-dafit}-~\ref{ss-ire_mod}), but the
spectroscopic modelling provides an additional wealth of information. 

\begin{table}
\setlength{\tabcolsep}{0.9ex}
\caption{\label{t-qsopm} Objects with quasar-like infrared colours
($H-K\geq0.627$), but large ($\geq10$mas/yr) and statistically significant
($>3\sigma$) proper motions. The significance of the proper motions is
listed as $\sigma_\mathrm{p.m.}$. Classifications are given for the
objects that have SDSS spectra. The comment ``BG object'' refers to a
second resolved source being seen in the UKIDSS images. These are most
likely background galaxies which would significantly affect the $H-K$
colour. For full coordinates, refer to the online spectroscopic and
photometric-only tables available via CDS.}
\begin{tabular}{llllll}
\hline \hline
Name & $H-K$ & p.m. (mas/yr) & $\sigma_\mathrm{p.m.}$ & Class & Comment \\
\hline
0043+0005 & $0.79\pm0.06$ & $13.68\pm3.06$ & $4.5$ & QSO &  \\
0753+2447 & $0.64\pm0.24$ & $32.16\pm3.20$ & $10.0$ & DA &  \\
0858+0938 & $0.88\pm0.03$ & $13.13\pm3.01$ & $4.4$ & QSO &  \\
0959-0200 & $0.65\pm0.15$ & $30.25\pm3.10$ & $9.8$ & - &  \\
1031+0341 & $1.05\pm0.08$ & $12.25\pm3.92$ & $3.1$ & QSO &  \\
1142+1347 & $1.02\pm0.05$ & $10.10\pm2.81$ & $3.6$ & QSO &  \\
1212+0136 & $1.00\pm0.06$ & $69.19\pm3.21$ & $21.6$ & DAH &  $(1)$\\
1244+0402 & $0.88\pm0.17$ & $11.80\pm3.09$ & $3.8$ & DA & BG object \\
1250+1549 & $1.49\pm0.02$ & $93.14\pm3.12$ & $29.9$ & DAH &  \\
1342+0522 & $0.64\pm0.27$ & $51.53\pm3.24$ & $15.9$ & DZ &  \\
1427-0054 & $0.68\pm0.04$ & $15.99\pm3.25$ & $4.9$ & QSO &  \\
1440+1223 & $1.43\pm0.08$ & $88.83\pm2.65$ & $33.5$ & - & BG object \\
1443+0910 & $0.96\pm0.02$ & $12.46\pm2.98$ & $4.2$ & QSO &  \\
1509+0539 & $0.98\pm0.25$ & $48.29\pm3.07$ & $15.7$ & - & BG object \\
1514+0744 & $0.76\pm0.07$ & $102.64\pm5.06$ & $20.3$ & DAH &  \\
1553+0718 & $0.80\pm0.01$ & $14.64\pm2.75$ & $5.3$ & QSO &  \\
1557+0916 & $0.70\pm0.21$ & $25.48\pm2.85$ & $8.9$ & DA &  \\
1557+2646 & $0.84\pm0.04$ & $10.07\pm3.21$ & $3.1$ & QSO &  \\
\hline
\end{tabular}
(1)~~A magnetic cataclysmic variable in a low state
\citep{schmidtetal05-3, debesetal06-1, burleighetal06-1,
  farihietal08-2, howelletal08-1, linnelletal10-1}.
\end{table}

\subsection{Overall Numbers}
\label{ss-sum}

A summary of the numbers of objects at each stage of the analysis is given
in Table\,\ref{t-nums}. The total numbers of white dwarfs with near-IR
excess are broken down according to the spectral type of their companions
in Table\,\ref{t-lmc}.

We find that $3.3\%$ (42 of \UKDA) of the SDSS spectroscopically confirmed
DA white dwarfs with at least one of the $YJHK$ UKIDSS magnitudes have an
IR excess and are therefore candidates for having a companion or a debris
disc. However, this does not take account of the fact that the sample of
white dwarfs is incomplete even within UKIDSS DR8 because for many only
subsets of the IR magnitudes are available. Thus we are limited by UKIDSS
coverage and the real number is higher. $2.0\%$ of the spectroscopic DA
white dwarfs are candidates for having a companion of type L0 or later,
i.e. have brown dwarf companions. Similarly, $0.5\%$ are promising disc
candidates, having an excess compatible with a companion type of L7 or
later. If we only discuss the objects with a detection in the
K-band (required for detecting a disk), $1.2\%$ are disk candidates, and
only including objects where we are confident of the IR excess (not
``DAire:'' in Table\,\ref{t-spa}; see Section\,\ref{ss-ire_det} and
\ref{ss-ire_mod}), a lower limit of $0.8\%$ of DA white dwarfs have a
brown dwarf companion.

For the photometric-only sample, where we have fitted the SDSS and UKIDSS
photometry, $5.9\%$ are IR excess candidates. However, this number will be
affected by the efficiency of white dwarf selection (\Effi\%) and the
efficiency of removing contaminants. Assuming that we remove all the
obvious photometric-only quasar contaminants (38, see online QSO table)
from the IR excess objects, the remaining 67 photometric-only IR excess
candidates will be either DA white dwarfs or NLHS. As discussed
previously, we estimate that $\sim37$ of these are NLHS
(Section\,\ref{sss-nlhs}). This number is very similar to the amount of
``bad fits'' found in from the photometric method and so we believe the
majority of ``bad fits'' come about from NLHS contaminants in the
photometric-only sample. Therefore, we expect $\sim30$ ($2.7\%$ of
$\sim1103$) genuine DA white dwarfs with IR excess among the
photometric-only DA white dwarf candidates. These infrared excess
candidates show a similar distribution in companion type when compared to
the spectroscopic sample. Considering only the objects we are confident of
the excess (not ``DA:ire:'' in Table\,\ref{t-poa}; see
Section\,\ref{ss-ire_det} and \ref{ss-ire_mod}), $1.8\%$ (19) of the
photometric-only DA white dwarfs candidates are likely to have a brown
dwarf companion.

\begin{table}
\caption{\label{t-nums} Summary of numbers at each stage of the
processing. The columns are split for the spectroscopic and photometric
methods. The spectroscopic sample is also further split by the
classification of the optical spectra.}
\begin{tabular}{lll}
\hline \hline
Constraint  & Spectroscopic & Photometric-only \\
 & Objects & Objects \\ \hline
Objects Satisfying SDSS &  &  \\
Colour Cuts & \SDS & \SDP \\ \hline
Spectroscopically Confirmed &  & \\
\,\,\,\,DA white dwarfs & 4636 & \\
\,\,\,\,QSO & 1280 & \\
\,\,\,\,NLHS & 840 & \\
\,\,\,\,Other white dwarfs & 661 & \\
\,\,\,\,Other Objects & 27 & \\ \hline
Objects cross matched with & & \\
UKIDSS with detection in & & \\
any of $Y$, $J$, $H$ or $K$ & \UKS & \UKP \\ \hline
Spectroscopically Confirmed &  & \\
\,\,\,\,DA white dwarfs & \UKDA & \\
\,\,\,\,QSO & 328 & \\
\,\,\,\,NLHS & 209 & \\
\,\,\,\,Other white dwarfs & 172 & \\
\,\,\,\,Other Objects & 6 & \\ \hline
Objects with Detections in &  &  \\
$H$ \& $K$ & 1075 & 809 \\
$K$ & 1108 & 840 \\ \hline
Spectroscopically Confirmed &  & \\
\,\,\,\,DA white dwarfs & 571 & \\
\,\,\,\,QSO & 316 & \\
\,\,\,\,NLHS & 124 & \\
\,\,\,\,Other white dwarfs & 94 & \\
\,\,\,\,Other Objects & 3 & \\ \hline
IR Excess Objects & 42 & 105 \\
\hline
\end{tabular}
\end{table}

\section{Notes on individual Objects}
\label{s-wdwire}

Some objects of particular interest identified in the SDSS / UKIDSS
cross-correlation are discussed below. We separate those into objects
already known to host a disc or companion (Section\,\ref{ss-pwd}), notable
objects from the spectroscopic sample (Section\,\ref{ss-sire}), and the
same from the photometric-only sample (Section\,\ref{ss-pire}).

\subsection{Prototypical White Dwarfs}
\label{ss-pwd}

Our spectroscopic sample contains two white dwarfs that were known
to host debris discs and one magnetic white dwarf with a substellar
companion, and they serve hence as a benchmark for our selection
procedures. 

\subsubsection{SDSS\,J1228+1040}
\label{ss-wd1228}

\begin{figure*}
\includegraphics[width=\columnwidth]{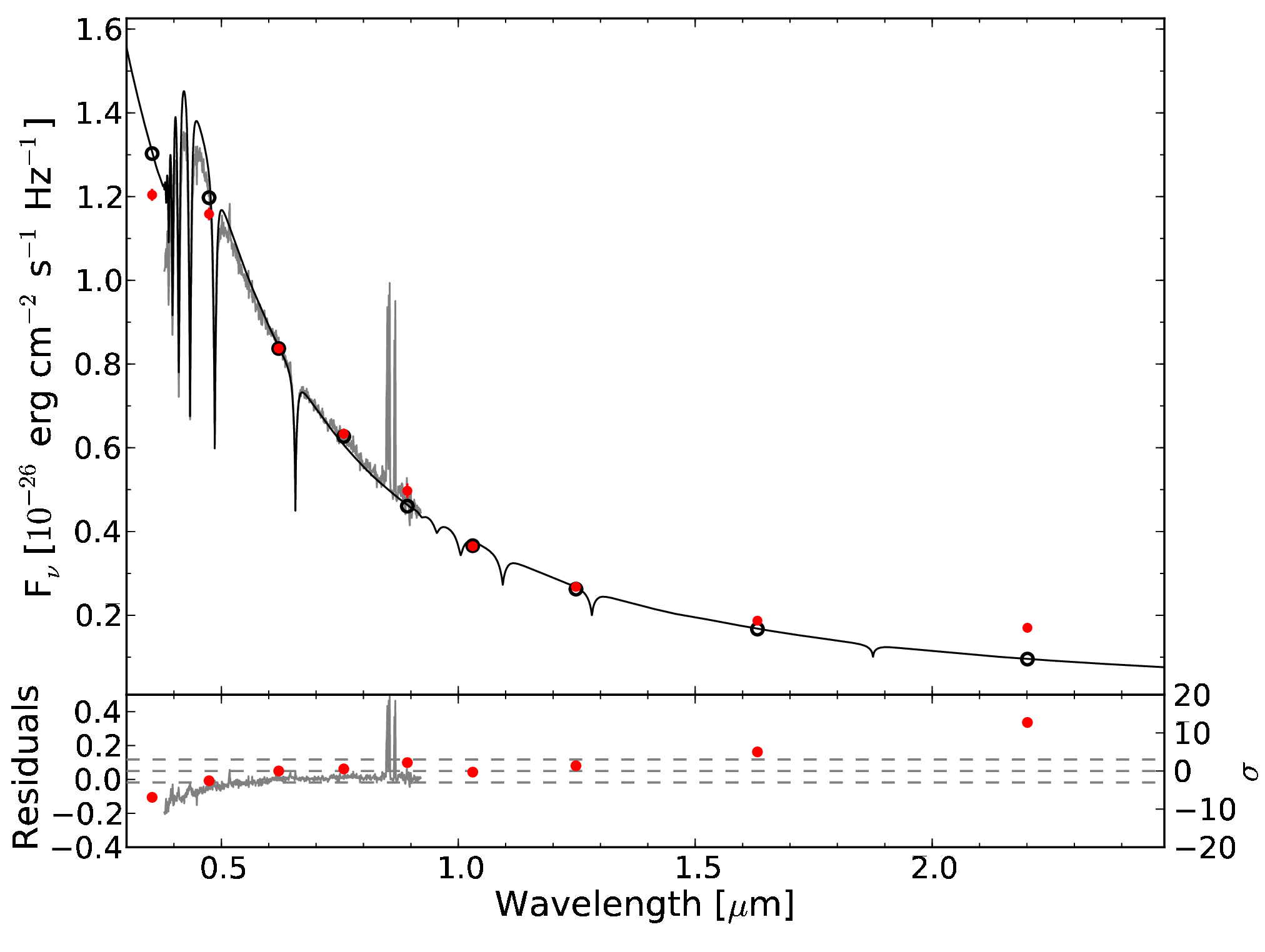}
\includegraphics[width=\columnwidth]{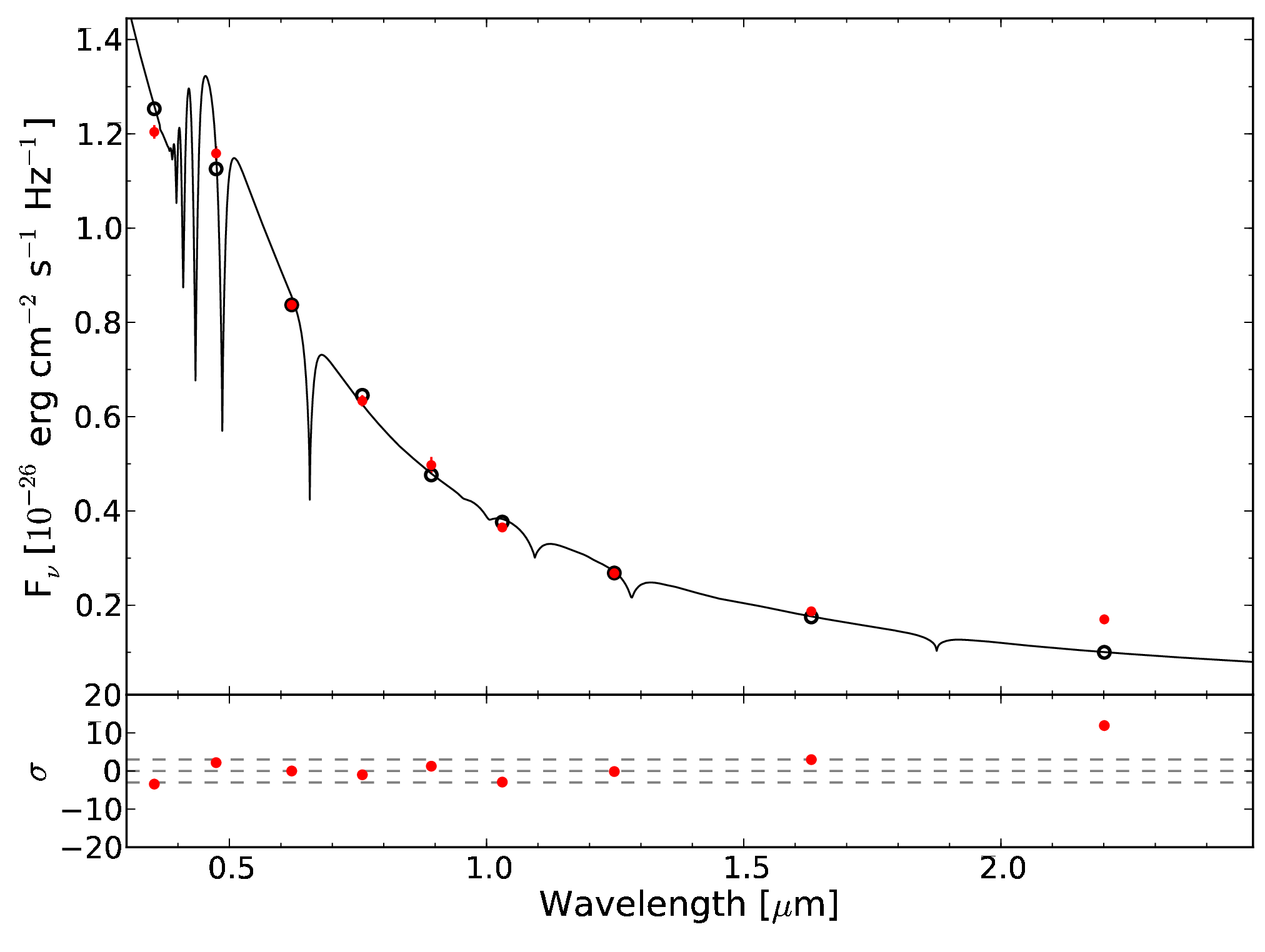}
\caption{\label{f-wdire1228} SDSS\,J1228+1040 is one of two DA white
  dwarfs in our SDSS/UKIDSS sample that are known to have a gaseous
  debris disc \citep{gaensickeetal06-3}. It also exhibits a
  substantial infrared excess in VLT/ISAAC and \textit{Spitzer}
  observations \citep{brinkworthetal09-1}, and hence serves as a
  general benchmark for our method. Fitting the SDSS spectrum (left
  panel) results in $\Teff = 22037\pm199$K, $\Logg=8.19\pm0.04$,
  implying $\Mwd=0.74\pm0.02\Msun$, and a distance of $134\pm3$pc,
  consistent with the parameters derived by
  \citet{gaensickeetal06-3}. The SDSS spectrum is shown in gray, the
  best-fit model in black, observed $ugrizYJHK$ fluxes in red, and
  model fluxes in these bands as black circles. The lower panel shows
  the residuals from the fit (in flux units on the left axis, and
  statistical significance on the right axis). The corresponding fit
  to the $ugri$ photometry gives a somewhat lower temperature, $\Teff
  = 20000\pm^{10}_{10}$K, $\Logg = 8.5\pm^{0.03}_{0.13}$K, which has
  no noticeable effect on the detection of the infrared excess (right
  panel).}
\end{figure*}

SDSS\,J1228+1040 is one of two DA white dwarfs in our SDSS/UKIDSS
sample known to have a debris disc. The disc was initially identified
because of the highly unusual emission lines of the Ca\,II
8200\,\AA\ triplet \citep{gaensickeetal06-3}, whose double-peaked
shaped can only be explained by metal-rich gas orbiting the white
dwarf within its tidal disruption radius
\citep{gaensickeetal06-3}. Near- and mid-IR observations revealed a
substantial infrared excess over the white dwarf, unambiguously
identifying a dusty component of the debris disc, in addition to the
gaseous one \citep{brinkworthetal09-1}. Our fits to the SDSS
spectroscopy and photometry are shown in Fig.\,\ref{f-wdire1228}. The
two fits differ in \Teff\ by 2000\,K. However, this has very little
effect on the extrapolated infrared magnitudes of the white dwarf, and
the object shows a $3\sigma$ excess in $H$ and $12\sigma$ in $K$,
independent of the method used for fitting. At the temperature of
SDSS\,J1228+1040, \Logg\ is not well constrained from fitting the
$ugri$ photometry alone, however, this primarily effects the widths of
the hydrogen lines, and has a negligible effect on the spectral slope
of the white dwarf model. This demonstrates that the detection of a
genuine infrared excess is robust and independent of whether an
optical spectrum is available.

Taking the infrared measurements at face value and ignoring our
knowledge about this star, we have modelled the SDSS/UKIDSS spectral
energy distribution as a white dwarf plus low-mass companion, which
results in a most likely spectral type of $\ge$L6 for the
companion. Based on the SDSS/UKIDSS data alone, it is impossible to
distinguish between a low-mass companion and a dusty disc, but
mid-infrared data data can break this degeneracy. We therefore
classify infrared excess candidates that require a companion later
than L7--8 as brown dwarf / dusty debris disc candidates

SDSS\,J1228+1040 exhibits a very red $H-K$ colour in the $(z-H, H-K$)
colour-colour diagram shown in Fig.\,\ref{f-zhhk}, where it is clearly
separated from the white dwarf model sequence. This region of the
colour-colour space is therefore likely to harbour white dwarfs with a
strong $K$-band excess. A significant number of the IR excess candidates
in Sect.~\ref{ss-ire_det} also lie in this region, as well as somewhat
below (corresponding to a $H$ and $K$-band excess). SDSS\,J1228+1040 is
also prominent in Fig.\,\ref{f-pm} thanks to its relatively high proper
motion and red $H-K$ colour. In summary, it is encouraging that this
benchmark system indeed stands out in the various diagnostics we have
considered.

\subsubsection{SDSS\,J1043+0855}
\label{ss-wd1043}

\begin{figure*}
\includegraphics[width=\columnwidth]{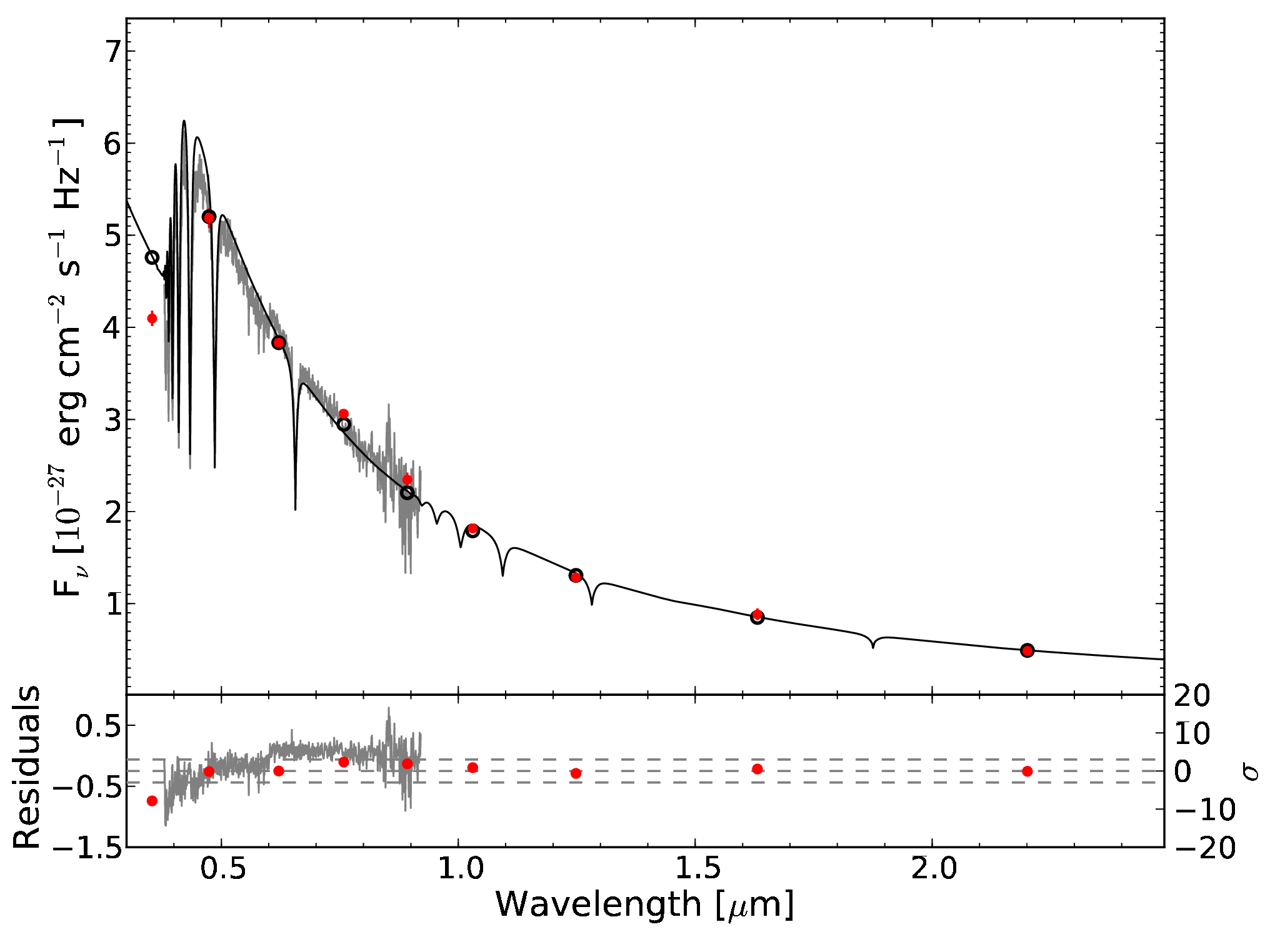}
\includegraphics[width=\columnwidth]{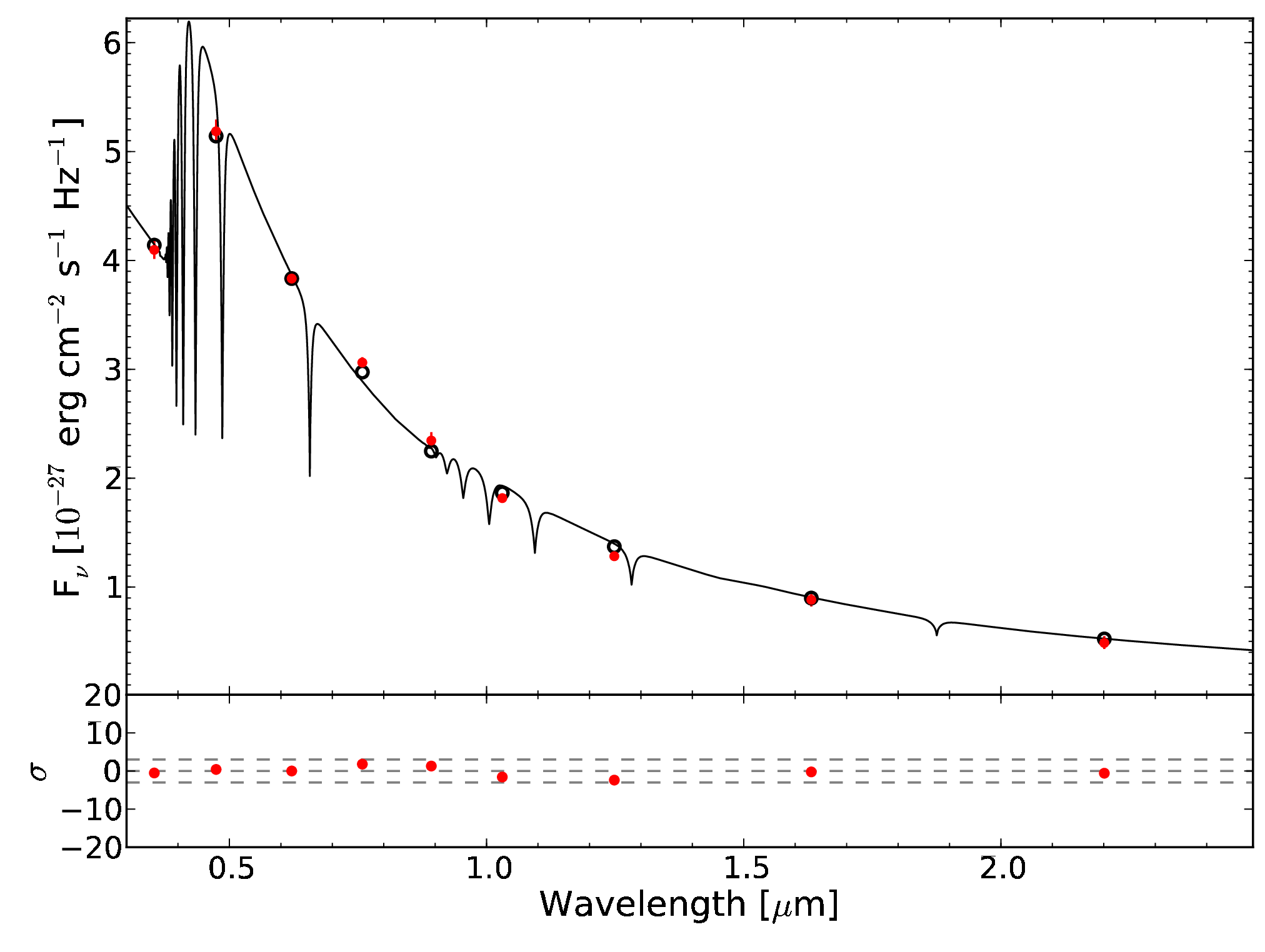}
\caption{\label{f-wd1043} SDSS\,J1043+0855 is the second of only two
  DA white dwarfs in our SDSS/UKIDSS sample that is known to have a
  dusty debris disc. This object was discovered through optical Ca
  emission \citep{gaensickeetal07-1} and has at best a marginal flux
  excess in the \textit{Spitzer} mid-IR \citep[][Brinkworth et
  al. 2011, in prep.]{melisetal10-1}. Fitting the SDSS spectrum results in
  $\Teff = 17912\pm360$K, $\Logg=8.07\pm0.08$, the corresponding fit to
  the $ugriz$ photometry results in $\Teff = 16000\pm^{170}_{100}$K,
  $\Logg = 7.5\pm^{0.42}_{0.8}$K (right panel). No IR excess is detected
  in the near-IR using either method, consistent with the inconspicuous
  location of SDSS\,J1043+0855 in the $(z-H, H-K)$ and proper motion vs
  $H-K$ diagrams (Figs.~\ref{f-zhhk} and \ref{f-pm}, respectively).}
\end{figure*}

SDSS\,J1043+0855 is the second white dwarf in our sample known to have a
gaseous debris disc \citep{gaensickeetal07-1}, though the evidence for an
infrared excess in the near- and mid-IR is marginal \citep[][Brinkworth et
al. 2011, in prep.]{melisetal10-1}. Based on the published results on this
object, we would not expect to detect any excess in the UKIDSS photometry.
In our analysis of the SDSS spectrum, we find $\Teff = 17912\pm360$\,K and
$\Logg=8.07\pm0.08$, consistent within the errors with the parameters in
\citet{gaensickeetal07-1}. The corresponding model is an excellent fit to
both the SDSS and UKIDSS photometry, with no detection of an IR excess in
any of the near-IR bands (Fig.\,\ref{f-wd1043}). Adopting the $K_s$ flux
from \citet{melisetal10-1} rather than the UKIDSS measurement gives a
$2\sigma$ excess above our model, which would again not classify as an IR
excess candidate within our classification scheme
(Sect.~\ref{ss-ire_det}). This conclusion is confirmed by the fact that
SDSS\,J1043+0855 falls very close to the model white dwarf sequence in
Fig.\,\ref{f-zhhk}.  This system is an example where there either is no
dusty disc, or where the dusty disc is too faint, e.g. edge-on, to be
detected in the near-IR.

\subsubsection{SDSS\,J1212+0136}
\label{ss-wd1212}

\begin{figure}
\includegraphics[width=\columnwidth]{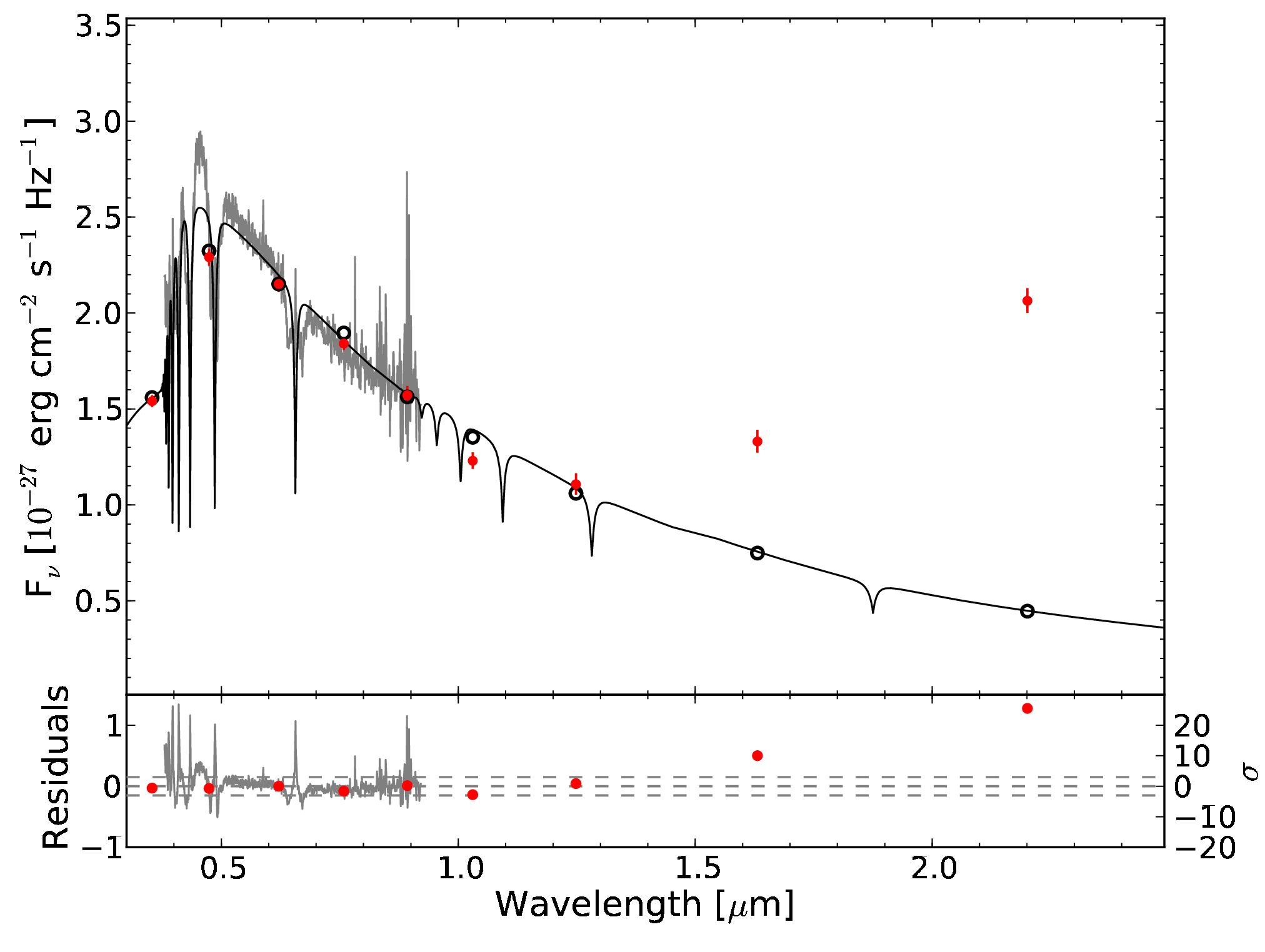}
\caption{\label{f-1212} SDSS\,J1212+0136, a short-period binary
  containing a magnetic DA(H) white dwarf plus a brown dwarf companion
  \citep{schmidtetal05-3}. The system is undergoing weak mass
  transfer, producing cyclotron emission that contributes to the
  observed near-IR excess \citep{debesetal06-1, burleighetal06-1,
    farihietal08-2}. SDSS\,J1212+0136 was picked up by our fit to all
  the photometric objects satisfying the colour cuts designed to find
  DA white dwarfs (Table\,\ref{e-poly}).}
\end{figure}

SDSS\,J1212+0136 is one of three magnetic (DAH) white dwarfs that were
selected by our colour cut (Table\,\ref{e-poly}) and that, when
fitting their $ugri$ photometry, show a substantial IR flux excess
(Fig.\,\ref{f-1212}, Sect.~\ref{sss-nlhs}). This white dwarf was first
reported to have a magnetic field of $\simeq13$\,MG by
\citet{schmidtetal03-1}. \citet{schmidtetal05-3} subsequently detected
a weak H$\alpha$ emission line, from which they measured an orbital
period of $\sim90$\,min. Based on the $J$ band magnitude of
SDSSJ1212+0136, \citet{schmidtetal05-3} concluded that the companion
to the white dwarf is a brown dwarf. Additional studies in the near-IR
confirmed the brown dwarf to have a spectral type in the range L5--L8,
and detected variable cyclotron emission, indicative of ongoing
accretion onto the magnetic white dwarf \citep{debesetal06-1,
  farihietal08-2, howelletal08-1}.  Observations at blue and
ultraviolet wavelengths show a quasi-sinusoidal flux modulation
interpreted as emission from an accretion-heated polecap
\citep{burleighetal06-1, linnelletal10-1} which is typical of strongly
magnetic cataclysmic variables \citep{gaensickeetal95-1,
  araujo-betancoretal05-2, gaensickeetal06-2}. Despite the fact that
no state of high accretion activity has been observed in
SDSS\,J1212+0136, all observational evidence suggests that is nearly a
twin of the prototypical magnetic cataclysmic variable EF\,Eri
\citep[e.g.][]{beuermannetal00-1}.

Taken on its own, the very red $H-K$ colour of SDSS\,J1212+0136 would
suggests it to be a quasar (Sect.~\ref{sss-qso},
Fig.~\ref{f-zhhk}). However, its stellar nature is confirmed by the
detection of a significant and large proper motion
(Fig.\,~\ref{f-pm}). Recovering this DAH white dwarf with a brown
dwarf companion shows that our methods to identify DA white dwarfs
with IR-excess are sufficiently robust to also find non-DA white
dwarfs with genuine IR excess.

\subsection{Example Spectroscopic IR Excess Candidates}
\label{ss-sire}

\subsubsection{SDSS\,J0135+1445}

SDSS\,J0135+1445 is a clear candidate for being a cool white dwarf
with a late-type stellar companion or brown dwarf. An excess is seen
to extend over all the UKIDSS bands in Fig.\,\ref{f-0135}. Modelling
of the companion suggests that its spectral type is between L7 and
L8. Spectral fitting implies a white dwarf of $\Teff=7467\pm18$K
at a distance of $69\pm2$pc. A $\Teff=8000\pm^{10}_{20}$K is
calculated from fitting of the photometry, broadly similar to the
recently discovered (resolved) DA plus brown dwarf binary PHL\,5038
\citep{steeleetal09-1}.

\subsubsection{SDSS\,J0753+2447}

\begin{figure}
\includegraphics[width=\columnwidth]{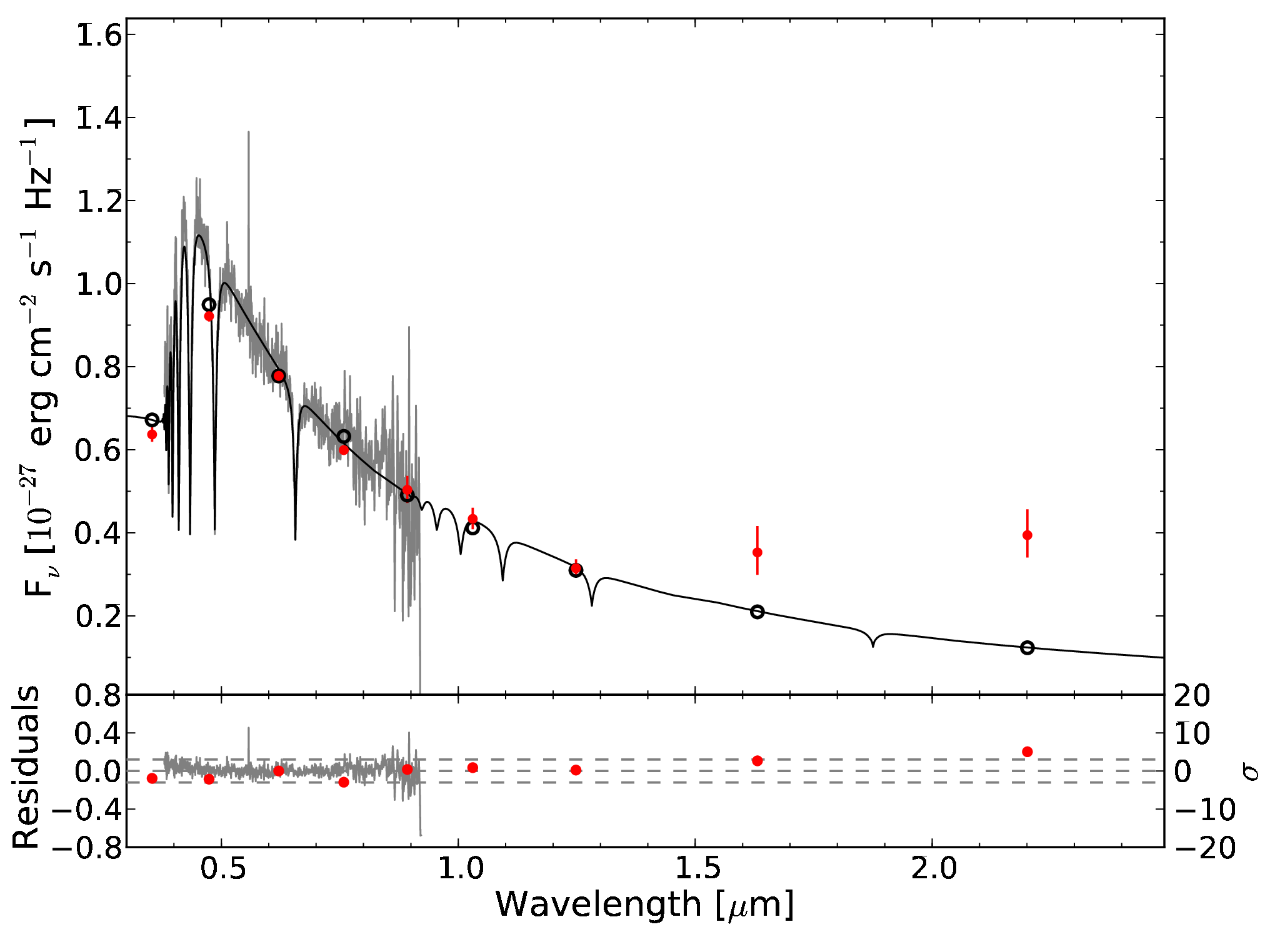}
\caption{\label{f-0753} SDSS\,J0753+2447, a DA white dwarf plus dusty
  disc or low-mass companion candidate. The best-fit to the SDSS
  spectrum (black line) gives $\Teff=13432\pm710$K and $\Logg=7.81\pm0.15$
  at a distance of $349\pm32$pc. The implied mass of the
  white dwarf is $0.50\pm0.08\Msun$. Fitting of the $4\sigma$ $K$-band
  excess with main sequence star models proved inconclusive, however,
  considering the similarity to SDSS\,1228+1040
  (Fig.\,~\ref{f-wdire1228}), this is an excellent candidate for
  having a brown dwarf companion or debris disc. Fitting the $ugri$
  photometry leads to $\Teff=12000\pm^{1130}_{290}$K and
  $\Logg=7.75\pm^{0.45}_{0.41}$ at $327^{18}_{14}$pc, with no change to
the
  conclusions as to the nature of the excess.}
\end{figure}

SDSS\,J0753+2447 is a very strong candidate for being a DA white dwarf
with a late-type brown dwarf or debris disc (Fig.\,\ref{f-0753}). The
fit to the SDSS spectrum implies a $\Teff=13432\pm710$K,
$\Logg=7.81\pm0.15$, with an implied distance $d=349\pm32$pc and white
dwarf mass $\Mwd=0.50\pm0.08\Msun$.  Fitting of the companion type was
inconclusive, but the shape of the SED is similar to that of the
benchmark object, SDSS\,J1228+1040. Therefore SDSS\,J0753+2447 is
classified as a brown dwarf or disc candidate.

\subsubsection{SDSS\,J1247+1035}

SDSS\,J1247+1035 is a candidate for having a brown dwarf companion or
dusty debris disk, however the UKIDSS $K$-band is only in excess by
$\sim3\sigma$ over the white dwarf model. Far-IR photometry of the object
is required to confirm the IR excess.

\subsubsection{SDSS\,J1557+0916}

UKIDSS photometry of SDSS\,J1557+0916 shows a $4\sigma$ $K$ band
excess for both the spectroscopic and photometric fitting methods. The
spectroscopy and photometric \Teff\ differ by 3800K, however, this
does not significantly affect the result. This is a good example of
where reddening is probably reducing the blue flux. Reddening the
white dwarf model spectrum by $E(B-V)\simeq0.05$ brings the overall
SED in line with the SDSS optical spectrum. Modelling of the companion
object proved inconclusive as to its spectral type
(Table\,\ref{t-spa}). SDSS\,J1557+0916 is a good candidate for having
a dusty disc or low-mass companion based on its IR spectral shape.

\subsubsection{SDSS\,J2220$-$0041}

PHL\,5038 (\,=\,SDSS\,J2220$-$0041) is a wide ($0.94\arcsec$) binary
containing a cool ($\sim8000$\,K) white dwarf with an $\sim$L8 companion,
only the fourth white dwarf plus brown dwarf binary known
\citep{steeleetal09-1}.

\subsection{Example Photometric IR Excess Candidates}
\label{ss-pire}

As described in Sect.\,\ref{sss-dafitphot}, we fitted \textit{all}
photometric objects satisfying our DA white dwarf constraint set
(Table\,\ref{e-poly}) with DA model spectra, independent on whether
they also have an SDSS spectrum. 

\subsubsection{DA white dwarf candidates}

\paragraph{\textit{SDSS\,J0959$-$0200}}

SDSS\,J0959$-$0200 is a photometric-only DA candidate with $\Teff =
12000\pm^{1160}_{500}$\,K and $\Logg = 8.00\pm^{1.20}_{0.22}$. The
UKIDSS $K$-band magnitude shows a large ($5\sigma$) excess over the
white dwarf model, no excess is seen at shorter wavelengths. This
object is a strong candidate for having a very late type brown dwarf
companion, or a dusty debris disc.

\begin{figure}
\includegraphics[width=\columnwidth]{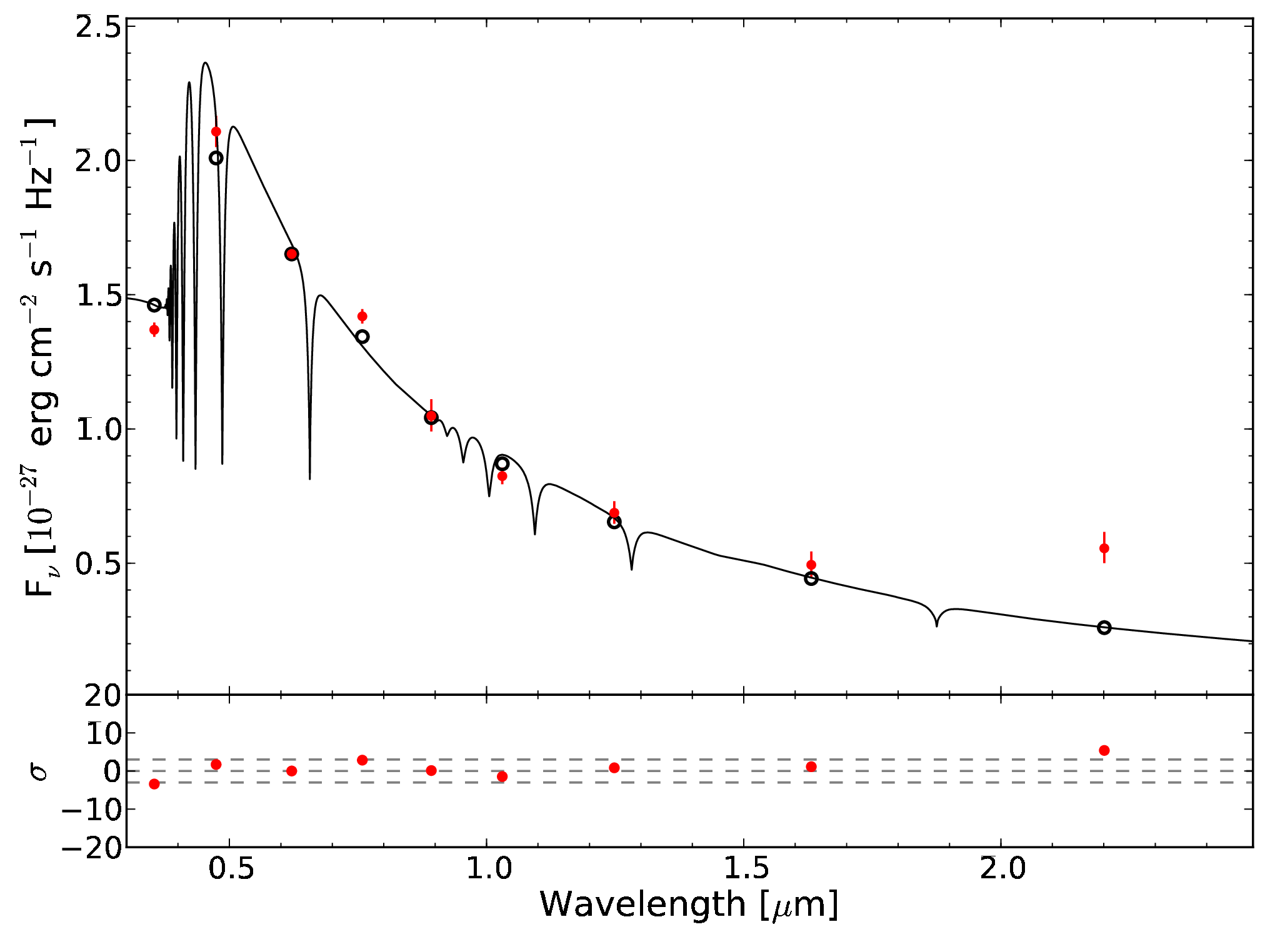}
\caption{\label{f-0959} SDSS\,J0959$-$0200, a photometric-only DA white
  dwarf candidate ($\Teff = 12000\pm^{1160}_{500} $K, $\Logg =
  8.00\pm^{1.20}_{0.55}$) that exhibits a strong $K$-band excess,
  making it a strong candidate for having either a late-type brown
  dwarf companion or a dusty debris disc.}
\end{figure}

\paragraph{\textit{SDSS\,J1221+1245}}

A second interesting photometric-only DA white dwarf candidate is
SDSS\,J1221+1245. The white dwarf is best fitted by a model with system
parameters $\Teff = 12000\pm^{1110}_{270}$\,K and $\Logg =
8.00\pm^{0.51}_{0.34}$. The UKIDSS $K$-band magnitude shows a borderline
excess over the white dwarf model, but no excess is seen at shorter
wavelengths. This object is again a good candidate for having a very late
type brown dwarf companion, or a dusty debris disc.

\begin{figure}
\includegraphics[width=\columnwidth]{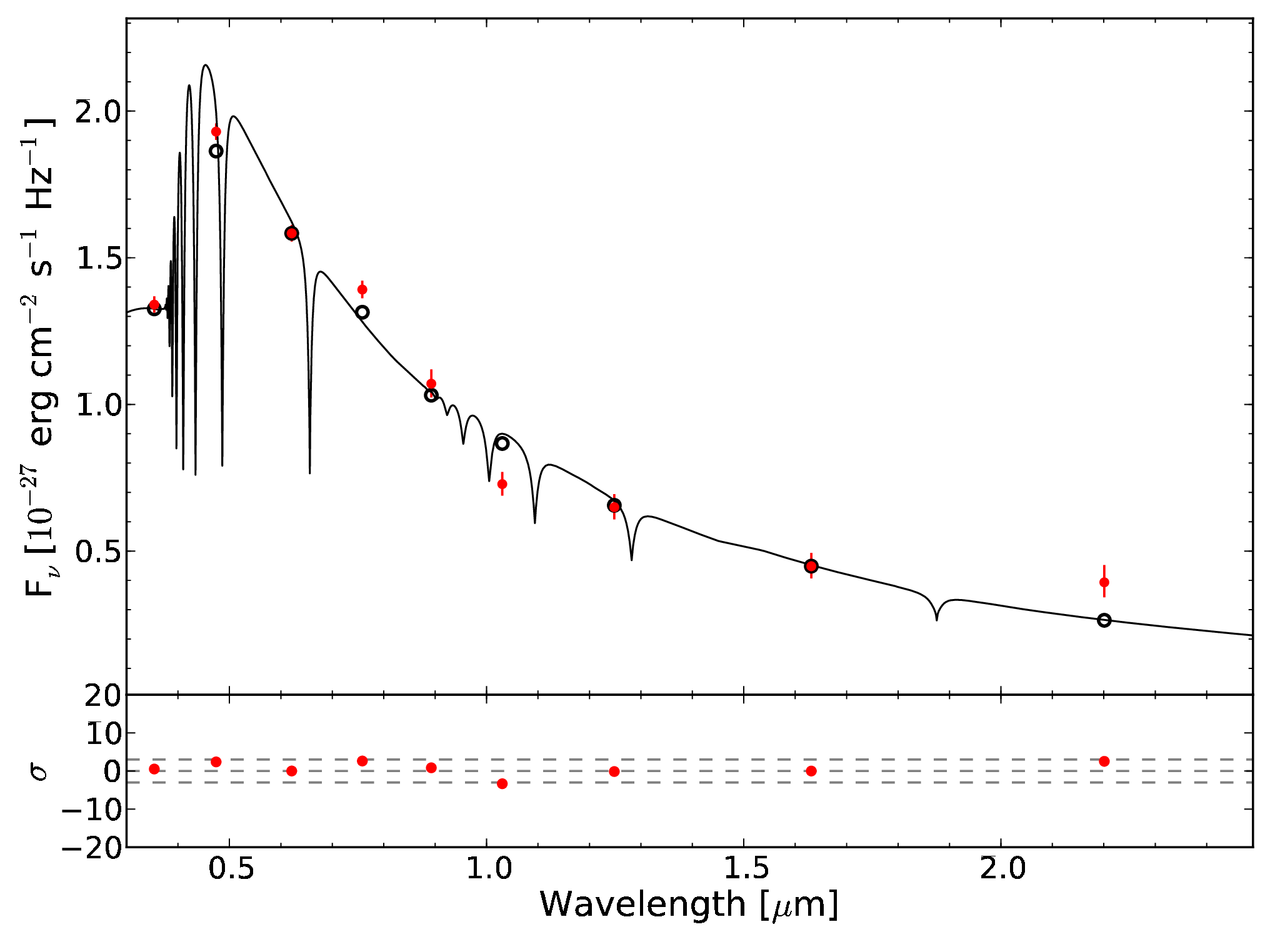}
\caption{\label{f-1221} SDSS\,J1221+1245, a photometric-only DA white
  dwarf candidate ($\Teff = 12000\pm^{1110}_{270} $K, $\Logg =
  8.00\pm^{0.51}_{0.34}$) that exhibits a borderline $K$-band excess. It
  is good candidate for having either a late-type brown
  dwarf companion or a dusty debris disc.}
\end{figure}

\subsubsection{Other Composite Objects}

Inspection of Simbad reveals that four of the photometric-only DA
candidates with IR excess (DA:ire and DA:ire:, Table\,\ref{t-poa}) are
previously known (pre-)white dwarf binaries and one pulsating subdwarf,
which provides a preview on the mixture of objects that can be expected
within this sample. It also underlines that the method efficiently
identifies genuine infrared-excess white dwarfs.

\paragraph{\textit{SDSS\,J0016+0704\,=\,PG\,0014+067}}
\label{s-pg0014+067}

\citet{brassardetal01-1} identified PG\,0014+067 as a pulsating sdB
with $\Teff=33550\pm380$K and $\Logg=5.77\pm0.10$. Fitting the $ugri$
photometry with DA model atmospheres results in
$\Teff=24000\pm^{2100}_{600}$K with the surface gravity fixed to
$\Logg=8$, and reveals a clear IR excess. At such high temperatures,
the slope of the optical and near-IR SED of this object is close to a
Rayleigh-Jeans distribution, and while modelling the photometric data
with DA models may not be perfect, we believe that PG\,0014+067 does
exhibit a genuine IR excess. In their asteroseismological analysis,
\citet{brassardetal01-1} found that the pulsation frequency spectrum
of PG\,0014+067 exhibits fine structure that they tentatively
interpreted as a rotational period of $29.2\pm0.9$\,h, revised later by
\citep{jeffreyetal05-1} to $\sim4$\,d. One possibility is that
PG\,0014+067 has a close low-mass binary companion with an orbital period
of a few days and that the sdB is tidally locked, rotating at the same
period. This hypothesis can be tested by a radial velocity study of this
subdwarf.

\paragraph{\textit{Cataclysmic Variables}}

BK\,Lyn (\,=\,SDSS\,J0920+3356) and HS\,0139+0559 (\,=\,SDSS0141+0614)
are novalike variables \citep{dobrzycka+howell92-1,
aungwerojwitetal05-1} with optically thick accretion discs, and
their optical colours are similar to that of hot white dwarfs or
subdwarfs. However, their companion stars and cooler outer regions of
the accretion discs start to contributed noticeably in the near-IR.

\paragraph{\textit{Detached Binaries}}

Abell\,31 (\,=\,SDSS\,J0854+0853, PN\,G219.1+31.2) is a planetary
nebula with a nearby ($0.26\arcsec$) M-dwarf, both stars are most
likely an associated wide binary \citep{ciardulloetal99-1}.

GK\,Vir (\,=\,SDSS\,J1415+0117) is an eclipsing binary containing a hot
($\simeq48800$\,K) white dwarf plus an $\sim$M4V companion with an orbital
period of 0.344\,d \citep{greenetal78-1, fulbrightetal93-1,
parsonsetal10-2}.

\section{Comparison with SDSS DR6 White Dwarf--Main Sequence Binaries}
\label{s-wdms}

\begin{figure*}
\includegraphics[width=2\columnwidth]{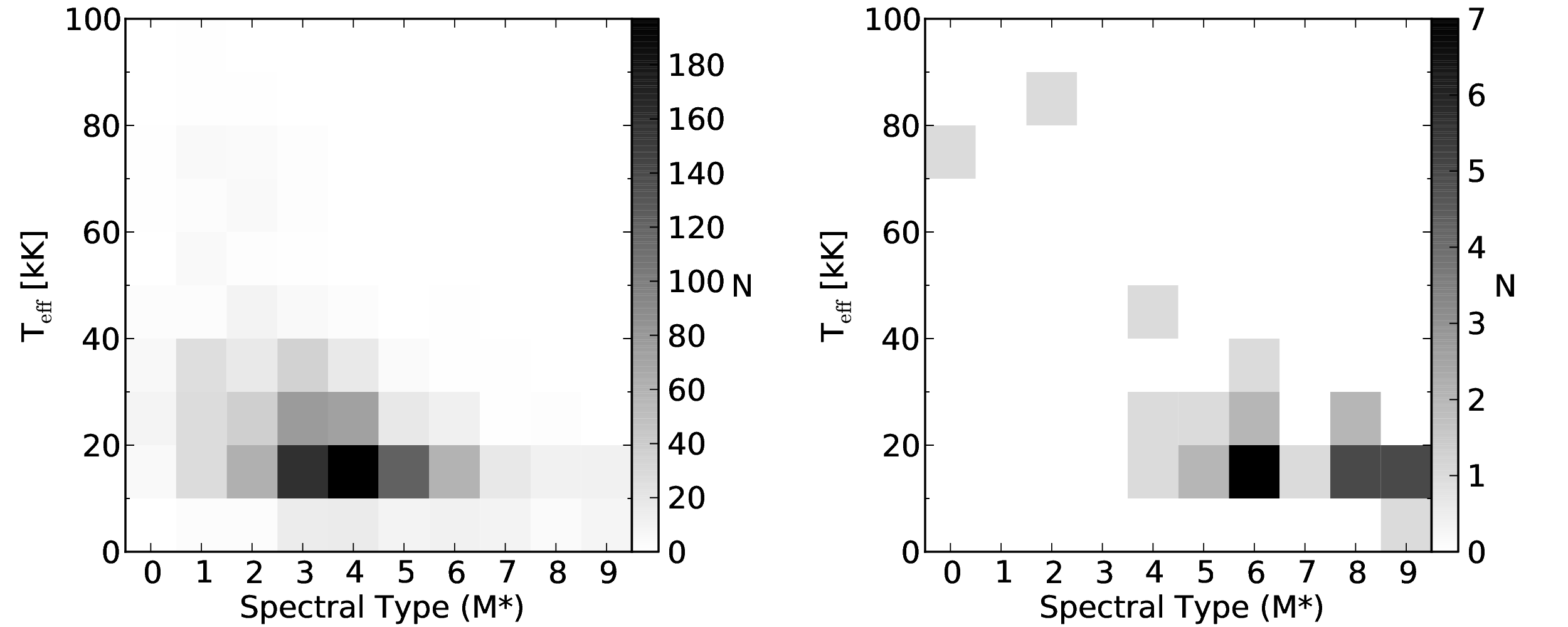}
\caption{\label{f-wdms} The distribution of the WDMS binaries from the
  catalogue of \citet{rebassa-mansergasetal10-1} as a function of
  effective temperature of the white dwarf and spectral type of the
  companion star. The left hand panel shows all the 1173 objects with
  values for both \Teff\ and \Logg.  53 of these are contained within the
  colour-colour region defined in Table\,\ref{e-poly}. Excluding spatially
  resolved binaries leaves 32 objects satisfying all the criteria of our
  DA white dwarf selection, which are shown in the right hand panel.}
\end{figure*}

\citet[][hereafter RM10]{rebassa-mansergasetal10-1} compiled a catalogue
of white dwarf--main sequence (WDMS) binaries from all spectroscopic
objects within SDSS DR6. Given that their detection method was based on
optical data alone, RM10 were primarily sensitive to white dwarfs with
M-type companions. The distribution of their WDMS binaries as a function
of effective temperature and companion star spectral type shows a clear
concentration of $\Teff=10-20$kK white dwarfs with $\sim$M4-type
companions (Fig.\,\ref{f-wdms}, left panel). The large luminosity of hot
white dwarfs prevents the identification of low-mass companions around
them, explaining the relative dearth of late spectral types at higher
temperatures. The decreasing number of very late M-dwarfs ($>$M6) could
also be affected to some degree by the same contrast problem, however, it
is known that the companion mass distribution of WDMS binaries is dropping
towards the low-mass end of the main sequence \citep{farihietal05-1}. 

The sample of WDMS binaries of RM10 provides a natural comparison for the
work done here. We have subjected their entire sample of WDMS binaries to
our DA colour cuts (Table\,\ref{e-poly}), finding that only 93 of the 1602
systems fall within the colour cuts. This small number is not too
surprising, as the red flux from the M-dwarf companions moves the
majority of RM10's WDMS binaries out of our colour selection. RM10 list WD
temperature \textit{and} companion spectral type for 53 of these 93
systems. In contrast to our work here (Sect.\,\ref{ss-ire_det}), the study
of RM10 included (partially) resolved systems. Consequently, we removed 21
that appeared resolved in the SDSS (or UKIDSS images where available),
which leaves us with 32 objects in common. Figure\,\ref{f-wdms} shows that
the two samples only overlap for systems where the companion is relatively
faint in the optical compared to the white dwarf, which is expected as
our DA selection needs the white dwarf to dominate. Finally, 10 of the 32
objects are in the UKIDSS footprint.

We cross-correlated the white dwarfs with an IR excess from our
spectroscopic and photometric samples (Table\,\ref{t-spa} and
\ref{t-poa}), and the WDMS catalogue of RM10 and found 19 objects in
common (Table\,\ref{t-wdms}). This is comprised of the 10 objects above
that we expect to be in the sample, along with 9 others where no spectral
type is listed in RM10's catalogue. There is in general a good agreement
between the white dwarf system parameters, however the underestimation of
photometric temperatures is highlighted again. The spectroscopic sample
and the WDMS catalogue largely find companion types within two spectral
types of each other. All of these are M-type companions as expected from
the sensitivity of the WDMS catalogue. Five objects are found to
have an excess in the photometric method, but were rejected from the
spectroscopic sample because signatures of a main sequence star companion
can be seen in the optical spectra, whereas the spectroscopic sample only
contains pure DA white dwarfs. Those are marked as such in
Table\,\ref{t-wdms}, and suffer from the same companion type biases
discussed in Section\,\ref{sss-da}.

\section{Confirmation of IR excess candidates in WISE}
\label{s-wise}

We cross correlated all spectroscopic and photometric-only IR excess
candidates (from Table\,\ref{t-spa} and \ref{t-poa} respectively) with
WISE PDR within $2.5\arcsec$. Of the 42 IR excess objects from the
spectroscopic method, 3 had a detection in at least one WISE band.
Similarly 14 of the 67 photometric-only objects were detected and are
listed in the online master table.

The three spectroscopically confirmed DA white dwarfs with IR excesses
and WISE data: SDSS\,J0236$-$0103, SDSS\,J0847+2831 and SDSS\,J1448+0713,
are all confirmed to have an excess in the WISE $3.4$ and $4.6\mu$m
bands. They are however all predicted to have late M-type companions and
are not brown dwarf or debris disk candidates.

\begin{figure}
\includegraphics[width=\columnwidth]{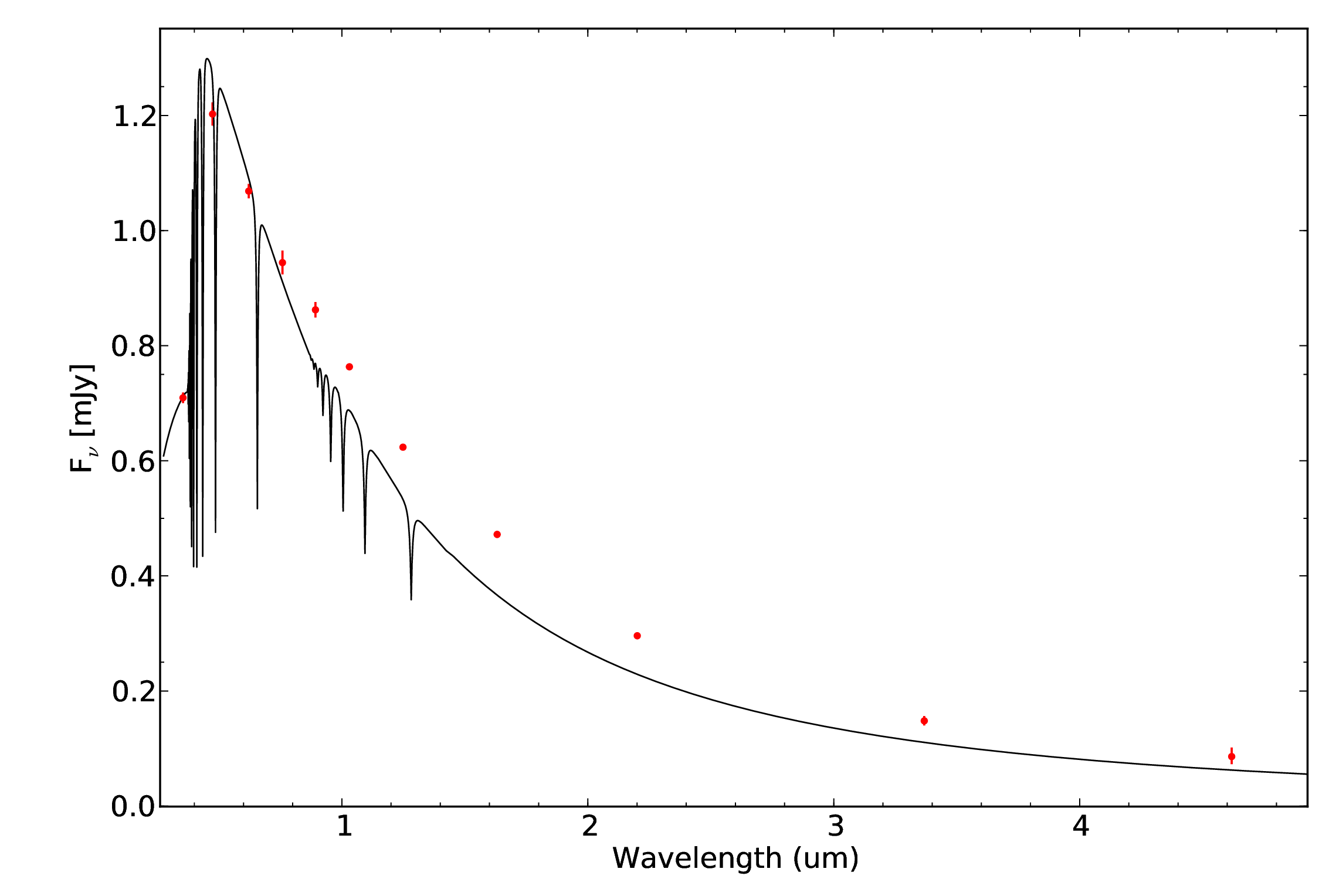}
\caption{\label{f-wise1} SED of SDSS\,J1538+0644. The SDSS, UKIDSS and
WISE $3.4$ and $4.6\mu$m fluxes are shown as red circles. The best fit
white dwarf model to the SDSS photometry is shown as a black line. The
shape of the IR excess is not consistent with any companion or disk. It is
more likely that the excess is an artifact of overestimating the
white dwarf effective temperature.}
\end{figure}

Of the 14 photometric-only IR excess candidates with WISE data,
SDSS\,J1524$-$0128 and SDSS\,J1549+0325 are \textit{not} found to have an
excess in the $3.4$ and $4.6\mu$m bands. Similarly,
because we do not trust the effective temperature of the white dwarf fit,
we also do not believe the far-IR excess found for SDSS\,J0841+0501,
SDSS\,J1441+0137, SDSS\,J1538+0644 and SDSS\,J1551$-$0118. This is
indicated by a flat, constant excess over the white dwarf model and is
continued into the far-IR WISE data in these cases (see
Fig.\,\ref{f-wise1}).

\begin{figure*}
\includegraphics[width=0.98\columnwidth]{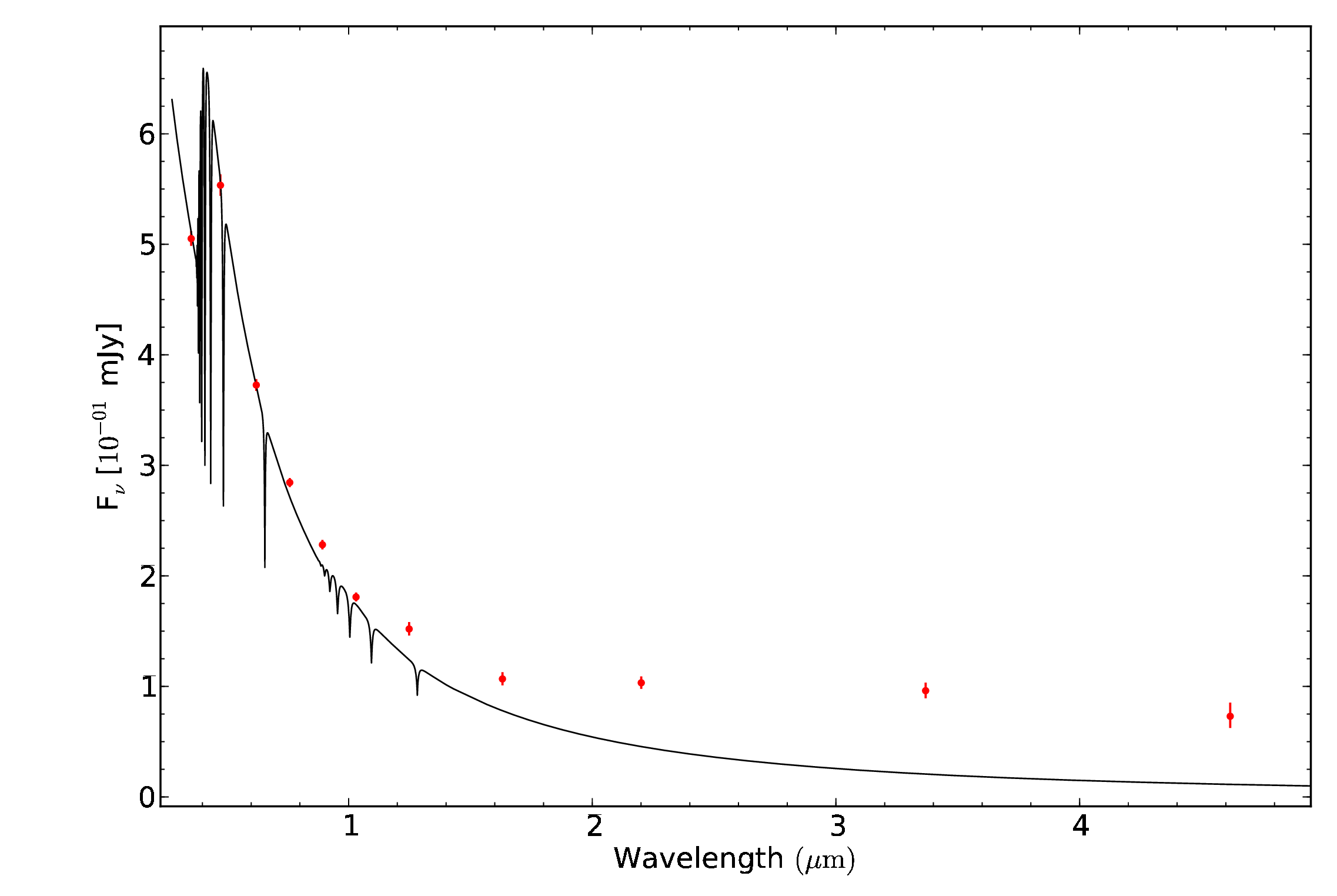}
\includegraphics[width=0.98\columnwidth]{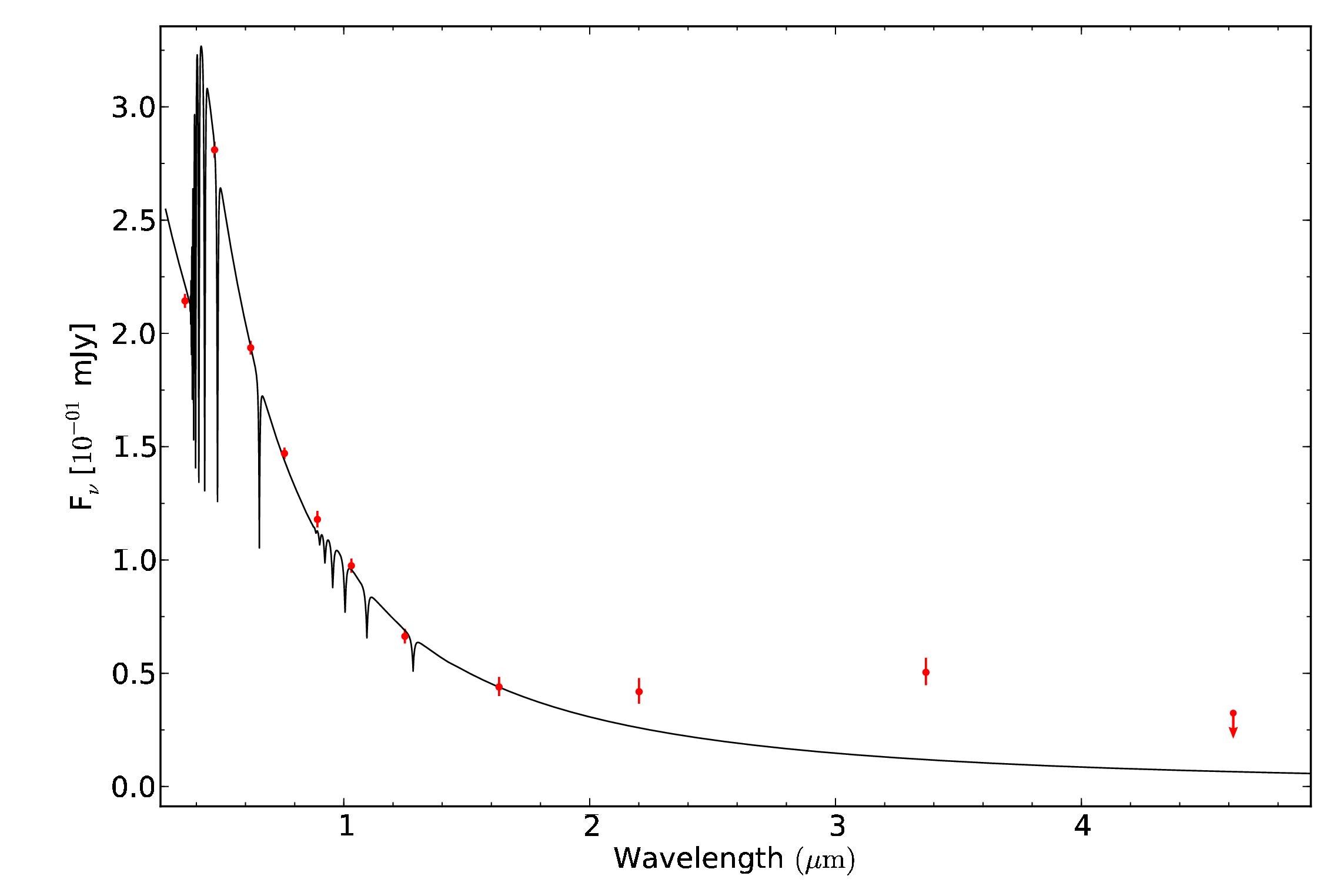}
\caption{\label{f-wise2} SEDs of SDSS\,J1538+2957 (left) and
SDSS\,J1635+2912 (right). The plot follows the same format as
Fig.\,\ref{f-wise1}. SDSS\,J1538+2957 and SDSS\,J1635+2912 are predicted
herein to have an M7: and an L6:-type companion or potential debris disk
respectively.}
\end{figure*}

The remaining 8 objects (SDSS\,J0207+0715, SDSS\,J0742+2857,
SDSS\,J0751+2002, SDSS\,J0920+3356, SDSS\,J1448+0812, SDSS\,J1456+1040,
SDSS\,J1538+2957, and SDSS\,J1635+2912) are confirmed to have an IR excess
consistent with a late type companion in the WISE far-IR data. Some
interesting examples of these are shown in Fig.\,\ref{f-wise2}.
SDSS\,J1538+2957 (Fig.\,\ref{f-wise2}) is predicted to have an M8-type
companion from the photometric method. However, the spectral shape of the
excess is found to be inconsistent with such an early type companion. The
excess in UKIDSS, and now WISE, is more indicative of a later type brown
dwarf companion or dusty debris disk. This mismatch is most likely caused
by over estimating the distance to the white dwarf in the photometric
method. SDSS\,J1635+2912 (Fig.\,\ref{f-wise2}) is one of the
photometric-only debris disk candidate systems. The WISE $3.5\mu$m flux
confirms the infrared excess, but still leaves the origin of the excess,
brown dwarf or disk, open.

\section{Discussion}
\label{s-dis}

\subsection{Selection of DA white dwarfs from optical photometry}

We have developed dedicated colour-colour cuts to select DA white
dwarfs only from their optical photometry (Table\,\ref{e-poly},
Fig.\,\ref{f-colsel}). This method can easily be optimised to
prioritise for completeness, efficiency or a compromise of both.  We
have demonstrated that a high completeness (\Comp\%) can be obtained
with a reasonable efficiency (\Effi\%) based on the $ugriz$ data from
SDSS\,DR7. The strengths of this approach, is that it provides
substantially larger and statistically better characterised white dwarf
samples (Sect.\,\ref{s-comp}) when compared to spectroscopic catalogues
such as \citet{mccook+sion99-1} and \citet{eisensteinetal06-1}. We have
also investigated methods to account for the contamination of the DA white
dwarf candidate sample by quasars and NLHS. This method can be adapted to
obtain large uniform samples of white dwarfs from other multicolour
optical photometric surveys, such as e.g. SkyMapper
\citep{kelleretal07-1}.

\subsection{Implications for the brown dwarf desert and the numbers of white
dwarfs with dusty debris discs}

Fitting the optical spectroscopy or photometry and probing for IR flux
excess above the best-fit white dwarf, as previously done by
\citet{tremblay+bergeron07-1}, has proven to be an efficient and robust
approach. It also allows one to identify hot white dwarfs with low-level
excesses which have IR colours that are very similar to the bulk
population of white dwarfs (Fig.\,\ref{f-zhhk}). Our search is sensitive
to companions as late as $\sim$L8, and to warm dusty debris discs.

Of the 1275 spectroscopically confirmed DA white dwarfs with at least one
UKIDSS magnitude, 26 ($2.0\%$) are found (or are candidates) to have an IR
excess consistent with a L0-type companion or later, i.e. a brown dwarf.
The exact cut off of where the brown dwarf sequence starts is age
dependent and can vary from mid-M to mid-L-type. Taking the white dwarf
mass distribution and the initial-final mass relation we can calculate the
white dwarf progenitor masses. This with the white dwarf cooling age
allows us to estimate the total age of the system. The average total age
of the companion is a few Gyr, and in this regime L0-type is a suitable
cut-off for being a brown dwarf. Taking only the systems where we are
confident of the excess (not ``DAire:'' in Table\,\ref{t-spa}) gives a
lower limit for the number of white dwarfs with brown dwarf type
companions of $0.8\%$. This is compatible with previous estimates. An
adaptive optics imaging survey of 266 solar-like stars by
\citet{metchev+hillenbrand09-1} found a wide (28--1590\,AU) sub-stellar
companion fraction of $3.2^{+3.1}_{-2.7}\%$. \citet{farihietal05-1}
carried out an extensive near-IR imaging survey for both wide and
unresolved low-mass companions to 394 known white dwarfs, finding an
overall stellar companion fraction of 22\%, and a brown dwarf companion
fraction of $<0.5$\%.

When using the photometric fitting method, 105 of the \UKP\
photometric-only DA white dwarf candidates with UKIDSS data exhibit an
IR excess, of which we eliminate 38 likely quasars
(Sect.\,\ref{sss-qso} and see online QSO table). Taking into account our
estimate for the contamination by NLHS (Sect.\,\ref{sss-nlhs}), we
find that $\sim2.7\%$ of white dwarfs in the photometric-only sample
have an IR excess, of which $1.8\%$ are candidates for having a brown
dwarf companion. This is consistent with the frequencies found in the
spectroscopic sample, though not as secure.

Taking only the objects with a detection in the $K$-band (required for
detecting a disk), of the 571 spectroscopically confirmed DA white dwarfs,
7 ($1.2\%$) are found to have an IR excess compatible with a companion
spectral type later or equal to L8, and are hence viable disc candidates.
Similarly, $0.5\%$ (4/840) of the photometric-only sample are disc
candidates, or, scaling for the \Effi\% efficiency of our DA white dwarf
selection, $0.8\%$ (4/523). This is consistent with the results of
\citet{farihietal09-1}, who estimated that the frequency of white dwarfs
with dusty debris discs is at least $1\%$. Any search for debris discs
based on $K$-band data is biased towards warm, bright circumstellar dust
rather than faint discs or narrow rings. Of the known white dwarfs with a
dusty debris disc, only about half reveal themselves shortward of $3\,\mu
m$ \citep{kilicetal06-1, farihietal09-1}.

To confirm the IR excess candidates we have cross matched all the
spectroscopic (Table\,\ref{t-spa}) and photometric-only
(Table\,\ref{t-poa}) candidates with the far-IR WISE PDR. This has
provided $3.4$ and $4.6\mu$m fluxes for $7\%$ (3/42) and $21\%$ (14/64) of
the spectroscopic and photometric-only IR excess candidates respectively.
We find that all three of the spectroscopic IR excess candidates also have
an excess in the far-IR.

However, in the photometric-only sample, a total of six objects are
found not to have a definite IR excess at $>3\mu$m. The remaining 8 of 14
white dwarf candidates have a real IR excess in the WISE fluxes. We
therefore find that $\sim60\%$ of the photometric-only IR excess
candidates have real IR excesses consistent with a MS star companion,
brown dwarf companion or dusty debris disk.

Determining the true nature of the white dwarfs with an excess in $K$ only
will require far-IR data to distinguish between a late-type brown dwarf
companion or a dusty debris disc. More generally, deeper near-IR and/or
far-IR follow up observations will be required to verify, or refute, the
IR excess candidates (DAire: and DA:ire:), as some of these candidates are
likely to be spurious. Similarly, while our colour selection
(Table\,\ref{e-poly}) has a high efficiency, optical spectroscopy of the
white dwarf candidates (DA:ire and DA:ire:) will be necessary to confirm
their (DA) white dwarf nature.

\section{Conclusions}
\label{s-con}

We have developed a detailed method to select DA white dwarfs in
$ugriz$ colour space, and applying this method to SDSS\,DR7, we have
identified \SDS\ DA white dwarfs with $g\le19$ and SDSS spectroscopy,
approximately $70\%$ more than the corresponding number from DR4
\citep{eisensteinetal06-1}, and \SDP\ photometric-only DA
candidates. Using this sample, we estimate the spectroscopic
completeness of DA white dwarfs with $T\ga8000$\,K in SDSS\,DR7 to be
$\simeq44$\%.

We cross-correlated these spectroscopic and photometric DA samples
with UKIDSS\,DR8 to carry out the currently largest and deepest
untargeted search for low-mass companions to and debris discs around
DA white dwarfs. This search led to the identification of a
significant number of DA white dwarfs with low-mass companions,
including several brown dwarf and dusty debris disc
candidates. Similar studies to the one presented here making use of
the full UKIDSS area, as well as the corresponding surveys in the
southern hemisphere, e.g. Skymapper, VISTA and VHS, promise to further
increase the sample sizes of these types of systems.

\vspace*{-3ex}
\section*{Note}

Before submission the authors became aware of the study of Steele et al.
(2011, in prep.) who have carried out an independent study, cross-matching
the spectroscopically confirmed white dwarfs from \citet{mccook+sion99-1}
and \citet{eisensteinetal06-1} with UKIDSS DR8. Comparison of their
results with our spectroscopic sample shows that both studies lead to
broadly consistent results.

\vspace*{-3ex}
\section*{Acknowledgements}
Funding for the Sloan Digital Sky Survey (SDSS) and SDSS-II has been provided by
the Alfred P. Sloan Foundation, the Participating Institutions, the National
Science Foundation, the U.S. Department of Energy, the National Aeronautics and
Space Administration, the Japanese Monbukagakusho, and the Max Planck Society,
and the Higher Education Funding Council for England. The SDSS Web site is
http://www.sdss.org/.

D. Steeghs acknowledges a STFC Advanced Fellowship.

This publication makes use of data products from the Wide-field Infrared
Survey Explorer, which is a joint project of the University of California,
Los Angeles, and the Jet Propulsion Laboratory/California Institute of
Technology, funded by the National Aeronautics and Space Administration.

\bibliographystyle{mn_new}
\bibliography{aamnem99,aabib,proceedings,submitted,jon}

\bsp


\begin{landscape}
\begin{table}
\begin{minipage}{180mm}
\caption{\label{t-spa} 42 DA white dwarf IR excess candidates from the
spectroscopic method. In all columns, ':' indicates an uncertain classification.
The notes column indicates the level of certainty of the IR excess. The key to
the notes2 column is: bg (background contamination), disc (disc candidate),
moreIR (more IR data required), resolved (resolved or partially resolved
binary). The ``Eisenstein`` column is the classifications according to the
\citet{eisensteinetal06-1} catalogue. The column ''JHK`` firstly indicates
whether the object has a UKIDSS $J$, $H$ and/or $K$ magnitude. Secondly, if the
corresponding letter is bold, it means an excess was found in that band.}
\begin{tabular}{lllllllllll}
\hline\hline
Coord & \Teff $(K)$ & \Logg & \Mwd (\Msun) & \Rwd $(10^8\,cm)$ & $d\,(pc)$ &
Companion & Notes & Notes2 & Eisenstein & JHK \\ \hline
\input{spa}
\hline
\end{tabular}
\end{minipage}
\end{table}
\end{landscape}

\begin{landscape}
\begin{table}
\begin{minipage}{180mm}
\caption{\label{t-poa} 67 photometric-only IR excess candidates which are
found to have an IR excess in the photometric method. The columns follow
the same format as that in Table\,\ref{t-spa}. The key to the ''Notes2``
column is: bf (visually bad fit), bg (background contamination), disc
(disc candidate), hotWD (high \Teff\ white dwarf and so small changes in,
e.g. flux calibration, remove the excess), moreIR (more IR data required),
resolved (resolved or partially resolved binary), qso (quasar), ?
(unknown). Also, the ``Simbad Name'' and ``Simbad Class'' columns give an
indication of what is already known about the objects. If the surface
gravity could not be constrained, it was assigned the canonical value of
8.0 and therefore no error could be defined. We do not quote the distances
to the systems because the uncertainty is so large. The two objects marked
as ''QSO:`` are borderline cases of the objects found in the online QSO
table and are therefore left as candidates for being DA white dwarfs with
an IR excess.}
\begin{tabular}{lllllllll}
\hline\hline
Coord & \Teff $(K)$ & \Logg & Companion & Notes & Notes2 & JHK &
Simbad Name & Simbad Class \\
\hline
\input{poa}
\hline
\end{tabular}
\end{minipage}
\end{table}
\end{landscape}

\begin{landscape}
\begin{table}
\begin{minipage}{180mm}
\caption{\label{t-s+p} IR excess candidates which are found to have an excess in
either the spectroscopic or photometric-only methods, but all have an SDSS
spectrum for classification. The table follows the same format as
Table\,\ref{t-spa} and \ref{t-poa}, with addition that S and P in the titles
refers to the spectroscopic and photometric methods respectively. Fiber
and fit parameters are only shown if an IR excess was found in the
corresponding method.}
\begin{tabular}{llllllllllllll}
\hline\hline
 & S & S & P & P & S & P & & S & P & & Simbad & Simbad \\
Coord & \Teff $(K)$ & \Logg & \Teff $(K)$ & \Logg &
\multicolumn{2}{c}{Companion} & Notes & Notes & Notes & Classification &
Name & Class \\ \hline
\input{spa+poa}
\hline
\end{tabular}
\end{minipage}
\end{table}
\end{landscape}

\begin{table*}
\caption{\label{t-wdms} The 19 IR excess candidates from both the
spectroscopic and photometric fitting methods that are also found to be
white dwarf + main sequence binaries in \citet{rebassa-mansergasetal10-1}.
Objects from the spectroscopic and photometric IR excess candidates are
denoted as ''S`` and ''P`` in the ``Method'' column respectively.
The first set of system parameters are from the methods described
herein. Where both spectroscopic and photometric IR excesses were found,
the quoted parameters are those from the spectroscopic method, the second
from \citeauthor{rebassa-mansergasetal10-1}. In some cases, such as
SDSS\,J0207$+$0702, the two spectroscopic fitting techniques disagree
significantly, this is where a different hot/cold fit solution was chosen.
Photometric temperatures may also disagree significantly from the
\citet{rebassa-mansergasetal10-1} temperatures as discussed in
Section\,\ref{sss-dafitphot}. The columns labelled with a ''R-M`` prefix
are the best fit taken from Table\,7 of \citet{rebassa-mansergasetal10-1}.
R-M\,Sp=-1 indicates that no spectral type could be associated.}
\begin{tabular}{lllrrlllr}
\hline\hline
Name & \Teff & \Logg & Method & Companion & Notes & R-M \Teff & R-M \Logg
& R-M
Sp \\
\hline
\input{wdms}
\hline
\end{tabular}
\end{table*}

\end{document}

%% file: spa.tex
0032$+$0739 & $21045\pm249$ & $7.42\pm0.05$ & $0.37\pm0.02$ & $1.37\pm0.05$ & $398\pm15$ & M6 & DAire: & resolved,bg: & DA & \textbf{J}\textbf{H}\textbf{K} \\
0039$-$0030 & $12392\pm544$ & $7.35\pm0.16$ & $0.32\pm0.05$ & $1.37\pm0.15$ & $417\pm44$ & M9 & DAire: &  & DA\_auto & \textbf{J}\textbf{H}\textbf{K} \\
0135$+$1445 & $7467\pm18$ & $7.34\pm0.04$ & $0.29\pm0.01$ & $1.33\pm0.03$ & $69\pm2$ & L6 & DAire &  & DA & \textbf{J}\textbf{H}\textbf{K} \\
0207$+$0702 & $10073\pm77$ & $8.32\pm0.09$ & $0.80\pm0.06$ & $0.72\pm0.05$ & $162\pm10$ & L7 & DAire: &  &  & \textbf{J}\textbf{H}\textbf{K} \\
0236$-$0103 & $20566\pm498$ & $7.66\pm0.09$ & $0.46\pm0.04$ & $1.15\pm0.07$ & $411\pm25$ & M6 & DAire &  & DA$+$M: & \textbf{J}\textbf{H}\textbf{K} \\
0253$-$0027 & $18974\pm286$ & $7.71\pm0.06$ & $0.47\pm0.03$ & $1.11\pm0.05$ & $380\pm15$ & L4 & DAire & moreIR & DA\_auto & JH \\
0748$+$2058 & $86726\pm7732$ & $7.15\pm0.27$ & $0.53\pm0.06$ & $2.23\pm1.37$ & $2021\pm469$ & M3 & DAire & resolved: & DA$+$M: & \textbf{J}\textbf{H}\textbf{K} \\
0753$+$2447 & $13432\pm710$ & $7.81\pm0.15$ & $0.50\pm0.08$ & $1.02\pm0.10$ & $349\pm32$ & L5: & DAire & disk & DA\_auto & JH\textbf{K} \\
0847$+$2831 & $12828\pm930$ & $7.76\pm0.20$ & $0.48\pm0.11$ & $1.05\pm0.14$ & $357\pm42$ & M6 & DAire &  & DA$+$M: & \textbf{J}\textbf{H}\textbf{K} \\
0907$+$0536 & $19416\pm458$ & $7.78\pm0.09$ & $0.51\pm0.05$ & $1.06\pm0.06$ & $367\pm21$ & L6 & DAire: & moreIR & DA\_auto & JH\textbf{K} \\
0933$+$3200 & $11565\pm381$ & $8.36\pm0.20$ & $0.83\pm0.13$ & $0.70\pm0.11$ & $229\pm34$ & L4 & DAire & resolved: &  & \textbf{J}\textbf{H}\textbf{K} \\
0950$+$0115 & $21785\pm365$ & $7.89\pm0.06$ & $0.57\pm0.03$ & $0.99\pm0.04$ & $329\pm14$ & L8 & DAire: & moreIR & DA & J\textbf{H} \\
1002$+$0939 & $21785\pm808$ & $7.92\pm0.14$ & $0.58\pm0.08$ & $0.97\pm0.09$ & $515\pm45$ & L0 & DAire &  & DA\_auto & \textbf{J}\textbf{H}\textbf{K} \\
1010$+$0407 & $13588\pm668$ & $7.76\pm0.11$ & $0.48\pm0.05$ & $1.05\pm0.07$ & $280\pm18$ & L8: & DAire: & moreIR & DA\_auto & J\textbf{H} \\
1015$+$0425 & $34526\pm86$ & $7.38\pm0.07$ & $0.41\pm0.02$ & $1.51\pm0.08$ & $691\pm39$ & L4 & DAire: & moreIR & DA\_auto & JH\textbf{K} \\
1016$+$0020 & $21045\pm703$ & $8.48\pm0.12$ & $0.92\pm0.07$ & $0.64\pm0.06$ & $325\pm32$ & L6 & DAire: & moreIR & DA & \textbf{H} \\
1037$+$0139 & $11433\pm208$ & $8.38\pm0.13$ & $0.84\pm0.08$ & $0.69\pm0.07$ & $151\pm14$ & L5: & DAire &  & DA & \textbf{J}\textbf{H}\textbf{K} \\
1120$+$0639 & $20331\pm609$ & $7.69\pm0.11$ & $0.47\pm0.05$ & $1.13\pm0.09$ & $517\pm38$ & L3 & DAire: & resolved,bg: & DA\_auto & \textbf{J} \\
1141$+$0420 & $11835\pm818$ & $7.37\pm0.22$ & $0.32\pm0.08$ & $1.35\pm0.21$ & $317\pm46$ & M7 & DAire &  & DA$+$M: & \textbf{J}\textbf{H}\textbf{K} \\
1208$+$0610 & $23076\pm1125$ & $7.64\pm0.17$ & $0.46\pm0.08$ & $1.18\pm0.14$ & $725\pm81$ & M7 & DAire &  & DA\_auto & \textbf{J}\textbf{H} \\
1218$+$0042 & $11173\pm180$ & $8.30\pm0.11$ & $0.79\pm0.07$ & $0.73\pm0.06$ & $171\pm13$ & L8 & DAire: & moreIR & DA & J\textbf{H} \\
1228$+$1040 & $22037\pm199$ & $8.19\pm0.04$ & $0.74\pm0.02$ & $0.80\pm0.02$ & $134\pm3$ & L6: & DAire & disk & DA\_auto & JH\textbf{K} \\
1246$+$0707 & $10793\pm189$ & $8.03\pm0.16$ & $0.62\pm0.10$ & $0.88\pm0.10$ & $200\pm21$ & L2 & DAire &  &  & \textbf{H}\textbf{K} \\
1247$+$1035 & $17912\pm159$ & $7.82\pm0.04$ & $0.52\pm0.02$ & $1.03\pm0.03$ & $394\pm9$ & L6: & DAire: & bg:,resolved: &  & JH\textbf{K} \\
1314$+$0057 & $17707\pm163$ & $7.80\pm0.04$ & $0.51\pm0.02$ & $1.04\pm0.03$ & $312\pm8$ & L6 & DAire: & disk: &  & H\textbf{K} \\
1320$+$0018 & $19193\pm213$ & $8.40\pm0.04$ & $0.87\pm0.03$ & $0.68\pm0.02$ & $142\pm5$ & L8 & DAire: & disk: & DA & JH\textbf{K} \\
1329$+$1230 & $13432\pm272$ & $7.42\pm0.05$ & $0.34\pm0.02$ & $1.32\pm0.05$ & $209\pm8$ & M7 & DAire &  & DA\_auto & \textbf{J}\textbf{H}\textbf{K} \\
1331$+$0040 & $15964\pm714$ & $7.72\pm0.17$ & $0.47\pm0.08$ & $1.09\pm0.12$ & $465\pm48$ & ? & DAire: & moreIR & DA &  \\
1341$+$0056 & $18330\pm436$ & $7.99\pm0.10$ & $0.61\pm0.06$ & $0.92\pm0.06$ & $360\pm22$ & M7 & DAire &  &  & \textbf{J}\textbf{H}\textbf{K} \\
1352$+$0910 & $36154\pm722$ & $7.49\pm0.11$ & $0.45\pm0.04$ & $1.39\pm0.13$ & $849\pm76$ & M5 & DAire &  &  & \textbf{J}\textbf{H}\textbf{K} \\
1410$+$0225 & $11565\pm246$ & $8.63\pm0.13$ & $1.00\pm0.07$ & $0.56\pm0.06$ & $152\pm16$ & $\geq L8$ & DAire: & moreIR & DA & J\textbf{H} \\
1425$-$0013 & $10670\pm169$ & $7.97\pm0.17$ & $0.58\pm0.10$ & $0.91\pm0.10$ & $242\pm26$ & L8: & DAire: & moreIR & DA & JH\textbf{K} \\
1448$+$0240 & $14728\pm247$ & $7.41\pm0.06$ & $0.34\pm0.02$ & $1.34\pm0.06$ & $292\pm13$ & L4 & DAire & moreIR & DA & \textbf{J}\textbf{H} \\
1448$+$0713 & $12250\pm256$ & $7.87\pm0.09$ & $0.53\pm0.05$ & $0.98\pm0.06$ & $143\pm8$ & M8 & DAire &  &  & \textbf{J}\textbf{H}\textbf{K} \\
1450$+$0000 & $17106\pm175$ & $7.93\pm0.05$ & $0.58\pm0.03$ & $0.95\pm0.03$ & $311\pm9$ & L8: & DAire: & moreIR &  & JH\textbf{K} \\
1557$+$0916 & $22811\pm734$ & $7.75\pm0.11$ & $0.50\pm0.05$ & $1.09\pm0.08$ & $488\pm35$ & L1: & DAire &  & DA\_auto & \textbf{J}H\textbf{K} \\
1609$+$2905 & $8773\pm91$ & $8.26\pm0.14$ & $0.76\pm0.09$ & $0.75\pm0.08$ & $122\pm12$ & $\geq L8$ & DAire &  &  & JH\textbf{K} \\
1619$+$2533 & $25595\pm1139$ & $7.21\pm0.15$ & $0.33\pm0.04$ & $1.65\pm0.20$ & $879\pm101$ & M5 & DAire &  &  & \textbf{J}\textbf{H}\textbf{K} \\
1625$+$3026 & $72136\pm4550$ & $7.77\pm0.19$ & $0.64\pm0.08$ & $1.21\pm0.21$ & $739\pm112$ & M5 & DAire &  & DA\_auto & \textbf{J}\textbf{H}\textbf{K} \\
2220$-$0041 & $7467\pm21$ & $7.66\pm0.12$ & $0.41\pm0.06$ & $1.10\pm0.08$ & $69\pm5$ & $\geq L8$ & DAire: & resolved & DA\_auto & J\textbf{H}\textbf{K} \\
2225$+$0016 & $11045\pm235$ & $8.44\pm0.16$ & $0.88\pm0.10$ & $0.65\pm0.08$ & $167\pm20$ & L6 & DAire: & resolved: & DA\_auto & \textbf{J}\textbf{H}\textbf{K} \\
2331$+$1342 & $11173\pm284$ & $8.79\pm0.15$ & $1.09\pm0.08$ & $0.49\pm0.07$ & $141\pm19$ & L6 & DAire &  & DA$+$M: & \textbf{J}\textbf{H}\textbf{K} \\

%% file: poa.tex
0016$+$0704 & $24000\pm^{2100}_{600}$ & $8.00$ & L0 & DA:ire &  & \textbf{J}\textbf{H}\textbf{K} & V* EK Psc & Variable Star \\
0104$+$1459 & $12000\pm^{110}_{50}$ & $7.75\pm^{0.12}_{0.05}$ & L5 & DA:ire & moreIR & J\textbf{H} &  &  \\
0141$+$0614 & $11000\pm^{70}_{1040}$ & $8.50\pm^{0.08}_{0.77}$ & L8: & DA:ire: & bf & \textbf{J}\textbf{H}\textbf{K} & 2MASS & Nova \\
0207$+$0715 & $14000\pm^{2140}_{1940}$ & $9.25\pm^{1.37}_{0.25}$ & L6: & DA:ire & bf & \textbf{J}\textbf{H}\textbf{K} & PB6668 & Star \\
0742$+$2857 & $9000\pm^{50}_{120}$ & $7.25\pm^{0.12}_{0.28}$ & L6 & DA:ire &  & \textbf{J}\textbf{H}\textbf{K} &  &  \\
0751$+$2002 & $20000\pm^{2460}_{1140}$ & $8.25\pm^{3.25}_{1.25}$ & L5 & DA:ire &  & \textbf{J}\textbf{H}\textbf{K} &  &  \\
0758$+$2406 & $13000\pm^{660}_{80}$ & $8.00\pm^{0.29}_{0.12}$ & L2: & DA:ire &  & \textbf{J}\textbf{H}\textbf{K} &  &  \\
0841$+$0501 & $8000\pm^{100}_{300}$ & $8.00$ & $\geq$L8: & DA:ire: & bf & \textbf{J}\textbf{H}\textbf{K} &  &  \\
0842$+$0004 & $22000\pm^{100}_{220}$ & $8.00$ & M9 & DA:ire &  & \textbf{J}\textbf{H} &  &  \\
0854$+$0853 & $60000\pm^{6890}_{7210}$ & $8.00$ & M6 & DA:ire &  & \textbf{J}\textbf{H}\textbf{K} & PN\_A66\_31 & PN \\
0906$+$0001 & $7000\pm^{10}_{0}$ & $8.00$ & $\geq$L8: & DA:ire: & bf & JH\textbf{K} &  &  \\
0920$+$3356 & $12000\pm^{90}_{110}$ & $9.00\pm^{0.28}_{0.15}$ & L8: & DA:ire: & bf & \textbf{J}\textbf{H}\textbf{K} & V* BK Lyn & Nova \\
0923$+$0652 & $7000\pm^{10}_{0}$ & $8.00$ & $\geq$L8: & DA:ire: & bf,bg & \textbf{J}\textbf{H}\textbf{K} &  &  \\
0925$-$0140 & $14000\pm^{80}_{140}$ & $7.50\pm^{0.28}_{0.37}$ & M9 & DA:ire &  & \textbf{J}\textbf{H}\textbf{K} &  &  \\
0952$+$1205 & $26000\pm^{1540}_{2310}$ & $8.00$ & $\geq$L8: & DA:ire: & bf & \textbf{H}\textbf{K} &  &  \\
0959$-$0200 & $12000\pm^{1160}_{500}$ & $8.00\pm^{1.20}_{0.22}$ & L6 & DA:ire & disk & JH\textbf{K} &  &  \\
0959$+$0330 & $32000\pm^{2010}_{2570}$ & $8.00$ & L0 & DA:ire &  & \textbf{J}\textbf{H}\textbf{K} & PG0957$+$037 & UV \\
1005$-$0225 & $14000\pm^{50}_{120}$ & $7.75\pm^{0.45}_{0.13}$ & L5: & DA:ire: & bf & \textbf{J}\textbf{H}\textbf{K} &  &  \\
1006$+$0032 & $20000\pm^{470}_{830}$ & $9.50\pm^{0.91}_{}$ & L8 & DA:ire & bf: & \textbf{J}\textbf{H}\textbf{K} & PG1004$+$008 & UV \\
1038$+$1100 & $10000\pm^{170}_{90}$ & $7.75\pm^{0.17}_{0.27}$ & L3 & DA:ire &  & \textbf{J}\textbf{H}\textbf{K} &  &  \\
1057$+$0628 & $7000\pm^{30}_{10}$ & $8.00$ & L8: & DA:ire: & QSO: & J\textbf{H}\textbf{K} &  &  \\
1103$+$0101 & $7000\pm^{10}_{0}$ & $8.00$ & $\geq$L8: & DA:ire: & bf: & J\textbf{H}\textbf{K} & [VV2006]J110336.3$+$010141 & Quasar \\
1103$+$1100 & $34000\pm^{260}_{2010}$ & $8.00$ & L4 & DA:ire: & hotWD,moreIR & \textbf{J} & PG1101$+$113 & SD \\
1116$+$0755 & $22000\pm^{2040}_{110}$ & $8.00$ & M9 & DA:ire &  & \textbf{J}\textbf{H}\textbf{K} &  &  \\
1153$+$0048 & $15000\pm^{1180}_{560}$ & $9.50\pm^{0.36}_{}$ & $\geq$L8: & DA:ire: & bf & \textbf{J}\textbf{H}\textbf{K} & LBQS1151$+$0104 & Star \\
1221$+$1245 & $12000\pm^{1110}_{270}$ & $8.00\pm^{0.51}_{0.34}$ & L8: & DA:ire: & disk: & JH\textbf{K} &  &  \\
1224$+$0907 & $7000\pm^{30}_{30}$ & $8.00$ & L7 & DA:ire: & bg:,resolved:,bf: & \textbf{J}\textbf{H}\textbf{K} &  &  \\
1237$-$0151 & $20000\pm^{40}_{520}$ & $8.75\pm^{0.32}_{0.16}$ & L4: & DA:ire: & resolved,bg: & \textbf{J}\textbf{H}\textbf{K} &  &  \\
1246$+$1600 & $19000\pm^{610}_{530}$ & $9.50\pm^{0.40}_{}$ & $\geq$L8: & DA:ire: & bf: & \textbf{J}\textbf{H}\textbf{K} &  &  \\
1315$+$0245 & $28000\pm^{1500}_{1790}$ & $8.00$ & L1: & DA:ire: & resolved,bg: & \textbf{J}\textbf{H}\textbf{K} &  &  \\
1319$+$0152 & $16000\pm^{150}_{60}$ & $8.25\pm^{0.36}_{0.28}$ & $\geq$L8: & DA:ire: & bf & \textbf{H}\textbf{K} & WD1317$+$021 & DA \\
1323$+$2615 & $17000\pm^{1070}_{520}$ & $9.50\pm^{0.89}_{}$ & $\geq$L8: & DA:ire: & bf & \textbf{H}\textbf{K} &  &  \\
1334$+$0534 & $30000\pm^{2410}_{390}$ & $8.00$ & $\geq$L8: & DA:ire: & bf: & \textbf{H}\textbf{K} &  &  \\
1348$+$1100 & $22000\pm^{330}_{1180}$ & $8.00$ & L1 & DA:ire & moreIR & \textbf{J} &  &  \\
1355$+$1454 & $22000\pm^{2260}_{250}$ & $8.00$ & L1 & DA:ire &  & \textbf{J}\textbf{H}\textbf{K} & PB4150 & SD \\
1416$+$1352 & $38000\pm^{20800}_{6750}$ & $8.00$ & M6 & DA:ire &  & \textbf{J}\textbf{H}\textbf{K} & PK\_003$+$66\_1 & Galaxy \\
1423$-$0138 & $16000\pm^{380}_{60}$ & $8.00\pm^{0.21}_{0.41}$ & $\geq$L8: & DA:ire: & moreIR & J\textbf{H} &  &  \\
1441$+$0137 & $26000\pm^{60}_{420}$ & $8.00$ & L6: & DA:ire: & bf: & \textbf{J}\textbf{H}\textbf{K} & PG1438$+$018 & SD \\
\hline
\end{tabular}
\end{minipage}
\end{table}
\end{landscape}

\begin{landscape}
\begin{table}
\begin{minipage}{180mm}
\contcaption{}
\begin{tabular}{lllllllll}
\hline\hline
Coord & \Teff $(K)$ & \Logg & Companion & Notes & Notes2 & JHK &
Simbad Name & Simbad Class \\ \hline
1442$+$0910 & $20000\pm^{170}_{140}$ & $7.75\pm^{1.13}_{0.59}$ & M7 & DA:ire &  & \textbf{J}\textbf{H}\textbf{K} &  &  \\
1448$+$0812 & $24000\pm^{1150}_{2380}$ & $8.00$ & M7 & DA:ire &  & \textbf{J}\textbf{H}\textbf{K} & WD1446$+$028 & DA \\
1455$+$0458 & $30000\pm^{2380}_{240}$ & $8.00$ & M8 & DA:ire &  & \textbf{J}\textbf{H}\textbf{K} &  &  \\
1456$+$1040 & $14000\pm^{1430}_{1580}$ & $8.75\pm^{0.91}_{0.75}$ & L1: & DA:ire &  & \textbf{J}\textbf{H}\textbf{K} &  &  \\
1507$+$0724 & $26000\pm^{2400}_{2340}$ & $8.00$ & M7 & DA:ire &  & \textbf{J}\textbf{H}\textbf{K} &  &  \\
1523$+$0023 & $24000\pm^{1360}_{2330}$ & $8.00$ & M8 & DA:ire &  & \textbf{J}\textbf{H}\textbf{K} &  &  \\
1524$-$0128 & $18000\pm^{20}_{20}$ & $9.25\pm^{0.02}_{0.16}$ & $\geq$L8: & DA:ire: & bf & J\textbf{H}\textbf{K} & BPS\_CS\_22890$-$0079 & SD \\
1527$+$0802 & $20000\pm^{1530}_{830}$ & $9.50\pm^{0.63}_{}$ & $\geq$L8: & DA:ire: &  & \textbf{J}\textbf{H}\textbf{K} &  &  \\
1538$+$0644 & $10000\pm^{50}_{40}$ & $7.25\pm^{0.10}_{0.02}$ & L6 & DA:ire: & bf: & \textbf{J}\textbf{H}\textbf{K} &  &  \\
1538$+$2957 & $20000\pm^{40}_{120}$ & $5.75\pm^{0.75}_{0.84}$ & M7: & DA:ire &  & \textbf{J}\textbf{H}\textbf{K} &  &  \\
1540$+$2908 & $34000\pm^{4270}_{1240}$ & $8.00$ & M7 & DA:ire &  & J\textbf{H} &  &  \\
1548$+$0006 & $14000\pm^{50}_{50}$ & $8.50\pm^{0.08}_{0.04}$ & L6: & DA:ire: &  & \textbf{J}\textbf{H}K &  &  \\
1549$+$0325 & $13000\pm^{300}_{170}$ & $7.25\pm^{0.64}_{0.52}$ & L7: & DA:ire: & bf & \textbf{J}\textbf{H}\textbf{K} & PG1546$+$036 & UV \\
1551$-$0118 & $13000\pm^{70}_{1040}$ & $8.25\pm^{0.29}_{0.28}$ & $\geq$L8: & DA:ire: & bf & \textbf{J}\textbf{H}\textbf{K} &  &  \\
1554$+$0616 & $17000\pm^{630}_{200}$ & $7.25\pm^{1.57}_{0.88}$ & M6 & DA:ire: & resolved,bg: & \textbf{J}\textbf{H}\textbf{K} &  &  \\
1614$+$2235 & $20000\pm^{100}_{1090}$ & $8.50\pm^{0.56}_{0.74}$ & L6 & DA:ire &  & \textbf{J}\textbf{H}K & SDSS & SD \\
1635$+$2912 & $17000\pm^{100}_{1030}$ & $5.50\pm^{0.50}_{0.93}$ & L6: & DA:ire: & moreIR,disk:,bg: & JH\textbf{K} &  &  \\
2042$+$0055 & $28000\pm^{2160}_{350}$ & $8.00$ & $\geq$L8: & DA:ire: & bf: & \textbf{H}\textbf{K} &  &  \\
2049$-$0001 & $16000\pm^{1160}_{380}$ & $8.75\pm^{0.82}_{0.45}$ & $\geq$L8: & DA:ire: & bf: & \textbf{H}\textbf{K} &  &  \\
2052$+$0018 & $12000\pm^{1080}_{130}$ & $8.00\pm^{0.34}_{0.24}$ & $\geq$L8: & DA:ire: & bf: & \textbf{H}\textbf{K} &  &  \\
2117$-$0015 & $12000\pm^{120}_{140}$ & $8.00\pm^{0.15}_{0.05}$ & $\geq$L8: & DA:ire: & bf: & \textbf{H}\textbf{K} &  &  \\
2135$-$0031 & $30000\pm^{1390}_{2590}$ & $8.00$ & M7 & DA:ire &  & \textbf{J}\textbf{H}\textbf{K} &  &  \\
2147$-$0112 & $26000\pm^{2190}_{390}$ & $8.00$ & L1: & DA:ire & resolved: & \textbf{J}\textbf{H}\textbf{K} & 2MASS & Blue \\
2308$+$0658 & $28000\pm^{1260}_{2140}$ & $8.00$ & M6 & DA:ire &  & \textbf{J}\textbf{H}\textbf{K} &  &  \\
2326$+$1230 & $18000\pm^{200}_{180}$ & $8.75\pm^{0.58}_{0.29}$ & $\geq$L8: & DA:ire: & bf: & \textbf{J}\textbf{H}\textbf{K} & 2MASS & Star \\
2330$+$1450 & $13000\pm^{1580}_{400}$ & $9.50\pm^{0.64}_{}$ & $\geq$L8: & DA:ire: & bf: & \textbf{J}H\textbf{K} &  &  \\
2335$+$1047 & $11000\pm^{520}_{450}$ & $7.75\pm^{0.55}_{0.46}$ & L4 & DA:ire: & resolved,bg: & J\textbf{H}\textbf{K} &  &  \\
2344$+$0817 & $8000\pm^{100}_{410}$ & $8.00$ & L8: & DA:ire: & bf: & \textbf{J}\textbf{H}\textbf{K} & PB5517 & Star \\
2348$+$0200 & $8000\pm^{30}_{10}$ & $8.00$ & L6: & DA:ire: & QSO: & \textbf{J}\textbf{H}\textbf{K} &  &  \\

%% file: spa+poa.tex
0018$+$0101 &  &  & $22000\pm^{220}_{50}$ & $8.00\pm^{}_{}$ &  & L1 & DA:ire &  &  & NLHS &  &  \\
0032$+$0739 & $21045\pm249$ & $7.42\pm0.05$ &  &  & M6 &  & DAire: & resolved,bg: &  & DA &  &  \\
0039$-$0030 & $12392\pm544$ & $7.35\pm0.16$ & $11000\pm^{50}_{20}$ & $7.75\pm^{0.10}_{0.03}$ & M9 & L2 & DAire: &  &  & DA &  &  \\
0118$-$0025 &  &  & $26000\pm^{880}_{780}$ & $8.00\pm^{}_{}$ &  & L3 & DA:ire &  & resolved & NLHS &  &  \\
0135$+$1445 & $7467\pm18$ & $7.34\pm0.04$ & $8000\pm^{10}_{20}$ & $8.00\pm^{}_{}$ & L6 & L8 & DAire &  &  & DA &  &  \\
0207$+$0702 & $10073\pm77$ & $8.32\pm0.09$ & $15000\pm^{1090}_{360}$ & $8.75\pm^{0.89}_{0.28}$ & L7 & ? & DAire: &  & bf & DA &  &  \\
0236$-$0103 & $20566\pm498$ & $7.66\pm0.09$ & $18000\pm^{1390}_{1200}$ & $8.75\pm^{2.48}_{0.75}$ & M6 & $\geq$L8: & DAire &  & bf & DA &  &  \\
0253$-$0027 & $18974\pm286$ & $7.71\pm0.06$ &  &  & L4 &  & DAire & moreIR &  & DA &  &  \\
0333$+$0020 &  &  & $22000\pm^{2360}_{390}$ & $8.00\pm^{}_{}$ &  & M9 & DA:ire &  &  & NLHS &  &  \\
0748$+$2058 & $86726\pm7732$ & $7.15\pm0.27$ & $40000\pm^{3640}_{3620}$ & $8.00\pm^{}_{}$ & M3 & M5 & DAire & resolved: & resolved: & DA &  &  \\
0753$+$2447 & $13432\pm710$ & $7.81\pm0.15$ & $12000\pm^{1130}_{290}$ & $7.75\pm^{0.45}_{0.41}$ & L5: & L6: & DAire & disk &  & DA &  &  \\
0814$+$2811 &  &  & $9000\pm^{590}_{40}$ & $8.50\pm^{0.43}_{0.53}$ &  & $\geq$L8: & DA:ire: &  & bf & NLHS &  &  \\
0847$+$2831 & $12828\pm930$ & $7.76\pm0.20$ & $12000\pm^{2090}_{1030}$ & $8.25\pm^{1.22}_{1.04}$ & M6 & M9 & DAire &  &  & DA &  &  \\
0851$+$0330 &  &  & $20000\pm^{2090}_{2740}$ & $9.50\pm^{1.55}_{}$ &  & L6 & DA:ire &  &  & WDMS &  &  \\
0907$+$0536 & $19416\pm458$ & $7.78\pm0.09$ &  &  & L6 &  & DAire: & moreIR &  & DA &  &  \\
0933$+$3200 & $11565\pm381$ & $8.36\pm0.20$ & $10000\pm^{1070}_{70}$ & $7.50\pm^{0.36}_{0.23}$ & L4 & L2 & DAire & resolved: & resolved: & DA &  &  \\
0950$+$0115 & $21785\pm365$ & $7.89\pm0.06$ &  &  & L8 &  & DAire: & moreIR &  & DA &  &  \\
0951$+$0347 &  &  & $24000\pm^{100}_{2040}$ & $8.00\pm^{}_{}$ &  & $\geq$L8: & DA:ire: &  & bf & NLHS &  &  \\
1002$+$0939 & $21785\pm808$ & $7.92\pm0.14$ & $20000\pm^{140}_{100}$ & $8.00\pm^{0.48}_{0.52}$ & L0 & L0 & DAire &  &  & DA &  &  \\
1010$+$0407 & $13588\pm668$ & $7.76\pm0.11$ &  &  & L8: &  & DAire: & moreIR &  & DA &  &  \\
1015$+$0425 & $34526\pm86$ & $7.38\pm0.07$ &  &  & L4 &  & DAire: & moreIR &  & DA &  &  \\
1016$+$0020 & $21045\pm703$ & $8.48\pm0.12$ &  &  & L6 &  & DAire: & moreIR &  & DA &  &  \\
1025$+$1200 &  &  & $22000\pm^{610}_{370}$ & $8.00\pm^{}_{}$ &  & L1 & DA:ire &  &  & NLHS &  &  \\
1037$+$0139 & $11433\pm208$ & $8.38\pm0.13$ & $15000\pm^{360}_{1270}$ & $8.50\pm^{0.34}_{0.66}$ & L5: & L1 & DAire &  &  & DA &  &  \\
1100$+$0346 &  &  & $32000\pm^{6920}_{3130}$ & $8.00\pm^{}_{}$ &  & M7 & DA:ire &  &  & NLHS &  &  \\
1106$+$0737 &  &  & $28000\pm^{3800}_{2060}$ & $8.00\pm^{}_{}$ &  & M5 & DA:ire &  &  & WDMS &  &  \\
1120$+$0639 & $20331\pm609$ & $7.69\pm0.11$ &  &  & L3 &  & DAire: & resolved,bg: &  & DA &  &  \\
1135$+$0731 &  &  & $8000\pm^{80}_{70}$ & $8.00\pm^{}_{}$ &  & L7: & DA:ire: &  & bf: & NLHS &  &  \\
1141$+$0420 & $11835\pm818$ & $7.37\pm0.22$ & $12000\pm^{230}_{1080}$ & $7.50\pm^{0.48}_{0.35}$ & M7 & M7 & DAire &  &  & DA &  &  \\
1148$+$0336 &  &  & $28000\pm^{250}_{2000}$ & $8.00\pm^{}_{}$ &  & L0 & DA:ire &  &  & NLHS &  &  \\
1148$+$0640 &  &  & $24000\pm^{3710}_{1380}$ & $8.00\pm^{}_{}$ &  & M8 & DA:ire &  &  & NLHS &  &  \\
1208$+$0610 & $23076\pm1125$ & $7.64\pm0.17$ & $20000\pm^{420}_{1120}$ & $8.75\pm^{1.03}_{0.75}$ & M7 & L3 & DAire &  &  & DA &  &  \\
1211$+$1437 &  &  & $24000\pm^{740}_{2470}$ & $8.00\pm^{}_{}$ &  & M9 & DA:ire &  &  & NLHS &  &  \\
1212$+$0136 &  &  & $10000\pm^{80}_{70}$ & $7.75\pm^{0.25}_{0.06}$ &  & ? & DA:ire: &  & bf & MWD &  &  \\
1215$+$1351 &  &  & $24000\pm^{350}_{2180}$ & $8.00\pm^{}_{}$ &  & M6 & DA:ire: &  & resolved: & NLHS &  &  \\
1218$+$0042 & $11173\pm180$ & $8.30\pm0.11$ & $10000\pm^{130}_{30}$ & $7.75\pm^{0.07}_{0.08}$ & L8 & L8: & DAire: & moreIR & moreIR & DA &  &  \\
1219$+$1244 &  &  & $10000\pm^{330}_{530}$ & $7.50\pm^{0.50}_{0.34}$ &  & L8: & DA:ire: & No$-$Xs & bf: & DA &  &  \\
1228$+$1040 & $22037\pm199$ & $8.19\pm0.04$ & $20000\pm^{10}_{10}$ & $8.50\pm^{0.03}_{0.13}$ & L6: & L8: & DAire & disk & disk & DA &  &  \\
1246$+$0707 & $10793\pm189$ & $8.03\pm0.16$ & $9000\pm^{1010}_{50}$ & $7.00\pm^{0.15}_{0.35}$ & L2 & M8 & DAire &  &  & DA &  &  \\
1247$+$1035 & $17912\pm159$ & $7.82\pm0.04$ &  &  & L6: &  & DAire: & bg:,resolved: &  & DA &  &  \\
1249$+$0422 &  &  & $11000\pm^{10}_{10}$ & $8.00\pm^{0.03}_{0.01}$ &  & L8: & DA:ire: & No$-$Xs & moreIR & DA &  &  \\
1250$+$1549 &  &  & $9000\pm^{1060}_{90}$ & $8.50\pm^{0.62}_{0.72}$ &  & L3: & DA:ire: &  & bf & MWD &  &  \\
1300$+$0057 &  &  & $26000\pm^{2000}_{680}$ & $8.00\pm^{}_{}$ &  & $\geq$L8: & DA:ire: &  & bf: & NLHS &  &  \\
1310$+$0233 &  &  & $26000\pm^{240}_{340}$ & $8.00\pm^{}_{}$ &  & M8 & DA:ire &  &  & NLHS &  &  \\
\hline
\end{tabular}
\end{minipage}
\end{table}
\end{landscape}

\begin{landscape}
\begin{table}
\begin{minipage}{180mm}
\contcaption{}
\begin{tabular}{llllllllllllll}
\hline\hline
 & S & S & P & P & S & P & & S & P & & Simbad & Simbad \\
Coord & \Teff $(K)$ & \Logg & \Teff $(K)$ & \Logg &
\multicolumn{2}{c}{Companion} & Notes & Notes & Notes & Classification &
Name & Class \\ \hline
1314$+$0057 & $17707\pm163$ & $7.80\pm0.04$ & $17000\pm^{1050}_{510}$ & $8.25\pm^{1.21}_{0.60}$ & L6 & $\geq$L8: & DAire: & disk: & bf & DA &  &  \\
1320$+$0018 & $19193\pm213$ & $8.40\pm0.04$ & $19000\pm^{580}_{370}$ & $8.75\pm^{0.61}_{0.64}$ & L8 & L8: & DAire: & disk: & disk: & DA &  &  \\
1329$+$1230 & $13432\pm272$ & $7.42\pm0.05$ & $12000\pm^{80}_{80}$ & $7.50\pm^{0.26}_{0.09}$ & M7 & M8 & DAire &  &  & DA &  &  \\
1331$+$0040 & $15964\pm714$ & $7.72\pm0.17$ &  &  & ? &  & DAire: & moreIR &  & DA &  &  \\
1334$+$0647 &  &  & $8000\pm^{20}_{20}$ & $8.00\pm^{}_{}$ &  & $\geq$L8: & DA:ire: & No$-$Xs & moreIR & DA &  &  \\
1341$+$0056 & $18330\pm436$ & $7.99\pm0.10$ & $17000\pm^{530}_{1050}$ & $7.75\pm^{1.16}_{0.94}$ & M7 & M6: & DAire &  &  & DA &  &  \\
1352$+$0910 & $36154\pm722$ & $7.49\pm0.11$ & $28000\pm^{1080}_{2300}$ & $8.00\pm^{}_{}$ & M5 & M6 & DAire &  &  & DA &  &  \\
1410$+$0225 & $11565\pm246$ & $8.63\pm0.13$ & $11000\pm^{0}_{0}$ & $8.00\pm^{0.01}_{}$ & $\geq$L8 & L8 & DAire: & moreIR & moreIR & DA &  &  \\
1414$+$0212 &  &  & $8000\pm^{20}_{20}$ & $8.00\pm^{}_{}$ &  & $\geq$L8: & DA:ire: & No$-$Xs & moreIR,bf: & DA &  &  \\
1415$+$0117 &  &  & $32000\pm^{4360}_{2230}$ & $8.00\pm^{}_{}$ &  & M6 & DA:ire &  &  & WDMS &  &  \\
1422$+$0920 &  &  & $22000\pm^{950}_{630}$ & $8.00\pm^{}_{}$ &  & M9: & DA:ire &  &  & NLHS &  &  \\
1424$+$0319 &  &  & $30000\pm^{50}_{20}$ & $8.00\pm^{}_{}$ &  & L5 & DA:ire &  & hotWD & NLHS &  &  \\
1425$-$0013 & $10670\pm169$ & $7.97\pm0.17$ &  &  & L8: &  & DAire: & moreIR &  & DA &  &  \\
1436$+$0138 &  &  & $24000\pm^{590}_{90}$ & $8.00\pm^{}_{}$ &  & L4 & DA:ire: &  & moreIR & NLHS &  &  \\
1443$+$0931 &  &  & $26000\pm^{1600}_{890}$ & $8.00\pm^{}_{}$ &  & M9 & DA:ire &  &  & NLHS &  &  \\
1448$+$0240 & $14728\pm247$ & $7.41\pm0.06$ & $13000\pm^{1030}_{130}$ & $8.00\pm^{0.52}_{0.19}$ & L4 & L6 & DAire & moreIR & moreIR & DA &  &  \\
1448$+$0713 & $12250\pm256$ & $7.87\pm0.09$ & $11000\pm^{1130}_{440}$ & $7.75\pm^{0.55}_{0.62}$ & M8 & M9 & DAire &  &  & DA &  &  \\
1450$+$0000 & $17106\pm175$ & $7.93\pm0.05$ &  &  & L8: &  & DAire: & moreIR &  & DA &  &  \\
1500$+$0642 &  &  & $26000\pm^{110}_{2000}$ & $8.00\pm^{}_{}$ &  & L1: & DA:ire &  &  & NLHS &  &  \\
1510$+$0409 &  &  & $28000\pm^{560}_{2000}$ & $8.00\pm^{}_{}$ &  & M8 & DA:ire &  &  & NLHS &  &  \\
1514$+$0744 &  &  & $10000\pm^{60}_{230}$ & $8.50\pm^{0.30}_{0.42}$ &  & L5: & DA:ire: &  & bf & MWD &  &  \\
1519$+$0715 &  &  & $26000\pm^{390}_{80}$ & $8.00\pm^{}_{}$ &  & M7 & DA:ire &  & moreIR & NLHS &  &  \\
1525$+$0958 &  &  & $28000\pm^{2120}_{280}$ & $8.00\pm^{}_{}$ &  & M9 & DA:ire &  &  & NLHS &  &  \\
1539$+$2706 &  &  & $28000\pm^{330}_{670}$ & $8.00\pm^{}_{}$ &  & M9 & DA:ire & bf &  & DA &  &  \\
1541$+$0417 &  &  & $24000\pm^{230}_{240}$ & $8.00\pm^{}_{}$ &  & L2 & DA:ire &  &  & NLHS &  &  \\
1543$+$0012 &  &  & $17000\pm^{2110}_{2920}$ & $9.50\pm^{0.81}_{}$ &  & L8: & DA:ire: &  & bf & NLHS &  &  \\
1557$+$0916 & $22811\pm734$ & $7.75\pm0.11$ & $18000\pm^{1050}_{130}$ & $7.75\pm^{0.83}_{0.41}$ & L1: & L5 & DAire &  &  & DA &  &  \\
1609$+$2905 & $8773\pm91$ & $8.26\pm0.14$ & $9000\pm^{70}_{140}$ & $7.75\pm^{0.41}_{0.31}$ & $\geq$L8 & L8 & DAire &  &  & DA &  &  \\
1614$+$2616 &  &  & $24000\pm^{150}_{220}$ & $8.00\pm^{}_{}$ &  & L5 & DA:ire &  &  & NLHS &  &  \\
1619$+$2407 &  &  & $9000\pm^{180}_{210}$ & $8.75\pm^{1.02}_{0.75}$ &  & $\geq$L8: & DA:ire: &  & bf & NLHS &  &  \\
1619$+$2533 & $25595\pm1139$ & $7.21\pm0.15$ & $18000\pm^{2080}_{1390}$ & $9.50\pm^{1.89}_{}$ & M5 & L6: & DAire &  & bf & DA &  &  \\
1625$+$3026 & $72136\pm4550$ & $7.77\pm0.19$ & $40000\pm^{11380}_{4690}$ & $8.00\pm^{}_{}$ & M5 & M6 & DAire &  &  & DA &  &  \\
1637$+$3113 &  &  & $26000\pm^{400}_{270}$ & $8.00\pm^{}_{}$ &  & M9 & DA:ire &  &  & NLHS &  &  \\
1640$+$3117 &  &  & $26000\pm^{3230}_{2780}$ & $8.00\pm^{}_{}$ &  & M7 & DA:ire &  &  & NLHS &  &  \\
1645$+$3109 &  &  & $17000\pm^{2510}_{1970}$ & $9.50\pm^{1.24}_{}$ &  & $\geq$L8: & DA:ire: &  & bf: & NLHS &  &  \\
2038$+$0109 &  &  & $20000\pm^{2100}_{620}$ & $9.00\pm^{1.23}_{0.50}$ &  & $\geq$L8: & DA:ire: &  & bf: & NLHS &  &  \\
2117$-$0006 &  &  & $8000\pm^{20}_{10}$ & $8.00\pm^{}_{}$ &  & $\geq$L8: & DA:ire: &  & bf: & NLHS &  &  \\
2118$+$0028 &  &  & $15000\pm^{230}_{40}$ & $8.25\pm^{0.17}_{0.21}$ &  & $\geq$L8: & DA:ire: & No$-$Xs & bf:,resolved,bg: & DA &  &  \\
2220$-$0041 & $7467\pm21$ & $7.66\pm0.12$ & $8000\pm^{20}_{30}$ & $8.00\pm^{}_{}$ & $\geq$L8 & $\geq$L8 & DAire: & resolved & resolved,bg: & DA &  &  \\
2225$+$0016 & $11045\pm235$ & $8.44\pm0.16$ & $9000\pm^{20}_{20}$ & $7.00\pm^{0.26}_{0.08}$ & L6 & L5 & DAire: & resolved: & resolved:,bg: & DA &  &  \\
2255$-$0015 &  &  & $17000\pm^{170}_{620}$ & $9.00\pm^{0.25}_{0.37}$ &  & L4: & DA:ire: &  & bf:,resolved,bg: & WDMS &  &  \\
2331$+$1342 & $11173\pm284$ & $8.79\pm0.15$ & $15000\pm^{1230}_{1600}$ & $9.50\pm^{0.73}_{}$ & L6 & L8 & DAire &  &  & DA &  &  \\

%% file: wdms.tex
0032$+$0739 & $21045\pm249$ & $7.42\pm0.05$ & S & M6 & DAire: & $21045\pm371$ & $7.43\pm0.07$ & M2 \\
0039$-$0030 & $12392\pm544$ & $7.35\pm0.16$ & SP & M9 & DAire: & $12392\pm737$ & $7.39\pm0.22$ & $-$1 \\
0207$+$0702 & $10073\pm77$ & $8.32\pm0.09$ & SP & L7 & DAire: & $18756\pm502$ & $7.42\pm0.10$ & $-$1 \\
0236$-$0103 & $20566\pm498$ & $7.66\pm0.09$ & SP & M6 & DAire & $21289\pm723$ & $7.83\pm0.13$ & M8 \\
0748$+$2058 & $86726\pm7732$ & $7.15\pm0.27$ & SP & M3 & DAire & $87731\pm10728$ & $7.17\pm0.38$ & M2 \\
0847$+$2831 & $12828\pm930$ & $7.76\pm0.20$ & SP & M6 & DAire & $13588\pm923$ & $7.76\pm0.18$ & M6 \\
0851$+$0330 & $20000\pm^{2740}_{2090}$ & $9.50\pm^{0.00}_{1.55}$ & P & L6 & DA:ire & $22550\pm520$ & $7.40\pm0.08$ & M6 \\
1037$+$0139 & $11433\pm208$ & $8.38\pm0.13$ & SP & L5: & DAire & $16717\pm602$ & $7.78\pm0.14$ & M5 \\
1106$+$0737 & $28000\pm^{2060}_{3800}$ & $8.00$ & P & M5 & DA:ire & $36154\pm540$ & $7.74\pm0.09$ & M5 \\
1141$+$0420 & $11835\pm818$ & $7.37\pm0.22$ & SP & M7 & DAire & $12110\pm780$ & $7.39\pm0.26$ & $-$1 \\
1329$+$1230 & $13432\pm272$ & $7.42\pm0.05$ & SP & M7 & DAire & $12250\pm1036$ & $7.47\pm0.20$ & $-$1 \\
1341$+$0056 & $18330\pm436$ & $7.99\pm0.10$ & SP & M7 & DAire & $18330\pm608$ & $8.01\pm0.13$ & $-$1 \\
1352$+$0910 & $36154\pm722$ & $7.49\pm0.11$ & SP & M5 & DAire & $35331\pm919$ & $7.31\pm0.16$ & $-$1 \\
1415$+$0117 & $32000\pm^{2230}_{4360}$ & $8.00$ & P & M6 & DA:ire & $73816\pm4981$ & $8.43\pm0.18$ & M0 \\
1448$+$0713 & $12250\pm256$ & $7.87\pm0.09$ & SP & M8 & DAire & $12110\pm264$ & $7.84\pm0.13$ & M6 \\
1539$+$2706 & $28000\pm^{670}_{330}$ & $8.00$ & P & M9 & DA:ire & $36572\pm447$ & $7.31\pm0.07$ & $-$1 \\
1619$+$2533 & $25595\pm1139$ & $7.21\pm0.15$ & SP & M5 & DAire & $25891\pm1447$ & $7.30\pm0.20$ & $-$1 \\
2255$-$0015 & $17000\pm^{620}_{170}$ & $9.00\pm^{0.37}_{0.25}$ & P & L4: & DA:ire: & $22550\pm724$ & $7.88\pm0.12$ & M6 \\
2331$+$1342 & $11173\pm284$ & $8.79\pm0.15$ & SP & L6 & DAire & $18756\pm1020$ & $7.96\pm0.20$ & $-$1 \\